\documentclass[aps,prx,reprint,groupedaddress,superscriptaddress,floatfix,longbibliography]{revtex4-2}

\usepackage[T1]{fontenc}
\usepackage{newtxtext}
\usepackage{newtxmath}

\usepackage{amsmath}
\usepackage{mathtools}
\usepackage{amsfonts}
\usepackage{bbold}

\usepackage{color}
\usepackage{xcolor}
\usepackage[colorlinks,citecolor=blue,linkcolor=blue,urlcolor=blue]{hyperref}

\usepackage{graphicx}
\usepackage[all]{hypcap}

\usepackage{enumitem}

\usepackage{tabularx}
\usepackage{makecell}

\begin{document}

\title{Benchmarking Quantum Optimization for the Maximum-Cut Problem\\on a Superconducting Quantum Computer}

\author{Maxime Dupont}
\email[]{mdupont@rigetti.com}
\affiliation{Rigetti Computing, 775 Heinz Avenue, Berkeley, California 94710, USA}

\author{Bhuvanesh Sundar}
\affiliation{Rigetti Computing, 775 Heinz Avenue, Berkeley, California 94710, USA}

\author{Bram Evert}
\affiliation{Rigetti Computing, 775 Heinz Avenue, Berkeley, California 94710, USA}

\author{David E. Bernal Neira}
\affiliation{Davidson School of Chemical Engineering, Purdue University, West Lafayette, Indiana 47907, USA}
\affiliation{USRA Research Institute for Advanced Computer Science, Mountain View, California 94043, USA}
\affiliation{Quantum AI Laboratory, NASA Ames Research Center, Moffett Field, California 94035, USA}

\author{Zedong Peng}
\affiliation{Davidson School of Chemical Engineering, Purdue University, West Lafayette, Indiana 47907, USA}

\author{Stephen Jeffrey}
\affiliation{Rigetti Computing, 775 Heinz Avenue, Berkeley, California 94710, USA}

\author{Mark J. Hodson}
\affiliation{Rigetti Computing, 775 Heinz Avenue, Berkeley, California 94710, USA}

\begin{abstract}
    Achieving high-quality solutions faster than classical solvers on computationally hard problems is a challenge for quantum optimization to deliver utility. Using a superconducting quantum computer, we experimentally investigate the performance of a hybrid quantum-classical algorithm inspired by semidefinite programming approaches for solving the maximum-cut problem on 3-regular graphs up to several thousand variables. We leverage the structure of the input problems to address sizes beyond what current quantum machines can naively handle. We attain an average approximation ratio of 99\% over a random ensemble of thousands of problem instances. We benchmark the quantum solver against similarly high-performing classical heuristics, including the Gurobi optimizer, simulated annealing, and the Burer-Monteiro algorithm. A run-time analysis shows that the quantum solver on large-scale problems is competitive against Gurobi but short of others on a projected $100$-qubit quantum computer. We explore multiple leads to close the gap and discuss prospects for a practical quantum speedup.
\end{abstract}

\maketitle

\let\oldaddcontentsline\addcontentsline
\renewcommand{\addcontentsline}[3]{}
\section{Introduction}
\let\addcontentsline\oldaddcontentsline

Combinatorial optimization seeks a valid and optimal solution to a problem defined over discrete variables. It is a computationally demanding task in many applications, posing challenges in science and everyday situations~\cite{Kochenberger2014}. Using quantum machines to improve the performance of combinatorial optimization solvers is an appealing endeavor and is currently subject to intense research. The prospect of such performance improvements has been explored theoretically for over two decades via the development of varied quantum algorithms~\cite{Farhi2001,Farhi2014,RevModPhys.90.015002,BLEKOS20241}. 

However, executing these algorithms at scale on a large ensemble of problems to thoroughly benchmark them against classical solvers, and open avenues to outperform the latter, has faced significant challenges. First, the relatively small number of quantum bits (qubits) and high error rates in logical quantum operations of contemporary quantum hardware diminish the practical implementation of such algorithms to solve relevant problems. Nevertheless, experimental realizations are catching up, enabling the codesign of more efficient and performant quantum solvers, unleashing a quantum potential for combinatorial optimization. For instance, a scaling advantage has been observed for solving maximum-independent-set problems up to $289$ variables using individual atoms trapped in optical tweezers serving as qubits~\cite{Ebadi2022,PhysRevResearch.5.043277}. Quantum annealers have also been found to deliver a scaling advantage for solving spin-glass problems with up to 5,000 variables~\cite{King2023}.  Second, although a scaling speedup paves the way for a possible asymptotic quantum advantage in run-time, achieving a practical run-time speedup against state-of-the-art classical solvers at relevant scales requires additional theoretical and technological developments.

Here, we use a gate-based superconducting quantum computer featuring transmon qubits~\cite{PhysRevA.76.042319,PhysRevLett.111.080502} to extensively benchmark quantum optimization algorithms for solving thousands of maximum-cut problems with each having up to several thousand variables. The maximum-cut problem seeks the bipartition of a graph such that the number of edges connecting vertices from the two distinct groups, which we call the cut number, is maximized. Devising an algorithm that can guarantee a cut number at least $94.1\%$ of the optimal one on an arbitrary graph belongs to the computational complexity class NP-hard~\cite{sipser13,10.1145/502090.502098}. To date, the algorithm guaranteeing the highest cut number, at $87.9\%$ of the optimal one, comes from Goemans and Williamson~\cite{Goemans1995}. In practice, heuristic solvers with limited or no guarantees are used, as they usually yield solutions above those theoretical thresholds and often faster. Therefore, the main competitors for a practical utility are commercial solvers such as the Gurobi optimizer~\cite{gurobi}, state-of-the-art heuristics such as simulated annealing~\cite{doi:10.1126/science.220.4598.671}, and methods tailored toward specific applications, such as the Burer-Monteiro (BM) approach for the maximum-cut problem~\cite{doi:10.1137/S1052623400382467,Burer2003}. Quantum computers have been used to solve maximum-cut problems~\cite{Harrigan2021,PhysRevX.13.041057,PhysRevResearch.6.013223,Sciorilli2024}, but their results have lagged behind classical solvers in the performance-versus-absolute-run-time space to warrant a direct benchmark. Here, we pave the way for a practical quantum speedup by achieving an experimental cut number at $99\%$ of optimal up to several thousand variables in tens of seconds on a class of problems.

\let\oldaddcontentsline\addcontentsline
\renewcommand{\addcontentsline}[3]{}
\section{The Maximum-Cut Problem on a Quantum Computer}
\let\addcontentsline\oldaddcontentsline

\begin{figure*}[!ht]
    \centering
    \includegraphics[width=1\textwidth]{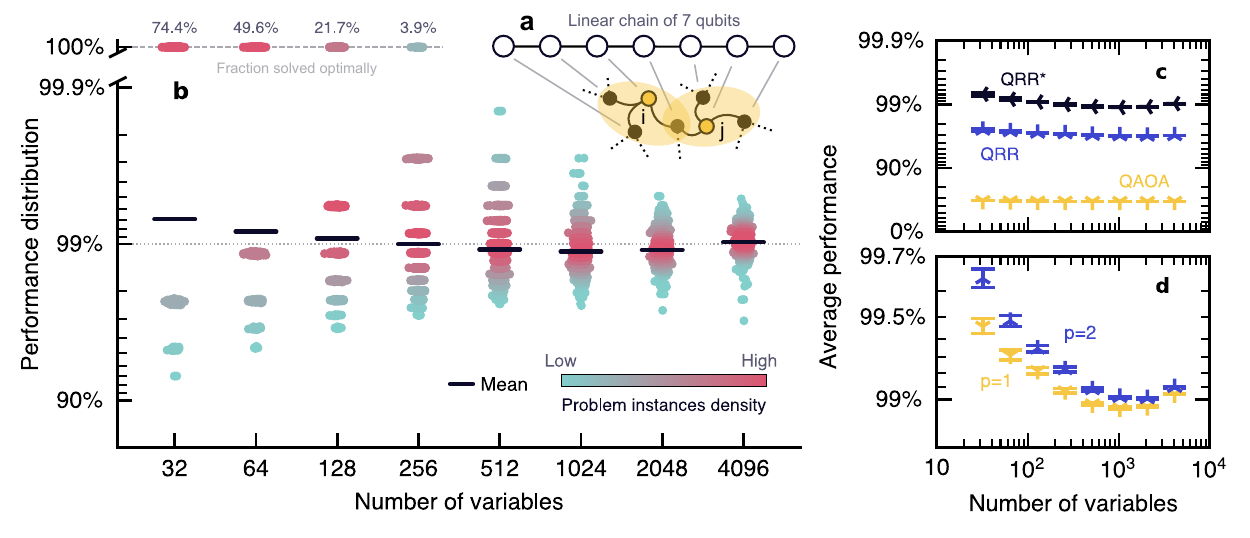}
    \caption{(a) The embedding of a light-cone-induced seven-vertex subgraph by the quantum approximate optimization algorithm (QAOA) at $p=1$ for a $3$-regular graph onto a linear chain of seven qubits. (b) Experimental performance distribution of the greedy-enhanced quantum relax-and-round (QRR*) solver with $p=1$ over $1,000$ randomly generated problem instances for various numbers of variables. (c) The average experimental performance as a function of the problem size for the QAOA, the quantum relax-and-round (QRR), and the QRR* solvers with $p=1$. (d) The simulated average performance of the QRR* solver as a function of the number of variables for $p=1$ and $p=2$ layers. The error bars indicate one standard deviation.}
    \label{fig:performance}
\end{figure*}

One seeks to maximize the cut number given by the objective function
\begin{equation}
    C(\boldsymbol{z})=\frac{1}{4}\sum\nolimits_{i,j=1}^N\mathsf{W}_{ij}\bigl(1-z_iz_j\bigr),
    \label{eq:maxcut_obj_func}
\end{equation}
defined over binary variables $z_i=\pm 1$ assigned to each of the $N$ graph vertices and labeling their bipartition assignment. The graph structure is encoded in its adjacency matrix $\mathsf{W}$, where $\mathsf{W}_{ij}=1$ if vertices $i$ and $j$ are connected by an edge and zero otherwise. Denoting $\boldsymbol{z}_{\textbf{opt}}$ as the optimal solution, $\alpha=C(\boldsymbol{z})/C(\boldsymbol{z}_{\textbf{opt}})$, known as the approximation ratio, measures the quality of a solution $\boldsymbol{z}$. 

A quantum algorithm will typically formulate an objective function, such as Eq.~\eqref{eq:maxcut_obj_func}, as a quantum many-body Hamiltonian by mapping each of the variables onto a qubit~\cite{lucas2014}. Quantum physical principles such as adiabatic time evolution~\cite{Farhi2001,RevModPhys.90.015002} or variational search~\cite{Farhi2014,BLEKOS20241} are then employed to find its ground state. In the context of gate-based quantum computing, the quantum approximate optimization algorithm (QAOA)~\cite{Farhi2014,BLEKOS20241}, based on such a variational search and parametrized by its number of layers $p$, is one of the most promising~\cite{Otterbach2017,Pagano2020,Harrigan2021}. However, even on an ideal quantum computer, the QAOA circuit needs a large depth to outperform classical approximation algorithms for several classes of problems~\cite{Wurtz2021,Basso2022},and no practical quantum speedup over state-of-the-art heuristics solvers has been suggested with the QAOA. On actual quantum hardware, the qubit count and error rate make the QAOA uncompetitive in practice~\cite{Harrigan2021,Maciejewski2023,Pelofske2024}.

To overcome those limitations, we embed the QAOA into a relax-and-round approach~\cite{Dupont2024}, which we complement with a guided greedy local search. We solve the maximum-cut problem on random $3$-regular graphs with thousands of nodes. We achieve an average approximation ratio of $99\%$ across thousands of problem instances. For reference, on such graphs, a perfect implementation of the QAOA at $p=1$ guarantees a cut of only $69.2\%$ of the optimal one~\cite{Farhi2014,3reggraphs}. Investigating $3$-regular graphs presents several advantages. First, they are mathematically proven hard to solve~\cite{3reggraphs}. Second, they are sparse, where each vertex is connected randomly to three others. Due to their topology and the relax-and-round strategy utilizing expectation values, we can leverage a light-cone technique to solve problem sizes with thousands of variables, exceeding what today's quantum machines can naively handle. Third, near-optimal variational QAOA parameters are tabulated~\cite{PhysRevA.103.042612,Wurtz2021}, significantly speeding up their solving in search of a quantum speedup. Finally, $3$-regular graphs are a standard benchmark problem in quantum optimization since the seminal QAOA work~\cite{Farhi2014}.

The QAOA with $p$ layers generates the quantum state
\begin{equation}
    \bigl\vert\Psi\bigr\rangle_p=\left[\prod\nolimits_{\ell=1}^pe^{-i\beta_\ell\sum_{j=1}^N\hat{X}_j}e^{i\gamma_\ell\hat{C}}\right]\hat{H}^{\otimes N}\vert{0}\rangle^{\otimes N},
    \label{eq:qaoa}
\end{equation}
where $\hat{H}$ is the one-qubit Hadamard gate, $\hat{X}_i$ is the Pauli $X$-operator on qubit $i$, and $\langle\boldsymbol{z}\vert\hat{C}\vert\boldsymbol{z'}\rangle=C(\boldsymbol{z})\delta_{\boldsymbol{z}\boldsymbol{z'}}$ is a diagonal unitary encoding the objective function. The parameters $\gamma_\ell$ and $\beta_\ell$ are variational and optimal when maximizing the expectation value of the objective function. Instead of considering sampled solutions $\{\boldsymbol{z}\}$ from the quantum computer directly, the quantum relax-and-round (QRR) algorithm~\cite{Dupont2024} works with the expectation value of two-body correlations $\langle\hat{Z}_i\hat{Z}_j\rangle_p$, where $\hat{Z}_i$ is the Pauli $Z$-operator on qubit $i$. Working with expectation values opens the door to a wide range of error-mitigation techniques~\cite{RevModPhys.95.045005} to be applied to optimization problems. From the point of view of qubit $i$, a $p$-layer QAOA circuit only induces quantum coherence with its $p$-nearest neighbors according to the graph topology. This collection of coherent qubits around $i$ is known as its light cone. To evaluate a two-body expectation value between two qubits $i$ and $j$, one only needs to consider the quantum circuit running on the subset of qubits made of the union of their respective light cones~\cite{Farhi2014}. The correlation of interest is nonzero only if the light cones intersect, leading to only $O(N)$ nontrivial correlations when $p\ll\ln N$ for $N$-variable $3$-regular graphs. Moreover, the sparse nature of $3$-regular graphs is such that subcircuits involve at most $1+6(2^p-1)$ qubits, independently of $N$. Therefore, one can trade a single QAOA execution on $N$ qubits for $O(N)$ smaller executions and access graph-problem sizes much larger than current hardware would normally support (see App.~\ref{sec:light_cone_sm}).

\let\oldaddcontentsline\addcontentsline
\renewcommand{\addcontentsline}[3]{}
\section{Quantum Optimization}
\let\addcontentsline\oldaddcontentsline

For the QAOA with $p=1$ on $3$-regular graphs, one needs seven qubits at most (Fig.~\ref{fig:performance}a). We use transmon qubits~\cite{PhysRevA.76.042319,PhysRevLett.111.080502} arranged linearly. We compile the original quantum logical operations of Eq.~\eqref{eq:qaoa} at the individual qubit level through single-qubit rotation gates $\texttt{Rz}$ (virtual, error-free) and $\texttt{Rx}$ (average error rate of $0.8\%$). We realize entangling operations, mediated by a tunable coupler~\cite{PhysRevApplied.15.064063,PhysRevApplied.16.024050} between nearest-neighboring qubits, in the form of $\texttt{ISWAP}$ gates (average error rate of $1.9\%$). The original quantum logical operations of Eq.~\eqref{eq:qaoa} between non-neighboring qubits are implemented via an extensive swap network~\cite{OGorman2019}. We employ readout-error mitigation~\cite{nachman_unfolding_2020}. We use fixed values for the variational parameters $\gamma_\ell$ and $\beta_\ell$, expected to produce a decent average performance on any $3$-regular graph, albeit suboptimal on a case-by-case basis~\cite{PhysRevA.103.042612,Wurtz2021}. It is a trade-off eliminating the need for variational optimization of the quantum circuit parameters~\cite{Bernal2024}. We have found that collecting $n\gtrsim 5\times 10^3$ bit strings from the quantum computer for each observable yields the best performance. In practice, we use $n=10^4$.

Once two-body correlations are collected for a given problem instance, the QRR algorithm requires the construction of a correlation matrix with entries $\mathsf{Z}_{ij}=(\delta_{ij} - 1)\langle\hat{Z}_i\hat{Z}_j\rangle_p$. A solution to the maximum-cut problem is then obtained by sign rounding one of its leading eigenvectors. In the guided greedy local search that we use to augment the relax-and-round approach, instead of sign rounding the selected eigenvector directly, we interpret the magnitude of its entries as a confidence measure of their sign. We leverage this information to implement a greedy local step that will revisit variables with low-confidence values. On such values, a sign flip is attempted and retained as part of the solution if it improves the cut number, as described in App.~\ref{sec:grdy_enh_qrr_alg_sm}. We refer to this variant of the solver as the greedy-enhanced quantum relax-and-round (QRR*) solver. Its performance distribution across thousands of problem instances from sizes $N=32$ to $N=4,096$, and based on the QAOA with $p=1$, is displayed in Fig.~\ref{fig:performance}b. We achieve an average approximation ratio of $99\%$. For problem sizes up to $N\leq 256$, we find the optimal solution for a finite fraction of the problems. Although the light-cone technique does not enable us to sample solutions from the underlying QAOA directly, we can estimate its expected performance by computing the expectation value of the objective function $\langle\hat{C}\rangle_p$. We report the corresponding average approximation ratio over all problem instances in Fig.~\ref{fig:performance}c. We find that the QRR algorithm enhances the average performance, relative to the QAOA, by an order of magnitude using the same underlying quantum resources, ($1-\alpha\simeq 30\%\to 3\%$). We observe that augmenting the QRR with a local greedy search (the QRR*) improves the average performance by a factor of $3$, ($1-\alpha\simeq 3\%\to 1\%$).

At $p=1$, the quantum algorithm gives the same result as a classical relax-and-round approach performed on the adjacency matrix of the graph problem $\mathsf{W}$, rather than the correlation matrix $\mathsf{Z}$. Indeed, one can show that they approximately share the same eigenvectors (and presumably the same leading eigenvectors), resulting in the same cut number (Appendix~\ref{app:eq_classical_quantum_rr}). By increasing the number of QAOA layers $p$, the classical and quantum relax-and-round differentiate, with the QRR converging asymptotically with $p$ to the optimal solution. In Fig.~\ref{fig:performance}d, we show the average performance of the QRR* algorithm from classically emulated quantum circuits at $p=1$ and $p=2$, displaying a performance increase with $p$. The apparent approximation-ratio increase at large $N$ is because we likely underestimate $C(\boldsymbol{z}_{\textbf{opt}})$ by not finding the optimal solution $\boldsymbol{z}_{\textbf{opt}}$; therefore the relative performance of the quantum solver with respect to the best classical solution that we have increases. The gap reduction in performance between the two is a finite-size effect similarly observed in the QAOA (App.~\ref{sec:comparison_solvers}). Nonetheless, the QRR* provides a very high-performance baseline, matching that of a classical algorithm (even with minimal quantum resources, $p=1$), and also a path toward improving the solution quality by increasing $p$.

\let\oldaddcontentsline\addcontentsline
\renewcommand{\addcontentsline}[3]{}
\section{Quantum Speedup}
\let\addcontentsline\oldaddcontentsline

The relatively high performance of the QRR* makes its solution meaningful for the maximum-cut problem on $3$-regular graphs. We turn our attention to a benchmark against state-of-the-art classical solvers and focus on three main competitors (as well as others in App.~\ref{sec:comparison_solvers}): simulated annealing (SA)~\cite{doi:10.1126/science.220.4598.671}, Gurobi~\cite{gurobi}, and the BM algorithm~\cite{doi:10.1137/S1052623400382467,Burer2003}. The first two are general-purpose solvers for combinatorial optimization, while BM is tailored for the maximum-cut problem and is currently the best algorithm for solving $3$-regular graphs~\cite{Lykov2023}. Whereas the QRR* is also not restricted to the maximum-cut problem, skipping the variational search by using fixed parameters~\cite{PhysRevA.103.042612,Wurtz2021} makes it special purpose. Neither Gurobi nor the BM have hyperparameters that need to be optimized for the solver to work best. On the other hand, SA requires a temperature schedule. Given that the quantum solver has no hyperparameters, we rely on a constant-time heuristic setting of the temperature schedule (App.~\ref{sec:sa_sm}).

A classical solver typically performs better given more run-time. The run-time is the only parameter given to solvers such as Gurobi and the BM, while the SA run-time is directly proportional to the number of sweeps (set by the user). The control parameter of the quantum algorithm is the number of layers $p$. Because imperfect quantum machines limit the practical size of quantum circuits that can be executed, one is constrained to a low-$p$ regime. All of the classical solvers tested here are randomized, thus returning a different solution for each run. We define $P(C(\boldsymbol{z}_{\textrm{cl}})\geq C(\boldsymbol{z}_{\textrm{QRR*}}))$ to be the probability for a classical solver with run-time $t$ to return a solution $\boldsymbol{z}_{\textrm{cl}}$ at
least matching that of the quantum algorithm $\boldsymbol{z}_{\textrm{QRR*}}$, so that
\begin{equation}
    t^* = t\bigr/P\Bigl(C(\boldsymbol{z}_{\textrm{cl}})\geq C(\boldsymbol{z}_{\textrm{QRR*}})\Bigr),
\end{equation}
is the average time for state-of-the-art classical solvers to match the performance of the QRR* algorithm. $P$ is intrinsically related to $t$ given that a longer run-time is more likely to return a better solution and, conversely, a smaller run-time $t$ is more likely to result in a lower value of $P$. Therefore, we seek the optimal time $t^*_\textrm{opt}$ for the classical solver to find a solution matching the quality of the QRR* solution, which concurrently minimizes $t$ and maximizes $P$. We compute on a per-problem-instance basis $t^*_\textrm{opt}$ for each of the three classical solvers (App.~\ref{sec:opt_run-time_cl_sm}).

We compare $t^*$ with the run-time for the QRR*, which we estimate as a function of the problem size $N$, the number of QAOA layers $p$, the total number of qubits available $M$ on a device, and the number of bit strings sampled $n$ (Appendix~\ref{app:run-time_qrr_solver}). There are four components to consider: executing the quantum circuits on the quantum computer, building the correlation matrix, finding its leading eigenvectors, and doing the greedy local search. The last three steps are purely classical. The execution time is based on the use of the light-cone technique on an $M$-qubit quantum computer. Correlation-induced subcircuits by the QAOA with $p$ layers on an $N$-variable $3$-regular graph problem are executed simultaneously, which may require multiple sequential runs if $M$ is too small to encapsulate them all at once (Appendix~\ref{app:run-time_qrr_solver}). The execution time of the subcircuits is based on their compilation onto a linear-chain topology of qubits using a worst-case-scenario swap network~\cite{OGorman2019} with hardware-native gates ($t_\texttt{Rx}=40$~ns and $t_\texttt{ISWAP}=122$~ns), measurement, and reset times ($t_\texttt{m+r}=6~\mu$s), all multiplied by $n=5\times 10^4$ (shot count), see Appendices~\ref{app:run-time_qrr_solver} and~\ref{app:eq_classical_quantum_rr}.

\begin{figure}[!t]
    \centering
    \includegraphics[width=1\columnwidth]{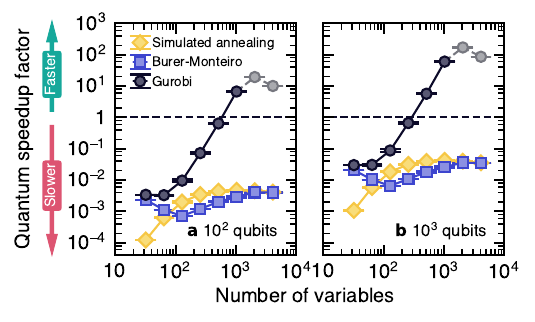}
    \caption{The optimal run-time for various classical solvers to match the performance of the QRR* solver with $p=1$, normalized by the run-time of the QRR* (Appendix~\ref{app:run-time_qrr_solver}), as a function of the number of variables. The Gurobi data points for the two largest sizes are lower bounds where no solutions matching that of the QRR* have been found in $600$ seconds for most of the problem instances. The quantum run-time component of QRR* is based on $t_\texttt{Rx}=40$ ns, $t_\texttt{ISWAP}=122$ ns, $t_\texttt{m+r}=6~\mu$s, and $n=5\times 10^4$ shots (Appendix~\ref{app:run-time_qrr_solver}). (a) Utilizing a quantum computer with $10^2$ qubits. (b) Utilizing a quantum computer with $10^3$ qubits. Each data point has been averaged over $1,000$ randomly generated problem instances for each number of variables. The error bars indicate one standard deviation.}
    \label{fig:runtime_to_match_qrr}
\end{figure}

We consider two scenarios in which a quantum computer has $M=10^2$ (Fig.~\ref{fig:runtime_to_match_qrr}a) and $10^3$ (Fig.~\ref{fig:runtime_to_match_qrr}b) qubits, enabling the parallel execution of more QAOA subcircuits. This range corresponds to what is already available and should become available within the next few years, respectively. We show the optimal time $t^*_\textrm{opt}$ for classical solvers to match the experimental performance of the QRR* with $p=1$, normalized by the run-time of the QRR* itself. In the case of Gurobi, this quantum speedup factor exceeds one for the largest problem sizes, meaning that the QRR* is faster. In fact, for the two largest sizes, we only provide a lower bound corresponding to a $10$-min run-time limit for Gurobi (grayed-out data symbols in Fig.~\ref{fig:runtime_to_match_qrr}), which has failed to match the performance of the QRR* for most of the $1,000$ problem instances (see App.~\ref{sec:opt_run-time_cl_sm}). Regardless, Gurobi provides additional valuable information, such as a bound on how close a solution is to optimality, and it can guarantee the optimality of a solution. The SA and BM solvers are faster than the QRR* at a fixed number of QAOA layers, with the gap converging to a finite value as the problem sizes increase. For SA, we have assumed knowledge of the optimal number of sweeps, which would otherwise need to be searched. Consequently, the quantum speedup factor for SA should be considered a lower bound.

The first path for improving the competitiveness of the quantum solver is to reduce its run-time. The classical steps could be optimized using standard software-engineering techniques and/or algorithmic improvements, and one could imagine the control system of the device estimating expectation values on the fly rather than collecting bit strings that require postprocessing. The execution time on the quantum computer could also be reduced through a more efficient compilation of the quantum circuit onto the hardware-native topology of the qubits, shorter logical gates, and the development of new gates, reducing the compilation overhead. The second path for improving competitiveness is to increase the performance of the QRR*. While additional classical processing is an option, increasing the number of layers $p$ in quantum algorithms remains the cornerstone of more performant quantum algorithms.

\let\oldaddcontentsline\addcontentsline
\renewcommand{\addcontentsline}[3]{}
\section{Towards a Quantum Optimization Utility}
\let\addcontentsline\oldaddcontentsline

\begin{figure}[!t]
    \centering
    \includegraphics[width=1\columnwidth]{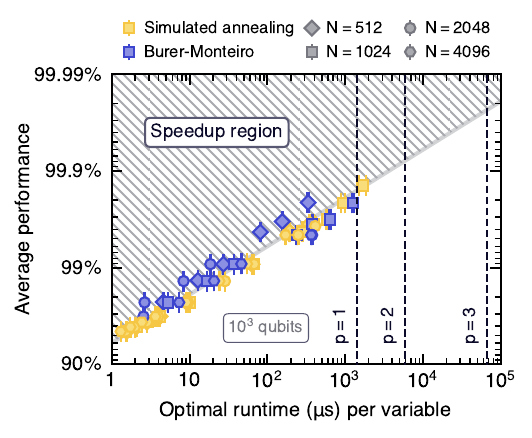}
    \caption{The optimal time $t^*_\textrm{opt}$, in seconds, normalized by the number of variables, to achieve a target performance $\alpha$ on a problem instance by problem-instance basis. We consider simulated annealing and the Burer-Monteiro solvers on large problem sizes. The solid line $1-\alpha\sim(t^*_\textrm{opt}/N)^{-1/2}$ is a guide to the eye. The vertical dashed lines indicate the estimated run-time of the QRR* algorithm with $p=1$, $2$, and $3$ layers, using $10^3$ qubits with $t_\texttt{Rx}=20$ ns, $t_\texttt{ISWAP}=50$ ns, $t_\texttt{m+r}=6~\mu$s, and a $p$-dependent optimal shot count (Appendix~\ref{app:number_of_bitstrings}). Each data point has been averaged over $1,000$ randomly generated problem instances for each number of variables. The error bars indicate one standard deviation.}
    \label{fig:runtime_vs_performance}
\end{figure}

We now establish run-time and performance goals to assess the potential quantum utility of the QRR* for the maximum-cut problem on random $3$-regular graphs. Using the same methodology as developed for Fig.~\ref{fig:runtime_to_match_qrr}, we estimate the optimal time $t^*_\textrm{opt}$ for the classical SA and BM solvers to achieve fictitious performance targets.

With the duration of a step proportional to the number of edges in the input problem, the rescaled variable $k=t^*_\textrm{opt}/N$ relates directly to the number of steps for solving $3$-regular graphs. We empirically observe in Fig.~\ref{fig:runtime_vs_performance} that in the regime considered, the distance to the optimal solution is reduced asymptotically as approximately $k^{-1/2}$ for both SA and the BM, independently of $N$. This power law in the context of SA is reminiscent of a Kibble-Zurek-type mechanism~\cite{Kibble1976,Zurek1985}, albeit different from the idealized statistical-physics scenario~\cite{PhysRevB.89.054307,PhysRevLett.114.147203,PhysRevA.86.052334,PhysRevE.81.050101}.

In the regime $p\ll\ln N$, the run-time of the QRR* is proportional to $N$ and moves toward $O(N^2)$ for $p\gg\ln N$, as the QAOA starts to generate global quantum coherence across the whole graph~\cite{Farhi2020}. Hence, in the low-$p$ regime, the QRR* has a linear algorithmic complexity with $N$ for a fixed performance, like its classical competitors. We display the run-time of the QRR* rescaled by $N$ in Fig.~\ref{fig:runtime_vs_performance}, where an $M=10^3$ qubit quantum computer is assumed with $t_\texttt{Rx}=20$ ns and $t_\texttt{ISWAP}=50$ ns. We observe that increasing $p$ substantially increases the run-time per variable because the light-cone technique yields circuits with sizes growing exponentially with $p$. Consequently, exponentially fewer such circuits can fit in parallel on $M$ qubits. The $p$-dependent run-time of the QRR*, together with the scaling of classical solvers shown in Fig.~\ref{fig:runtime_vs_performance}, set performance targets for the QRR* at fixed $p$ to achieve a quantum advantage for the maximum-cut problem on $3$-regular graphs. For example, the QRR* algorithm with $p=1$ achieves $99\%$ approximation ratio (Fig.~\ref{fig:performance}a). Under the assumptions of Fig.~\ref{fig:runtime_vs_performance}, QRR* should yield a fivefold average performance increase ($1-\alpha=1\%\to0.2\%$) to be competitive in run-time with the SA and BM algorithms. Moving forward, while the exponential run-time increase with $p$ seems daunting, it might be counterbalanced by an exponential convergence of the quantum algorithm with $p$, as has been observed for the QAOA~\cite{PhysRevX.10.021067}, where $1-\alpha\sim e^{-p}$, albeit on problem sizes up to $2$ orders of magnitude smaller than what we consider here, with the fate of larger problems remaining an open question.

\let\oldaddcontentsline\addcontentsline
\renewcommand{\addcontentsline}[3]{}
\section{Conclusions and Outlook}
\let\addcontentsline\oldaddcontentsline

Our results demonstrate the potential for quantum optimization to deliver high-quality solutions at speed on large-scale problems, comparable with state-of-the-art classical solvers. By leveraging the sparsity of 3-regular graphs, the light-cone technique has made it possible to address problem sizes beyond what current quantum machines can naively handle. We have developed the QRR*, a high-performant hybrid quantum-classical algorithm inspired by classical semidefinite programming methods~\cite{Dupont2024}. It finds optimal or near-optimal solutions with a single layer ($p=1$) of quantum logical operations and converges asymptotically to the optimal solution by increasing $p$~\cite{Dupont2024}.

The combined run-time of the quantum and classical components of the QRR* is of the order of tenths of seconds on problems with thousands of variables (Appendix~\ref{app:number_of_bitstrings}). This makes the approach competitive against Gurobi but short of others, such as the SA and BM algorithms in the low-$p$ regime considered. As presented, it is unclear whether the QRR* framework could lead to a quantum advantage given the narrow path offered by increasing $p$ in a run-time-versus-performance tradeoff. To overcome these challenges, one direction is to develop better and more efficient hybrid quantum-classical solvers~\cite{BLEKOS20241}. Another direction---not orthogonal to the other---is to investigate problems other than the maximum cut on $3$-regular graphs, which would potentially be more challenging for classical solvers and ideally easier for quantum ones. However, such problems are likely denser, rendering the light-cone technique inapplicable. It represents a challenge for superconducting platforms to encode a graph-problem structure in a gate-based fashion with shallow depth to mitigate the noise limitations of contemporary hardware. It would be interesting to investigate the run-time of quantum error correction~\cite{Devitt_2013} overhead and what it means for delivering a practical quantum speedup against state-of-the-art classical optimization solvers.

\let\oldaddcontentsline\addcontentsline
\renewcommand{\addcontentsline}[3]{}
\begin{acknowledgments}
    We gratefully acknowledge L. T. Brady, N. Didier, S. Ejtemaee, S. Grabbe, S. Hadfield, P. A. Lott, F. B. Maciejewski, M. J. Reagor, E. G. Rieffel, D. Venturelli, Z. Wang, and T. Wilson for discussions, input, and collaborations on related works. This work is supported by the Defense Advanced Research Projects Agency (DARPA) under Agreement No. HR00112090058 and IAA 8839, Annex 114. The authors from USRA also acknowledge support by NASA Academic Mission Services under Contract No. NNA16BD14C. This research used resources of the National Energy Research Scientific Computing Center (NERSC), a U.S. Department of Energy (DOE) Office of Science User Facility supported by the Office of Science of the DOE under Contract No. DE-AC02-05CH11231 using NERSC Awards DDR-ERCAP0024427 and No. ASCR-ERCAP0028951. The experiments were performed through Rigetti Computing Inc.'s Quantum Cloud Services QCS\textsuperscript{TM} on the Ankaa\textsuperscript{TM}-2 superconducting quantum processor developed, fabricated, and operated by Rigetti Computing Inc.
\end{acknowledgments}
\let\addcontentsline\oldaddcontentsline

\section*{Author contributions}

M.D. conceived and led the project with support from B.S. D.E.B.N. and Z.P. performed the numerical simulations with the classical solver Gurobi. M.J.H. performed the run-time analysis of the quantum computer. B.E., M.J.H., and S.J. developed software supporting the experimental runs with randomized compilation and readout-error mitigation. M.D. and B.E. performed the experiments and collected the data. M.D. carried out the rest of the simulations and analyses. M.D. wrote the manuscript, with contributions from all the coauthors. All coauthors contributed to the discussions leading to the completion of this project.

\section*{Competing interests}

M.D., B.E., M.H., S.J., and B.S. are, have been, or may in the future be participants in incentive stock plans at Rigetti Computing Inc. M.D. and B.S. are inventors on two pending patent applications related to this work (No. 63/631,643 and No. 63/632,079). The other authors declare that they have no competing interests.

\section*{Data availability}

The problem instances studied in this work, the cut number returned by the experimental runs of the QRR* algorithm, and the estimated optimal cut number used for computing the approximation ratio are publicly available at~\href{https://doi.org/10.5281/zenodo.11061744}{doi.org/10.5281/zenodo.11061744}.

\appendix


\section{Superconducting Quantum Processing Unit}

\subsection{Hardware Characteristics}

We performed experiments on the $84$-qubit superconducting quantum chip Rigetti Ankaa\textsuperscript{TM}-2 based on a square lattice topology with tunable couplers. We only used a subset of the device consisting of a continuous linear chain of seven qubits with characteristics reported in Tab.~\ref{tab:qubits_characteristics}. The device features parametric one-qubit gates, implementing rotations around the $x$ and $z$ axes, $\texttt{Rx}(\phi)=\exp(-i\hat{X}\phi/2)$ with $\phi=\pm\pi/2$, and $\texttt{Rz}(\phi\in\mathbb{R})=\exp(-i\hat{Z}\phi/2)$, respectively. The two-qubit gate between adjacent qubits is the $\texttt{ISWAP}$ gate defined as $\exp[i\pi(\hat{X}\otimes \hat{X}+\hat{Y}\otimes \hat{Y})/4]$. $\hat{X}$, $\hat{Y}$, and $\hat{Z}$ are Pauli operators. We discuss the compilation of relevant quantum logical gates using natives ones in App.~\ref{sec:compilation_sm}.

\begin{table*}[!ht]
    \centering
    \begin{tabular}{lcccccccc}
        \hline\hline\\[-0.8em]
        \makecell[l]{\textbf{Device}} & \multicolumn{7}{c}{\makecell[c]{\textbf{Qubits label}}} & \makecell[c]{\textbf{Average}}\\
        \makecell[l]{\textbf{characteristics}} & \makecell[c]{$\textbf{28}$} & \makecell[c]{$\textbf{29}$} & \makecell[c]{$\textbf{30}$} & \makecell[c]{$\textbf{31}$} & \makecell[c]{$\textbf{32}$} & \makecell[c]{$\textbf{33}$} & \makecell[c]{$\textbf{34}$} & \makecell[c]{\textbf{value}}\\
        \hline\\[-0.8em]
        \makecell[l]{Coherence time $T_1$ ($\mu$s)} & \makecell[c]{$9.8$} & \makecell[c]{$12.7$} & \makecell[c]{$5.3$} & \makecell[c]{$11.8$} & \makecell[c]{$14.8$} & \makecell[c]{$10.3$} & \makecell[c]{$16.2$} & \makecell[c]{$\textbf{11.6}$}\\
        \hline\\[-0.8em]
        \makecell[l]{One-qubit gate $\texttt{Rx}$ fidelity} & \makecell[c]{$99.7\%$} & \makecell[c]{$99.6\%$} & \makecell[c]{$99.0\%$} & \makecell[c]{$99.4\%$} & \makecell[c]{$99.6\%$} & \makecell[c]{$98.4\%$} & \makecell[c]{$98.7\%$} & \makecell[c]{$\textbf{99.2\%}$}\\
        \hline\\[-0.8em]
        \makecell[l]{One-qubit gate $\texttt{Rx}$ duration (ns)} & \makecell[c]{$40$} & \makecell[c]{$40$} & \makecell[c]{$40$} & \makecell[c]{$40$} & \makecell[c]{$40$} & \makecell[c]{$40$} & \makecell[c]{$40$} & \makecell[c]{$\textbf{40}$}\\
        \hline\\[-0.8em]
        \makecell[l]{Two-qubit gate} & \multicolumn{2}{c}{\makecell[c]{$98.9\%$}} & \multicolumn{2}{c}{\makecell[c]{$97.5\%$}} & \multicolumn{2}{c}{\makecell[c]{$97.7\%$}} & \makecell[c]{} & \makecell[c]{}\\
        \makecell[l]{$\texttt{ISWAP}$ fidelity} & \makecell[c]{} & \multicolumn{2}{c}{\makecell[c]{$98.7\%$}} & \multicolumn{2}{c}{\makecell[c]{$98.2\%$}} & \multicolumn{2}{c}{\makecell[c]{$97.8\%$}} & \makecell[c]{$\textbf{98.1\%}$}\\
        \hline\\[-0.8em]
        \makecell[l]{Two-qubit gate} & \multicolumn{2}{c}{\makecell[c]{$136$}} & \multicolumn{2}{c}{\makecell[c]{$128$}} & \multicolumn{2}{c}{\makecell[c]{$144$}} & \makecell[c]{} & \makecell[c]{}\\
        \makecell[l]{$\texttt{ISWAP}$ duration (ns)} & \makecell[c]{} & \multicolumn{2}{c}{\makecell[c]{$128$}} & \multicolumn{2}{c}{\makecell[c]{$152$}} & \multicolumn{2}{c}{\makecell[c]{$168$}} & \makecell[c]{$\textbf{122}$}\\
        \hline\\[-0.8em]
        \makecell[l]{Readout fidelity} & \makecell[c]{$94.6\%$} & \makecell[c]{$92.8\%$} & \makecell[c]{$94.7\%$} & \makecell[c]{$94.2\%$} & \makecell[c]{$95.1\%$} & \makecell[c]{$94.0\%$} & \makecell[c]{$96.2\%$} & \makecell[c]{$\textbf{94.5\%}$}\\
        \hline\hline\\[-0.8em]
    \end{tabular}
    \caption{Characteristics of a linear chain of seven qubits used for running the experiments on the Rigetti Ankaa\textsuperscript{TM}-2 superconducting chip. Reported fidelities were estimated using randomized benchmarking~\cite{PhysRevA.77.012307}.}
    \label{tab:qubits_characteristics}
\end{table*}

\subsection{Quantum Computer's Output}

\begin{figure*}[!t]
    \centering
    \includegraphics[width=1\textwidth]{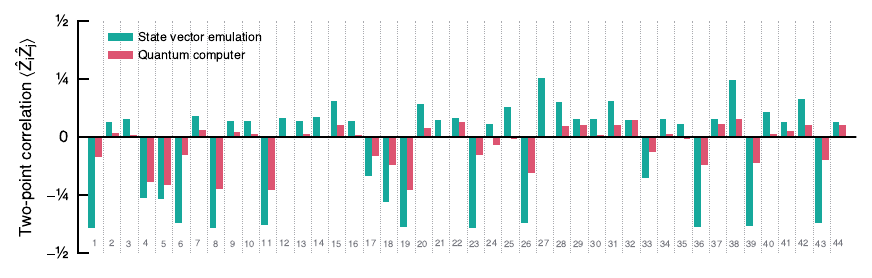}
    \caption{Expectation value of the two-point correlation function $\langle\hat{Z}_i\hat{Z}_j\rangle$ for two qubits $i$ and $j$ relevant for building the correlation matrices of all the graph problems based on the QAOA with $p=1$. Expectation values were estimated from $10^4$ bit strings (shots). There are $44$ entries corresponding to $44$ unique subgraphs according to Tab.~\ref{tab:isomorphism_db_size}. The data include noiseless state vector emulations and experimental data from the quantum computer using readout-error mitigation}
    \label{fig:experimental_data}
\end{figure*}

We compare the emulated and device results for problems solved using QAOA with $p=1$. As shown in Sections~\ref{sec:light_cone_sm} and~\ref{sec:graph_isomorphism_sm}, by leveraging the light cone induced by the QAOA with $p$ layers and graph isomorphism, we only need to evaluate a total of $44$ expectation values from $44$ different quantum circuits at $p=1$. The corresponding expectation values using readout error mitigation~\cite{nachman_unfolding_2020} (see also App.~\ref{sec:error_mit_sm}) are displayed in Fig.~\ref{fig:experimental_data} with a clear correlation between emulated and simulated data. The overall magnitude reduction of the experimental data is expected and understood under a simple depolarizing noise model~\cite{Dupont2024}, as developed in App.~\ref{sec:error_mit_sm}.

We note that graph isomorphism is just a practical convenience that enables us to execute a fixed number of circuits at fixed $p$ to get statistics over any number of problem instances. Its use is not assumed when estimating the run-time of the quantum algorithm.

\section{Combinatorial Optimization Solvers}

In this section, we describe combinatorial optimization solvers seeking to extremize an $N$-variable objective function of the form
\begin{equation}
    \tilde{C}(\boldsymbol{z})=\frac{1}{2}\sum\nolimits_{i,j=1}^N\mathsf{W}_{ij}z_iz_j,
    \label{eq:obj_func_sm}
\end{equation}
where $\mathsf{W}\in\mathbb{R}^{N\times N}$ is a symmetric matrix, i.e., $\mathsf{W}_{ij}=\mathsf{W}_{ji}$, encoding the problem of interest. The variables $z_i$ are subject to a binary constraint such that $\boldsymbol{z}\in\{\pm 1\}^N$. Specifically, we focus on minimizing the objective function of Eq.~\eqref{eq:obj_func_sm} where $\mathsf{W}$ is drawn uniformly from the ensemble of adjacency matrices of unit-weight $3$-regular graphs. As such, $\mathsf{W}$ is sparse with three nonzero entries per row and column, each equal to unity.

Up to an additional factor and overall multiplicative constant, minimizing Eq.~\eqref{eq:obj_func_sm} is similar to maximizing the objective function of the main text for the maximum-cut problem. When needed, we characterize the quality of a solution $\boldsymbol{z}$ by the approximation ratio
\begin{equation}
    \alpha = \frac{C(\boldsymbol{z})}{C(\boldsymbol{z}_\textrm{opt})} =\frac{3N-2\tilde{C}(\boldsymbol{z})}{3N-2\tilde{C}(\boldsymbol{z}_\textrm{opt})},
    \label{eq:app_ratio}
\end{equation}
where $C$ is the maximum cut objective function defined in the main text, $\tilde{C}$ the related objective function of Eq.~\eqref{eq:obj_func_sm}, and $\boldsymbol{z_\textrm{opt}}$ is the optimal solution.

\subsection{Classical}

\subsubsection{Markov Chain Monte-Carlo}

The objective function $\tilde{C}(\boldsymbol{z})$ of Eq.~\eqref{eq:obj_func_sm} describes an antiferromagnetic Ising model. The optimal solution $\boldsymbol{z}_\textrm{opt}$, which minimizes the objective function, is the ground state of the Ising model. Thus, in a canonical ensemble, the optimal solution $\boldsymbol{z}_\textrm{opt}$ would perfectly describe the state of the model at zero temperature, i.e., $T=0$. More generally, at temperature $T$, the probability for the model to be in the state $\boldsymbol{z}$ with energy $\tilde{C}(\boldsymbol{z})$ is given by a Boltzmann distribution
\begin{equation}
    p(T,\boldsymbol{z}) \propto e^{-\tilde{C}(\boldsymbol{z})/T}.
    \label{eq:boltz_dist_sm}
\end{equation}

The Metropolis-Hastings Markov chain Monte-Carlo algorithm~\cite{10.1063/1.1699114,10.1093/biomet/57.1.97}
is a celebrated approach for numerically evaluating the thermodynamic properties of an Ising model at temperature $T$. It is based on the detailed balance principle, which samples microstates $\boldsymbol{z}$ according to Eq.~\eqref{eq:boltz_dist_sm}, after a thermalization process. From a configuration $\boldsymbol{z}$, a new configuration $\boldsymbol{z'}$ is suggested. It is accepted with a Metropolis-Hastings probability
\begin{equation}
    p_\textrm{accept}\bigl(\boldsymbol{z}\leftarrow\boldsymbol{z'}\bigr)=\textrm{min}\Biggl\{1,e^{\bigl[\tilde{C}(\boldsymbol{z})-\tilde{C}(\boldsymbol{z'})\bigr]/T}\Biggr\}.
    \label{eq:metropolis_proba_sm}
\end{equation}
For instance, a strategy for suggesting a new configuration $\boldsymbol{z'}$ is to select uniformly at random one of the spins and flip its sign. This strategy, performed as many times as there are spins in the model, is known as a sweep. A Markov Chain Monte-Carlo simulation typically starts with a random configuration $\boldsymbol{z}$.

\subsubsection{Simulated Annealing}
\label{sec:sa_sm}

\paragraph{Description---} For the purpose of combinatorial optimization, one might be tempted to use directly the Markov chain Monte-Carlo algorithm by setting the temperature $T$ to a small value such that solutions $\boldsymbol{z}$ with a low energy $\tilde{C}(\boldsymbol{z})$ are sampled. However, for difficult problems such as spin glasses, the proposed configuration updates from $\boldsymbol{z}$ to $\boldsymbol{z'}$ might be inefficient as the algorithm effectively becomes stuck in a local minimum. In this situation, thermalization could be extremely slow, potentially to a point making the algorithm impractical. Simulated annealing~\cite{doi:10.1126/science.220.4598.671} seeks to circumvent this limitation by starting from a high temperature and reducing the temperature throughout the algorithm, thus ideally speeding up thermalization and avoiding local minima along the way.

In simulated annealing, a temperature schedule is defined as a function of the number of sweeps. One typically starts with a high temperature to enable fast mixing. Then, the temperature is reduced throughout the sweeps.

\paragraph{Numerical Implementation---}

We test both linear and geometric temperature schedules, controlled by the number of sweeps $k$ (at each sweep, the temperature is reduced). The initial hot temperature is noted $T_\textrm{hot}\equiv T_{\ell=1}$, the final cold temperature is noted $T_\textrm{cold}\equiv T_{\ell=k}$, and intermediate temperatures are noted by $T_\ell$. In the linear interpolation scheme, the intermediate temperatures are defined as
\begin{equation}
    T^{-1}_\ell=T_\textrm{hot}^{-1} + \ell\frac{T_\textrm{cold}^{-1} - T_\textrm{hot}^{-1}}{k}
    \label{eq:lin_temp_sched_sa_sm}
\end{equation}
where $\ell=1,~2,~\ldots,~k$. In the geometric interpolation scheme, the intermediate temperatures are defined as
\begin{equation}
    T^{-1}_\ell=\exp\Biggl(\ln T_\textrm{hot}^{-1} + \ell\frac{\ln T_\textrm{cold}^{-1} - \ln T_\textrm{hot}^{-1}}{k}\Biggr)
    \label{eq:geo_temp_sched_sa_sm}
\end{equation}
where $\ell=1,~2,~\ldots,~k$. We choose
\begin{equation}
    T_\textrm{hot}=6/\ln2\quad\textrm{and}\quad T_\textrm{cold}=2/\ln\bigl(100N\bigr).
    \label{eq:mcmc_temperatures_sm}
\end{equation}
The rationale behind those numbers is as follows. At first, we want fast mixing where all spins have a high probability of flipping. The three-fold connectivity of the $3$-regular graphs with $\textrm{max}(\mathsf{W}_{ij})=1$ is such that the largest energy difference between two configurations is $\tilde{C}(\boldsymbol{z})-\tilde{C}(\boldsymbol{z'})=6$. We want such a transition to be accepted half of the time. Solving Eq.~\eqref{eq:metropolis_proba_sm} with those numbers leads to $T_\textrm{hot}=6/\ln2$. Toward the end, in the last sweeps, the state should have a low probability of being excited as it should correspond to a local/global minimum. For unit-weight $3$-regular graphs, the smallest excitation, or smallest energy difference between two configurations is $\tilde{C}(\boldsymbol{z})-\tilde{C}(\boldsymbol{z'})=2$. We bound the probability to excite any of the $N$ spins by $1\%$, leading to the equation $1\%=Ne^{2/T_\textrm{cold}}$, which we solve for $T_\textrm{cold}=2/\ln(100N)$.

We report and discuss the relative performance of the linear and geometric temperature interpolation schemes in App.~\ref{sec:sa_lin_geo_sm}. We find that the geometric temperature schedule yields better results for the cases and regimes considered in this work.

\paragraph{Algorithmic Complexity---}

Our implementation of simulated annealing has an algorithmic complexity that is linear with the number of sweeps $k$. For each sweep, an update on each of the $N$ spins is attempted. Because the update is local and the $3$-regular graphs considered have an $O(1)$ connectivity between variables, the energy difference $\tilde{C}(\boldsymbol{z})-\tilde{C}(\boldsymbol{z'})$ between two configurations $\boldsymbol{z}$ and $\boldsymbol{z'}$ can be computed in $O(1)$ operations. Hence, the total complexity is $O(Nk)$.

\paragraph{Remarks---}

The temperature schedule can be fine-tuned for a specific problem to potentially yield better performance: Not only can the initial and finite temperatures be optimized, but also the shape of the schedule. Since we do not employ such fine-tuning for the quantum combinatorial optimization solvers, we stick with an $O(1)$ parameter setting heuristic. In any case, fine-tuning is impractical as we consider thousands of problem instances. Finally, we note that simulated annealing is guaranteed to converge to the optimal solution $\boldsymbol{z}_\textrm{opt}$~\cite{295910,10.1007/978-3-642-60744-8_32} when using a smooth temperature schedule with the number of steps scaling exponentially with the number of variables $N$. In a more practical setting, simulated annealing can get stuck in a local minimum, which one can mitigate by running the algorithm several times with different initial configurations. This leads to a tradeoff between increasing the number of sweeps for a better average solution versus running the algorithm several times.

\subsubsection{Parallel Tempering}

\paragraph{Description---}

Parallel tempering~\cite{doi:10.1143/JPSJ.65.1604} seeks to overcome the need to run simulated annealing several times to mitigate the risk of getting stuck in local minima. In parallel tempering, one runs $n$ Markov chain Monte-Carlo simulations in parallel, known as replicas, each at a different temperature $T_{\ell=1...n}$ sorted from low to high. After a desired number of sweeps performed individually on the $n$ replicas, one attempts to exchange the two configurations $\boldsymbol{z}_{\ell}$ and $\boldsymbol{z}_{\ell+1}$ from nearby replicas. The move is accepted with a Metropolis-Hastings probability
\begin{equation}
    p_\textrm{accept}\bigl(\boldsymbol{z}_{\ell}\leftarrow\boldsymbol{z}_{\ell+1}\bigr)=\textrm{min}\Biggl\{1,e^{\bigl[\tilde{C}(\boldsymbol{z}_{\ell})-\tilde{C}(\boldsymbol{z}_{\ell+1})\bigr]\bigl[T^{-1}_{\ell}-T^{-1}_{\ell+1}\bigr]}\Biggr\}.
    \label{eq:metropolis_pt_proba_sm}
\end{equation}
This step is performed successively from high to low temperatures. Then, more sweeps are performed in each of the replicas until one decides to perform the exchange step again.

The idea behind parallel tempering is that the high-temperature replicas will explore a much larger space of configurations, which enables lower temperatures to avoid local minima given that a better region of solutions to explore was found at higher temperatures.

\paragraph{Numerical Implementation---}

We employ the same cold and hot temperatures that were defined for simulated annealing in Eq.~\eqref{eq:mcmc_temperatures_sm}, based on the properties of unit-weight $3$-regular graphs. We fix the number of replicas to $n=10$ defined from $T_\textrm{hot}$ to $T_\textrm{cold}$ with intermediate temperatures defined using a geometric spacing. We perform one sweep on each individual replica before performing the replica exchange step, starting from the highest temperature $T_\textrm{hot}$.

\paragraph{Algorithmic Complexity---}

Our implementation of parallel tempering has an algorithmic complexity that is linear in the number of replicas $n$, linear in the number of sweeps $k$, and linear in the number of variables $N$. Indeed, for each sweep on each replica, an update on each of the $N$ spins is attempted. An update on an individual spin can be evaluated in $O(1)$ operations given that the spins have an $O(1)$ connectivity, thus enabling the computation of the energy difference $\tilde{C}(\boldsymbol{z})-\tilde{C}(\boldsymbol{z'})$ between two configurations $\boldsymbol{z}$ and $\boldsymbol{z'}$ in $O(1)$ operations. For the replica exchange step, computing the energy difference between two configurations has a more generic $O(N)$ complexity cost. Therefore, the total complexity of our parallel tempering implementation is $O(nNk)$.

\paragraph{Remarks---}

Both the number of replicas $n$ and the temperatures can be fine-tuned for a specific problem to potentially yield better performance. However, as we do not employ such fine-tuning for the quantum combinatorial optimization solvers, we stick with an $O(1)$ parameter setting heuristic.

\subsubsection{The Burer-Monteiro Solver: Nonlinear Optimization Heuristic with Local Search}
\label{sec:burer_monteiro_sm}

\paragraph{Description---}

The Burer-Monteiro solver is based on a relax-and-round scheme of the maximum cut objective function~\cite{doi:10.1137/S1052623400382467,Burer2003}. The relaxation is such that the resulting optimization problem is nonlinear with a nonconvex objective function. The nonconvexity may sound daunting since it means that, in general, the objective function has multiple local nonglobal minima. In practice, the Burer-Monteiro algorithm is found to return high-quality solutions, making it one of the best heuristics to date for the maximum-cut problem~\cite{DunningEtAl2018}.

\paragraph{Numerical Implementation---}

We rely on a numerical implementation of the Burer-Monteiro solver which is part of the MQLib library~\cite{DunningEtAl2018} under the name `Burer2002'. The fundamental control parameter of the Burer-Monteiro algorithm is its number of steps, corresponding to the number of random perturbations explored from the current minimum. However, the MQLib implementation only exposes a run-time limit where a number of steps will be performed until the run-time exceeds the limit. The algorithm always runs for at least one full step.

\paragraph{Algorithmic Complexity---}

Each step of the Burer-Monteiro solver has an algorithmic complexity scaling at most linearly with the number of edges in the input graph problem. Consequently, for $3$-regular graphs, where the number of edges is $3N/2$ for $N$ variables, the algorithm has a total complexity $O(Nk)$ for $k$ steps.

\subsubsection{The Commercial Solver Gurobi}
\label{sec:gurobi_sm}

\paragraph{Description---} 

Since Gurobi~\cite{gurobi} cannot directly handle an Ising model in the form of Eq.~\eqref{eq:obj_func_sm}, where variable $z_i$ takes values in $\{\pm 1\}$, we first reformulate it into a quadratic unconstrained binary optimization problem (QUBO) to minimize 
\begin{equation}
    \tilde{C}_\textrm{QUBO}(\boldsymbol{x})=\frac{1}{2}\sum\nolimits_{i,j=1}^N\mathsf{W}_{ij} \bigl(2x_i-1\bigr)\bigl(2x_j-1\bigr),
    \label{eq:obj_func_sm_qubo}
\end{equation}
where a binary variable $x_i\in \{0,1\}$ is introduced to replace a variable $z_i\in \{\pm 1\}$ such that $z_i=2x_i-1$. After the reformulation, we employ Gurobi to solve the QUBO problem directly. The routine to solve mixed-integer programs in Gurobi typically includes a presolve phase~\cite{achterberg2020presolve}, branch-and-cut method~\cite{conforti2014integer}, and heuristic methods~\cite{bixby1999mip}. The presolve phase eliminates redundant variables and constraints and strengthens continuous relaxations. For nonconvex quadratic problems, Gurobi applies the auxiliary variable method to reformulate the QUBO problem into the bilinear form. Then, the branch-and-cut method is performed, which begins with solving the quadratic programming relaxation problem by removing all integrality restrictions. If the result satisfies all the integrality restrictions, then this solution is an optimal solution of the original mixed integer programming. If not, Gurobi continues branching on the variables with fractional values in the quadratic programming relaxation solution. The corresponding quadratic programming relaxations will be solved to bound the objective value of the generated subproblems. In the branch-and-bound tree, nodes can be pruned by infeasibility, integrality and bound to reduce the search space. Therefore, the tightness of the dual bound is crucial, and it can be calculated by applying the cutting plane approach to the subproblems before branching. Additionally, heuristic methods are usually invoked both at the root node and after solving the subproblems to find good feasible solutions. After all nodes are explored or pruned, the branch-and-cut algorithm will terminate at the optimal solution.

\paragraph{Numerical Implementation---}

We employ Gurobi 11.0.0, with the run-time limit set to $600$ seconds and the $\texttt{NonConvex}$ parameter set to $2$, to solve nonconvex quadratic problems by translating them into bilinear form and applying spatial branching. All tests ran on a Linux cluster with 48 AMD EPYC 7643 2.3GHz CPUs and 1 TB RAM, with each test restricted to using only a single thread.

\paragraph{Algorithmic Complexity---}

Analyzing the complexity of the branch-and-cut method in Gurobi is challenging and is affected by branching, node selection, cut generation, and heuristic strategies. Generally, the time complexity of the branch-and-bound algorithm is measured as $O(b^d)$, where $b$ represents the branching factor, and $d$, the solution's depth, is equivalent to $N$ in the $3$-regular graphs problems of Eq.~\eqref{eq:obj_func_sm}. The space complexity is typically $O(bd)$, reflecting the maximum number of nodes stored in memory at any point. Despite the potential worst-case time complexity, the branch-and-cut method proves efficient in practice by using effective bounding techniques to prune a large section of the tree.

\subsubsection{Greedy Solver}
\label{sec:greedy_solver_sm}

\paragraph{Description---}

We employ a classical randomized greedy solver~\cite{Dupont2023}. It builds a solution $\boldsymbol{z}\in\{\pm 1\}^N$ iteratively. The vector is initially set to zero, i.e., $\boldsymbol{z}=\boldsymbol{0}$, and the algorithm runs as follows. One selects uniformly, at random, a variable $z_i$. One sets this variable to $+1$ and computes the corresponding objective value $\tilde{C}(\boldsymbol{z})$ according to Eq.~\eqref{eq:obj_func_sm}. One repeats this step by now setting the variable to $-1$. Finally, the variable is set once and for all to the value $z_i=\pm 1$ extremizing the objective function. Ties are broken down uniformly at random. The steps are repeated until all $N$ variables are set to $\pm 1$.

\paragraph{Algorithmic Complexity---}

The $N$-variable solution is built over $N$ iterations. Each of these iterations requires evaluating the objective function $\tilde{C}$ of Eq.~\eqref{eq:obj_func_sm} twice. At each iteration, one only needs to compute terms in $C$ involving the selected variable. For $3$-regular graph problems, there are only three such terms. Thus, the algorithmic complexity of evaluating the objective function is independent of $N$. As such, the global algorithmic complexity of the classical greedy solver is $O(N)$ for $3$-regular graph problems.

\subsubsection{Relax-and-Round Approach}
\label{sec:cl_relax_and_round_sm}

\paragraph{Description---}

This approach first relaxes the constraint on the solution space from $\boldsymbol{z}\in\{\pm 1\}^N$ to $\boldsymbol{z}\in\mathbb{R}^N$ with $\|\boldsymbol{z}\|=\textrm{constant}$. Thus, the extremization task of Eq.~\eqref{eq:obj_func_sm} becomes an eigenvalue problem for which the leading eigenvector of $\mathsf{W}$ is the optimal solution. The leading eigenvector is then sign-rounded entrywise to recover a valid solution. In practice, we compute the $k=8$ leading eigenvectors. Each of them is sign-rounded, and the one yielding the best objective value is returned as the solution from the solver.

\paragraph{Algorithmic Complexity---}

Finding $k\sim O(1)$ leading eigenvectors of a real and symmetric $N\times N$ matrix generally has a computational complexity of $O(N^2)$ using a power method such as the Lanczos algorithm based on matrix-vector multiplications. However, given that the input matrix is sparse with only $O(1)$ entries per row, the computational cost of the eigendecomposition is reduced to $O(N)$. Then, sign-rounding the eigenvectors entrywise has a complexity $O(N)$. Finally, computing the objective value of a candidate solution $\boldsymbol{z}$ involves summing $O(N)$ terms for $3$-regular graph problems. Therefore, the global algorithmic complexity of the classical relax-and-round approach is $O(N)$ for $3$-regular graphs.

\subsubsection{Goemans-Williamson Algorithm}
\label{sec:cl_gw_sm}

\paragraph{Description---}

We employ a formulation of the Goemans-Williamson algorithm~\cite{Goemans1995,williamson2011} based on an eigenvalue problem~\cite{Mohar1990,Delorme1993,DELORME1993145,POLJAK1995249} due to the quantum analog that we developed. The Goemans-Williamson algorithm seeks to solve the maximum-cut problem, corresponding to maximizing the objective function
\begin{equation}
    C(\boldsymbol{z})=\frac{1}{4}\sum\nolimits_{i,j=1}^N\mathsf{W}_{ij}\Bigl(1-z_iz_j\Bigr),
    \label{eq:obj_func_mc_sm}
\end{equation}
as defined in the main text. From a practical point of view, the tasks of maximizing Eq.~\eqref{eq:obj_func_mc_sm} and minimizing Eq.~\eqref{eq:obj_func_sm} lead to the same optimal solution $\boldsymbol{z}_\textrm{opt}\in\{\pm 1\}^N$, albeit with a different objective value $\tilde{C}(\boldsymbol{z}_\textrm{opt})\neq C(\boldsymbol{z}_\textrm{opt})$. Therefore, one can use the Goemans-Williamson algorithm for solving the original problem of interest in Eq.~\eqref{eq:obj_func_sm}, which has been rewritten in the form of Eq.~\eqref{eq:obj_func_mc_sm}.

We introduce the Laplacian matrix $\mathsf{L}=\mathsf{D}-\mathsf{W}$ where $\mathsf{D}_{ij}=\delta_{ij}\sum_k\mathsf{W}_{ik}$ is the degree matrix with $\delta_{ij}$ the Kronecker delta. For unit-weight $3$-regular graph problems, $\mathsf{D}=3\mathsf{I}$, where $\mathsf{I}$ is the identity matrix. Thus, the objective function of Eq.~\eqref{eq:obj_func_mc_sm} can be written as $C(\boldsymbol{z})=\sum\nolimits_{i,j=1}^N\mathsf{L}_{ij}z_iz_j/4$. First, the largest algebraic eigenvalue of $N\mathsf{L}/4$ is an upper bound to the optimal objective value $C(\boldsymbol{z}_\textrm{opt})$~\cite{Mohar1990,Delorme1993,DELORME1993145,POLJAK1995249}. Second, one notes that the objective function is invariant under the transformation $\mathsf{L}\to\mathsf{L}+\textrm{diag}(u)$, where $\textrm{diag}(u)$ is a traceless diagonal matrix formed by the vector $\boldsymbol{u}\in\mathbb{R}^N$, known as the correcting vector. This means that the optimal objective value $C(\boldsymbol{z}_\textrm{opt})$ is also bounded by the leading eigenvalue of $N[\mathsf{L}+\textrm{diag}(u)]/4$, given the constraint $\sum_iu_i=0$. Hence, one can leverage the additional degrees of freedom introduced by the correcting vector $\boldsymbol{u}$ to tighten the eigenvalue bound. In practice, we solve a relaxed version, i.e., $\boldsymbol{z}\in\{\pm 1\}^N\to\boldsymbol{z}\in\mathbb{R}^N$
\begin{equation}
    \textrm{min}_{\sum_iu_i=0}~\textrm{max}_{\|\boldsymbol{z}\|=1}\boldsymbol{z}^T\frac{N}{4}\Bigl[\mathsf{D}-\mathsf{W}+\textrm{diag}(u)\Bigr]\boldsymbol{z},
    \label{eq:obj_func_gw_sm}
\end{equation}
and then, sign-round the leading eigenvector, i.e., $\boldsymbol{z}\in\mathbb{R}^N\to\boldsymbol{z}\in\{\pm 1\}^N$, to recover a valid solution with respect to the original problem. This algorithm is equivalent to the celebrated Goemans-Williamson algorithm for the maximum-cut problem~\cite{Mohar1990,Delorme1993,DELORME1993145,POLJAK1995249,Goemans1995,williamson2011}.

\paragraph{Remarks---}

For unit-weight $3$-regular graph problems, the degree matrix is a constant diagonal matrix. Therefore, it is irrelevant for finding the leading eigenvector of Eq.~\eqref{eq:obj_func_gw_sm}. Now, if one sets $\boldsymbol{u}=\boldsymbol{0}$, the method maps to the relax-and-round approach developed in App.~\ref{sec:cl_relax_and_round_sm}. As such, it may seem like the relax-and-round approach of App.~\ref{sec:cl_relax_and_round_sm} gives a lower-bound to the solution of the Goemans-Williamson algorithm. However, the relevant solution is the sign-rounded one, which might not be better in the $\boldsymbol{u}$-augmented version.

Finding the optimal correcting vector $\boldsymbol{u}_\textrm{opt}$ is a convex optimization problem~\cite{Delorme1993,DELORME1993145,POLJAK1995249}, which we solve using a convex cone solver~\cite{ocpb:16,odonoghue:21}. Once $\boldsymbol{u}_\textrm{opt}$ is found, we compute the $k=8$ leading eigenvectors of $\mathsf{L}+\textrm{diag}(\boldsymbol{u}_\textrm{opt})$ and sign-round each of them entrywise. The sign-rounded eigenvector with the best objective value is the desired solution.

\subsection{Quantum}

We introduce quantum combinatorial optimization solvers based on quantum circuits. A quantum circuit is a series of logical quantum operations producing a quantum state $\vert\Psi\rangle$. Here, we employ a one-to-one encoding of the binary variables onto qubits such that the solution of an $N$-variable problem is encoded in an $N$-qubit quantum state $\vert\Psi\rangle\in\mathbb{C}^{2^N}$.

\subsubsection{Quantum Approximate Optimization Algorithm}
\label{sec:qaoa_sm}

\paragraph{Description---}

We employ the Quantum Approximate Optimization Algorithm (QAOA)~\cite{Farhi2014,Farhi2014b,Farhi2016,BLEKOS20241}. A first step is to translate the objective function of the problem of interest into a quantum operator
\begin{equation}
    \hat{\tilde{C}}=\frac{1}{2}\sum\nolimits_{i,j=1}^N\mathsf{W}_{ij}\hat{Z}_i\hat{Z}_j,
    \label{eq:qaoa_cost_sm}
\end{equation}
where $\hat{Z}_i$ is the Pauli operator on qubit $i$. The operator $\hat{C}$ is a diagonal matrix where each entry is the objective value of the associated bit string corresponding to a solution to the combinatorial optimization problem. The QAOA with $p\geq 1$ layers reads
\begin{equation}
    \bigl\vert\Psi\bigr\rangle_p=\left[\prod\nolimits_{\ell=1}^pe^{-i\beta_\ell\sum_{j=1}^N\hat{X}_j}e^{i\gamma_\ell\hat{\tilde{C}}}\right]\hat{H}^{\otimes N}\vert{0}\rangle^{\otimes N},
    \label{eq:qaoa_sm}
\end{equation}
where $\hat{X}_i$ is the Pauli operator on qubit $i$, $\hat{H}$ is the one-qubit Hadamard gate, the operator $\hat{\tilde{C}}$ is defined in Eq.~\eqref{eq:qaoa_cost_sm}, and $\{\gamma_\ell, \beta_\ell\}$ are real-valued angles. Ideally, these angles should be chosen such that they extremize the expectation value $\langle\hat{\tilde{C}}\rangle_p$ over the quantum state $\vert\Psi\rangle_p$ of Eq.~\eqref{eq:qaoa_sm}. The first term of the bracket is known as the mixer, and the second one as the phase separator.

\paragraph{Remarks---}

The adiabatic theorem~\cite{Farhi2014,Wurtz2022} guarantees that QAOA will return the optimal solution $\boldsymbol{z}_\textrm{opt}$ in the limit $p\to+\infty$. The performance guarantees of the QAOA at finite $p$ are much harder to establish. Specific problem instances benefit from performance guarantees, such as ring graphs with an approximation ratio $\alpha=(2p+1)/(2p+2)$~\cite{Farhi2014,PhysRevA.97.022304} or Sherrington-Kirkpatrick spin glasses~\cite{PhysRevLett.35.1792} for which the QAOA yields on average $\alpha\simeq 0.397$ at $p=1$, and up to $\alpha\simeq 0.901$ at $p=20$~\cite{Farhi2022,Basso2022}.

\paragraph{QAOA Angles---}

A difficulty commonly arising with the QAOA algorithm is finding the optimal angles $\{\boldsymbol{\gamma}, \boldsymbol{\beta}\}_p$, especially when the number of layers $p$ or qubits $N$ are large. These angles are typically optimized in a hybrid quantum-classical feedback loop. However, so-called barren plateaus~\cite{McClean2018}, a vanishing gradient phenomenon, may prevent efficient optimization of the angles. Heuristics have been advanced in an attempt to mitigate and circumvent this issue, e.g., Refs.~\cite{PhysRevX.10.021067,Lotshaw2021,Wurtz2021,Lee2022,Sud2022,Lee2023,Li2023}.

Here, we focus on unit-weight random $3$-regular graph problem instances. We rely on the fixed-angle conjecture of Refs.~\cite{PhysRevA.103.042612,Wurtz2021} to set the QAOA angles $\{\boldsymbol{\gamma}, \boldsymbol{\beta}\}_p$, thus strictly avoiding any optimization of the QAOA angles while maintaining near-optimality. We use the values reported in Ref.~\cite{Wurtz2021}, and reproduced in Tab.~\ref{tab:qaoa_angles}.

\begin{table*}[!ht]
    \centering
    \begin{tabular}{cclll}
        \hline\hline\\[-0.8em]
        \makecell[c]{\textbf{Number of}} & \makecell{\qquad\qquad\qquad\qquad} & \multicolumn{3}{c}{\makecell[c]{\textbf{QAOA Angles}}}\\
        \makecell[c]{\textbf{QAOA Layers}} & {} & \makecell{$\ell=1$} & \makecell[c]{$\ell=2$} & \makecell[c]{$\ell=3$}\\
        \hline\\[-0.8em]
        \makecell[c]{$p=1$} & \makecell{$\boldsymbol{\gamma}$\\$\boldsymbol{\beta}$} & \makecell[l]{$0.615533629$\\$0.3926720292447629$}\qquad\qquad & {}\qquad\qquad & {}\\[0.9em]
        \hline\\[-0.8em]
        \makecell[c]{$p=2$} & \makecell{$\boldsymbol{\gamma}$\\$\boldsymbol{\beta}$} & \makecell[l]{$0.4877097328$\\$0.5550603400685824$}\qquad\qquad & \makecell[l]{$0.8979876956$\\$0.29250781484335187$}\qquad\qquad & {}\\[0.9em]
        \hline\\[-0.8em]
        \makecell[c]{$p=3$} & \makecell{$\boldsymbol{\gamma}$\\$\boldsymbol{\beta}$} & \makecell[l]{$0.422084082$\\$0.608757260014991$}\qquad\qquad & \makecell[l]{$0.798412754$\\$0.45927530900125874$}\qquad\qquad & \makecell[l]{$0.9370887966$\\$0.23539562255067184$}\\[0.9em]
        \hline\hline\\[-0.8em]
    \end{tabular}
    \caption{QAOA angles for unit-weight $3$-regular graphs up to $p=3$ QAOA layers, as defined in Eq.~\eqref{eq:qaoa_sm}, extracted from Ref.~\cite{Wurtz2021}. We use these angles throughout this work, unless specified otherwise. Units are radians.}
    \label{tab:qaoa_angles}
\end{table*}

\paragraph{Algorithmic Complexity---}

The execution time of the QAOA to obtain $\vert\Psi\rangle_p$ from Eq.~\eqref{eq:qaoa_sm} is linear with the number of layers $p$. Sampling $n_\textrm{ex}$ bit strings from the resulting quantum state $\vert\Psi\rangle_p$ is also linear in $n_\textrm{ex}$. The algorithmic complexity scaling of the QAOA with the number of qubits $N$ depends on the implementation: quantum simulations versus classical emulations, and the nature of the emulator and quantum hardware, etc. We discuss the specific cases of interest in App.~\ref{sec:executing_circuits_sm}.

\subsubsection{Quantum-Flavored Relax-and-Round Algorithm}
\label{sec:qrr_sm}

\paragraph{Description---}

The quantum relax-and-round (QRR) algorithm was introduced in Ref.~\cite{Dupont2024}. It is based on the QAOA, from which one obtains a two-point correlation matrix with entries
\begin{equation}
    \mathsf{Z}_{ij}^{(p)} = \bigl(\delta_{ij} - 1\bigr)\bigl\langle\hat{Z}_i\hat{Z}_j\bigr\rangle_{p},
    \label{eq:corr_matrix_sm}
\end{equation}
where $\hat{Z}_i$ is the Pauli operator on qubit $i$, $\delta_{ij}$ is the Kronecker delta, and the expectation value $\langle\hat{Z}_i\hat{Z}_j\rangle_{p}$ is taken over the quantum state $\vert\Psi\rangle_p$ resulting from the QAOA of Eq.~\eqref{eq:qaoa_sm}. $\mathsf{Z}^{(p)}$ is an $N\times N$ real and symmetric matrix with null entries on its diagonal. The next step consists of running the classical relax-and-round step explained in App.~\ref{sec:cl_relax_and_round_sm} on the correlation matrix $\mathsf{Z}^{(p)}$ instead of the matrix $\mathsf{W}$ encoding the problem.

\paragraph{Remarks---}

As noted above, the QAOA yields the optimal solution $\boldsymbol{z}_\textrm{opt}$ in the infinite-$p$ limit. In Ref.~\cite{Dupont2024} it was shown that QRR also returns the optimal solution in the limit $p=\infty$, in which case the correlation matrix is
\begin{equation}
    \mathsf{Z}^{(p=\infty)} = \mathsf{I} - \boldsymbol{z}_\textrm{opt}\otimes\boldsymbol{z}_\textrm{opt},
    \label{eq:corr_matrix_pinf_sm}
\end{equation}
where $\otimes$ denotes the outer product. The first term of Eq.~\eqref{eq:corr_matrix_pinf_sm} is the identity matrix. The second term is a rank one matrix with eigenvector $\propto\boldsymbol{z}_\textrm{opt}$. A classical relax-and-round step on the correlation matrix of Eq.~\eqref{eq:corr_matrix_pinf_sm} will therefore return the optimal solution $\boldsymbol{z}_\textrm{opt}$. Ref.~\cite{Dupont2024} provides a strategy for dealing with degenerate optimal solutions.

In Ref.~\cite{Dupont2024}, it was shown analytically that at $p=1$, the performance of the QRR algorithm matches that of its classical counterpart for several problem instances, such as Sherrington-Kirkpatrick spin glasses~\cite{PhysRevLett.35.1792}, ring graphs, the Bethe lattice, the honeycomb lattice, the complete graph, and more generally, circulant graphs. The study included numerical simulations on several other graph problems, showing that even without theoretical performance bounds, the QRR algorithm remains a powerful heuristic, at least on par with its classical counterpart, for unit-weight random $3$-regular graphs. Moreover, the average performance increases with the number of layers $p$, until convergence to the optimal solution for $p\to+\infty$ layers.

\paragraph{Algorithmic Complexity---}

Compared to the classical relax-and-round algorithm of App.~\ref{sec:cl_relax_and_round_sm}, the algorithmic complexity of the quantum version has the additional cost of obtaining the correlation matrix $\mathsf{Z}^{(p)}$.

Moreover, the density of the correlation matrix is going to be larger than that of the adjacency matrix, $\mathsf{W}$. While in the limit of large $N$, one still expects $O(1)$ nonzero entries per row (see App.~\ref{sec:light_cone_sm}), at intermediate $N$, and depending on the number of layers $p$, the correlation matrix will get denser. As such, the scaling of the eigendecomposition will tend toward $O(N^2)$ instead of $O(N)$.

\subsubsection{Quantum-Flavored Goemans-Williamson Algorithm}
\label{sec:qgw_sm}

\paragraph{Description---}

Ref.~\cite{Dupont2024} introduced a quantum version of the classical Goemans-Williamson algorithm (QGW) outlined in App.~\ref{sec:cl_gw_sm}. The first step is to construct the correlation matrix $\mathsf{Z}^{(p)}$ of Eq.~\eqref{eq:corr_matrix_sm} using the QAOA with $p$ layers. Next, one substitutes the matrix $\mathsf{W}$ encoding the problem in Eq.~\eqref{eq:obj_func_gw_sm} with the correlation matrix $\mathsf{Z}^{(p)}$. The eigenvalue-based Goemans-Williamson algorithm of App.~\ref{sec:cl_gw_sm} is then executed as previously described.

\paragraph{Remarks---}

The QGW algorithm yields the optimal solution $\boldsymbol{z}_\textrm{opt}$ in the $p=\infty$ limit ~\cite{Dupont2024}. This can be shown by substituting the $p\to+\infty$ correlation matrix of Eq.~\eqref{eq:corr_matrix_pinf_sm} into Eq.~\eqref{eq:obj_func_gw_sm} and choosing $\mathsf{diag}(\boldsymbol{u})=\textrm{tr}(\mathsf{D})\mathsf{I}/N - \mathsf{D}$, thereby ensuring the leading eigenvector is $\propto\boldsymbol{z}_\textrm{opt}$. 
In ~\cite{Dupont2024} it was shown analytically that at $p=1$ the performance of the QGW algorithm was on par with that of the classical Goemans-Williamson algorithm for ring and complete graphs. Furthermore, numerical tests at $p=1$ on several instances of other nontrivial problems, including unit-weight random $3$-regular graphs, showed the quantum and classical algorithms to perform similarly. It should be noted that the average performance of the QGW algorithm increases with the number of layers $p$, until convergence to the optimal solution for $p\to+\infty$ layers.

\paragraph{Algorithmic Complexity---}

Compared to the classical Goemans-Williamson algorithm of App.~\ref{sec:cl_gw_sm}, the algorithmic complexity of the quantum version has the additional cost of obtaining the correlation matrix $\mathsf{Z}^{(p)}$.

\subsubsection{Greedy-Enhanced Relax-and-Round Algorithm}
\label{sec:grdy_enh_qrr_alg_sm}

\paragraph{Description---}

We introduce a greedy heuristic on top of the relax-and-round-based algorithms (either classical or quantum). Given the  
optimal sign-rounded solution returned by the relax-and-round algorithm, we perform opportunistic sign-flips on each variable. If the sign flip yields a better objective value, it is kept as part of the new solution. If not, one keeps the original sign of the variable and moves to another one.

The intuition behind such a greedy strategy is as simple: we have less confidence in the sign rounding of variables with values close to zero. It is, therefore, natural to attempt opportunistic local sign-flips of such variables. In fact, we leverage the corresponding non-rounded eigenvector $\boldsymbol{z}\in\mathbb{R}^N$ to visit variables $i$ with entry $z_i\in\mathbb{R}$ at random following the probability
\begin{equation}
    p_i= \vert{z_i}\vert^{-1}\Bigr/\sum\nolimits_{i=1}^N\vert{z_i}\vert^{-1},
\end{equation}
such that variables $z_i$ closer to zero have a larger chance of being visited. The choice of mapping between $p_i$ and $z_i$ is nonunique and could probably be fine-tuned. We set the number of variables visited as $fN$ with $f=10$ for an $N$-variable problem. We study the performance of this greedy heuristic as a function of $f$ in App.~\ref{sec:grdy_enh_qrr_sm}.

\paragraph{Remarks---}

We note again that this strategy is not specific to quantum-flavored relax-and-round solvers but could apply to any. Such a step is usually included in solvers, given its low algorithmic overhead. We investigate different sign-flip strategies in App.~\ref{sec:grdy_enh_qrr_sm}. In addition, we study the performance gain of the greedy-enhanced quantum relax-and-round solver over its original version in App.~\ref{sec:grdy_enh_qrr_sm}.

\paragraph{Algorithmic Complexity---}

The algorithmic complexity is that of the underlying relax-and-round algorithm plus an additional $O(N)$ cost related to sign-flip attempts on the solution. Indeed, computing the objective value of a sign-flipped solution can be performed in $O(1)$ time given the local structure of the $3$-regular graphs considered.

\section{Problem Instances}

\subsection{Unit-weight Random \texorpdfstring{$3$}{3}-Regular Graphs}
\label{sec:problem_instances_sm}

\begin{figure}[!t]
    \centering
    \includegraphics[width=0.5\textwidth]{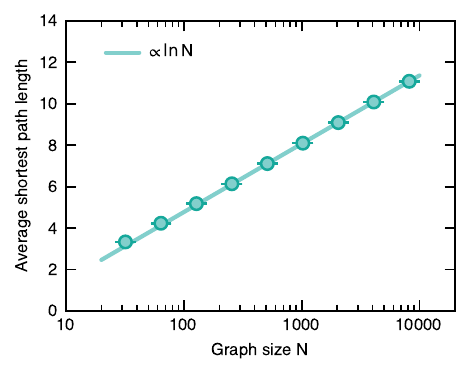}
    \caption{Average shortest path length in unit-weight $3$-regular random graphs as a function of the number of vertices $N$. Average taken over all pairs of vertices in a graph and over $50$ to $1,000$ randomly generated graphs. The solid line is a fit of the form $A\ln(N/B)$, where $A$ and $B$ are real-valued parameters obtained by least-square fitting. Error bars indicate one standard deviation.}
    \label{fig:avg_shortest_path_length}
\end{figure}

In this work, we focus on $N$-variable problem instances defined by Eq.~\eqref{eq:obj_func_sm}. Each instance is characterized by an $N\times N$ matrix $\mathsf{W}$, drawn uniformly from the ensemble of adjacency matrices of unit-weight $3$-regular graphs~\cite{STEGER_WORMALD_1999,10.1145/780542.780576}. $3$-regular graphs are such that their vertices are connected to three other unique vertices. Therefore, $\mathsf{W}$ is sparse, real, symmetric, and contains only three nonzero entries per row and column, all equal to unity, for a total of $3N$ entries. By definition, $N\geq 4$ and $N$ is even. The size of the ensemble at fixed $N$ is given by Ref.~\cite[A008277]{oeis}. We generate the problem instances using the Python package NetworkX~\cite{SciPyProceedings_11}. In the limit of large $N$, the average shortest path between two vertices grows as $\ln N$, see Fig.~\ref{fig:avg_shortest_path_length}. Moreover, the statistics on the cycles in a typical $3$-regular graph are known. The probability to have $K$ cycles of length $\ell$ is Poissonian and given by~\cite{WORMALD1981168,McKay2004,Lykov2023}
\begin{equation}
    p(K,\ell)=\frac{\lambda^Ke^{-\lambda}}{K!}\quad\textrm{where}~\lambda=\frac{2^\ell}{2\ell}.
\end{equation}
From there, one can compute the average number of cycles of length $\ell$
\begin{equation}
    \mathbb{E}\bigl[K_\ell\bigr]=\sum\nolimits_{K=1}^\infty Kp(K,\ell)=\lambda=\frac{2^\ell}{2\ell},
\end{equation}
which is independent of $N$. For example, the average number of triangles made of up of three connected vertices is $\mathbb{E}[K_{\ell=3}]=4/3$.

\subsection{Correlations are Constrained by the QAOA Light Cone}
\label{sec:light_cone_sm}

\subsubsection{The Light Cone Picture}

\begin{figure*}[!t]
    \centering
    \includegraphics[width=0.65\textwidth]{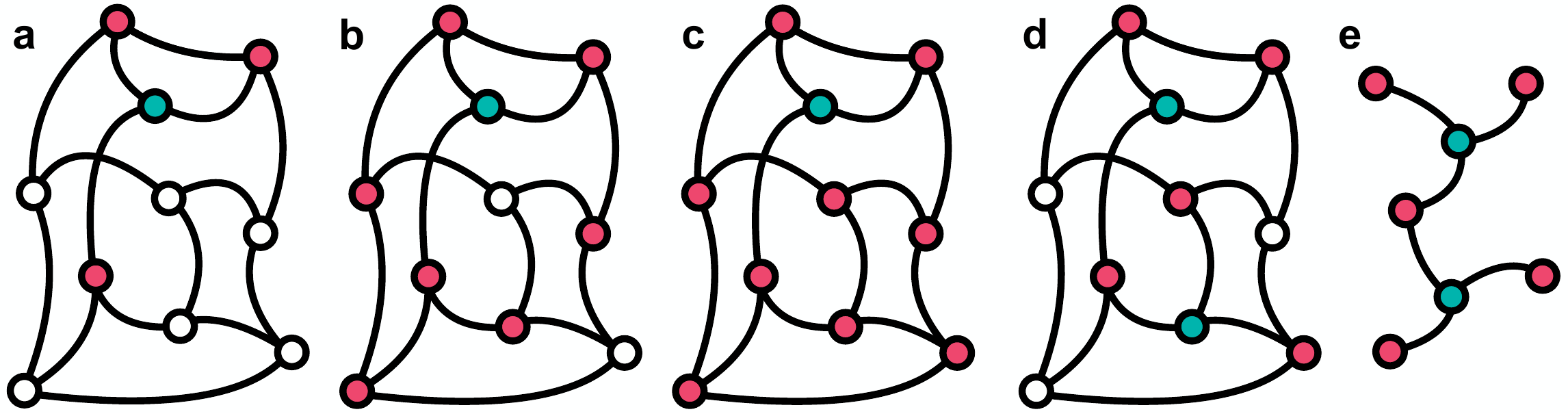}
    \caption{(a-d) An $N=10$ variable $3$-regular graph. An arbitrary reference qubit is colored in teal. Its $p$-nearest neighbors are colored in magenta. (a) $p=1$, (b) $p=2$, and (c) $p=3$. (d) Two arbitrary reference qubits with their nearest neighbors ($p=1$). These two qubits have intersecting light cones. (e) Qubits involved in two intersecting light cones made of two $3$-Cayley trees connected through a single common qubit.}
    \label{fig:light_cone}
\end{figure*}

The quantum algorithms that we leverage are based on the expectation value of two-point correlations $\langle\hat{Z}_i\hat{Z}_j\rangle_p$ between all problem variables $i$ and $j$ to build the correlation matrix $\mathsf{Z}^{(p)}$. Naively, this implies computing $N(N-1)/2\sim O(N^2)$ elements. A QAOA circuit with $p\geq 1$ layers, as defined in Eq.~\eqref{eq:qaoa_sm}, generates a light cone for each of the qubits centered around that qubit and including all its $p$-nearest neighbors, see Figs.~\ref{fig:light_cone}a,~\ref{fig:light_cone}b, and~\ref{fig:light_cone}c. Thus, for a $3$-regular graph, such a light cone will include at most
\begin{equation}
    1+3(2^p-1),
\end{equation}
qubits with a structure corresponding, for instance, to a $3$-Cayley tree with $p$ generations.

The existence of a light cone is such that the correlation $\langle\hat{Z}_i\hat{Z}_j\rangle_p$ is trivially zero if the light cones centered around qubits $i$ and $j$, do not intersect. In the limit of large $N$, because the average shortest path between two vertices grows as $\ln N$ (see Fig.~\ref{fig:avg_shortest_path_length}), this means that for a number of QAOA layers $p\ll\ln N$, one expects only $O(N)$ elements to be nonzero. For a given $3$-regular graph problem instance, these $O(N)$ pairs $(i,j)$ with a nonzero correlation can be found in $O(pN^2)$ operations. We show an example in Fig.~\ref{fig:light_cone}d of two intersecting light cones at $p=1$.

\subsubsection{Implications for the QAOA Runs}
\label{sec:implications_qaoa_sm}

To estimate the expectation value of two-point correlations $\langle\hat{Z}_i\hat{Z}_j\rangle_p$, one only needs to run the QAOA on the subset of the problem variables within the intersection of the light cones centered around variables $i$ and $j$. Thus, one only needs to run the QAOA on at most
\begin{equation}
    1+6(2^p-1)
    \label{eq:n_qubits_with_p}
\end{equation}
qubits to evaluate $\langle\hat{Z}_i\hat{Z}_j\rangle_p$. While this scales exponentially with $p$, it is independent of the size $N$ of the underlying graph problem instance. This has to be repeated $O(N)$ times to obtain the expectation value on all the pairs $(i,j)$ with nonzero correlation. Therefore, one can trade a QAOA run with $p$ layers on $N$ qubits for $O(N)$ QAOA runs with $p$ layers on $\exp(O(p))$ qubits.

In this work we leverage this property to solve unit-weight random $3$-regular graph problem instances with numbers of variables $N$ much larger than the number of qubits available in current devices.

\paragraph{Subgraph size distribution---}

\begin{figure*}[!t]
    \centering
    \includegraphics[width=1.0\textwidth]{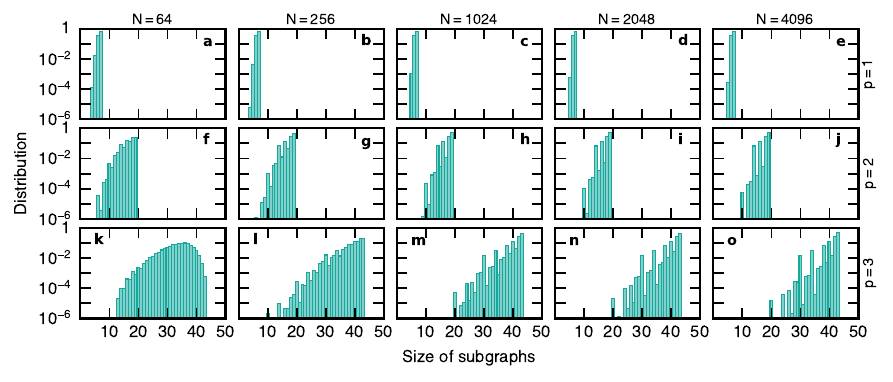}
    \caption{Distribution of the size of the subgraphs from intersecting light cones, for different problem sizes $N$. First row: $p=1$ QAOA layer. Second row: $p=2$ QAOA layers. Third row: $p=3$ QAOA layers. Columns from left to right correspond to: $N=64$, $N=256$, $N=1024$, $N=2048$, and $N=4096$. The maximum size is independent of the problem size $N$ at large $N$ and bounded by Eq.~\eqref{eq:n_qubits_with_p}.}
    \label{fig:statistics_subgraphs_size_distribution}
\end{figure*}

In Fig.~\ref{fig:statistics_subgraphs_size_distribution}, we show for different problem sizes $N$ the distribution of the size of the subgraphs generated from the intersection of the light cones on two variables. The maximum size depends on $p$ and is bounded by Eq.~\eqref{eq:n_qubits_with_p}. At large $N$, the distribution suggests that large subgraphs are predominant.

\paragraph{Subgraph treewidth distribution---}

\begin{figure*}[!t]
    \centering
    \includegraphics[width=1.0\textwidth]{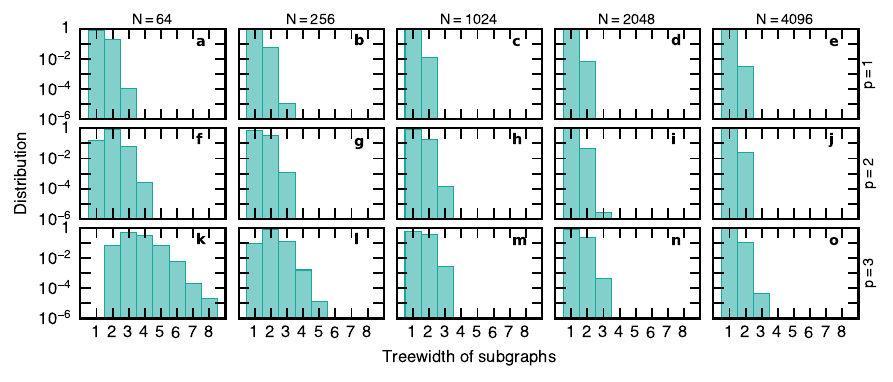}
    \caption{Distribution of the treewidth of the subgraphs from intersecting light cones, for different problem sizes $N$. First row: $p=1$ QAOA layer. Second row: $p=2$ QAOA layers. Third row: $p=3$ QAOA layers. Columns from left to right correspond to: $N=64$, $N=256$, $N=1024$, $N=2048$, and $N=4096$.}
    \label{fig:treewidth_subgraphs}
\end{figure*}

In addition to the size of the subgraphs, we consider their treewidth, which we compute using the minimum fill-in heuristic~\cite{BODLAENDER2010259,SciPyProceedings_11}. A treewidth of one means that the graph is a tree. As the treewidth increases, the graph becomes less tree-like and includes loops. We show in Fig.~\ref{fig:treewidth_subgraphs} the subgraph treewidth distribution for different number of QAOA layers $p$ and problem sizes $N$. As $N$ increases, we observe that the distribution narrows and peaks at a treewidth of one. This suggests that the majority of the subgraphs are trees.

Fig.~\ref{fig:statistics_subgraphs_size_distribution} shows that the majority of the subgraphs saturate the maximum size bound of Eq.~\eqref{eq:n_qubits_with_p}. We, therefore, conclude that most of the subgraphs consist of two $3$-Cayley trees connected by a single common vertex.

\subsection{Graph Isomorphism}
\label{sec:graph_isomorphism_sm}

\subsubsection{Connection to the Light Cone Picture from the QAOA}

When using the light cone approach (see App.~\ref{sec:light_cone_sm}), a single QAOA run on the complete graph is replaced by $O(N)$ QAOA runs on subgraphs formed by the intersecting light cones. Of the $O(N)$ subgraphs evaluated, we posit that the number of unique subgraphs is independent of $N$ in the large $N$ limit. Said differently, there are $O(N)$ isomorphic graphs formed by the intersecting light cones. Hence, determining the isomorphism can reduce to $O(1)$ the number of QAOA runs required to compute $\mathsf{Z}$ for a number of QAOA layers $p\ll\ln N$ in the limit of large $N$. This assertion is based on the tendency for subgraphs to become more tree-like with increasing $N$ (see the treewidth analysis in App.~\ref{sec:implications_qaoa_sm}) and supported by the numerical data presented in the remainder of this Section.

\paragraph{Working with an Isomorphism Database---}

In addition to using isomorphism to reduce the computational load for a given problem instance, we can use the property to share results between different problem instances. Hence, the result of a QAOA run on a given subgraph can be used multiple times for different problem instances having the same subgraph, thereby reducing the overall computing cost. This facilitates the collection of statistics used to evaluate the performance of quantum solvers over many problem instances.

We build an isomorphism database for a given number of QAOA layers $p$ as follows. We loop over problem instances and pairs $(i, j)$ with nonzero correlation. For each such pair, we obtain the subgraph consisting of the intersecting light cones centered around $i$ and $j$. We compute the Weisfeiler-Lehman subgraph hash~\cite{JMLR:v12:shervashidze11a}. This hash serves as the key for the entries of the isomorphism database. If the subgraph hash is not in the database, we add it together with the subgraph. If the subgraph hash is already in the database, we check that the subgraph is indeed isomorphic to the one stored in the database. If not, there is a hash collision. We append a counter to the hash and check whether the counter-augmented hash is in the database and, if so, if the subgraph is indeed isomorphic to the one stored in the database. We repeat this operation as many times as necessary. In practice, hash collisions are rare. If the subgraph is isomorphic to the one sharing the same hash in the database, we record the subgraph hash for the pair $(i,j)$ as well as the mapping of the vertices $i$ and $j$ onto that of the isomorphic subgraph already stored in the database.

\paragraph{Data---}

\begin{table*}[!ht]
    \centering
    \begin{tabular}{ccccccccc}
        \hline\hline\\[-0.8em]
        \makecell[c]{\textbf{Number of}\\\textbf{QAOA Layers}} & \makecell[c]{\qquad\qquad} & \makecell[l]{\textbf{Instance Size $N$}\\\textbf{(\# of Instances)}} & \makecell[c]{\qquad\qquad} & \makecell[c]{\textbf{Total \# of}\\\textbf{Entries}} & \makecell[c]{\qquad\qquad} & \makecell[c]{\textbf{Total \# of}\\\textbf{Nonzero Entries}} & \makecell[c]{\qquad\qquad} & \makecell[c]{\textbf{Total \# of}\\\textbf{Unique Entries}}\\
        \hline
        \makecell[c]{$p=1$} & {} & \makecell[l]{$32~(10^3)$, $64~(10^3)$,\\$128~(10^3)$, $256~(10^3)$,\\$512~(10^3)$, $1024~(10^3)$,\\$2048~(10^3)$, $4096~(10^3)$\\$\mathbf{Total: 8\times 10^3}$~\textbf{instances}} & {} & \makecell[c]{$1000\sum_{n=5}^{12}2^{n-1}\bigl(2^n-1\bigr)$\\$\simeq 10^{11}$} & {} & \makecell[c]{$\simeq 4\times 10^{7}$} & {} & \makecell[c]{$44$}\\[0.9em]
        \hline
        \makecell[c]{$p=2$} & {} & \makecell[l]{$32~(10^3)$, $64~(10^3)$, $128~(10^3)$\\$256~(10^3)$, $512~(10^3)$\\$1024~(10^3)$, $2048~(10^3)$\\$4096~(10^3)$\\$\mathbf{Total: 8\times 10^3}$~\textbf{instances}} & {} & \makecell[c]{$\simeq 10^{11}$} & {} & \makecell[c]{$\simeq 2\times 10^{8}$} & {} & \makecell[c]{$70,425$}\\[0.9em]
        \hline\hline\\[-0.8em]
    \end{tabular}
    \caption{Number of entries in the correlation matrix $\mathsf{Z}^{(p)}$ defined in Eq.~\eqref{eq:corr_matrix_sm}, summing all problem sizes and problem instances used in this work, for $p=1$, $p=2$, and $p=3$ QAOA layers. The total number of entries is the sum of $N(N-1)/2\times\textrm{number of instances}$, for each problem size $N$ considered (1000 problem instances were used for all problem sizes, and the sum over $n=5, ..., 12$ captures contributions from all problem sizes: $2^5 = 32, ..., 2^{12}=4096$. The total number of nonzero entries is obtained by eliminating all trivially zero entries which arise from nonintersecting light cones. The total number of unique entries accounts for isomorphism between the subgraphs made of intersecting light cones: this is the minimum number of QAOA runs required and the number used in this work.}
    \label{tab:isomorphism_db_size}
\end{table*}

We show in Tab.~\ref{tab:isomorphism_db_size} the total number of entries in the correlation matrix $\mathsf{Z}^{(p)}$ (defined in Eq.~\eqref{eq:corr_matrix_sm}), summed across all problem instances and sizes $N$ considered in this work. Using the fact that nonintersecting light cones yield a trivially zero correlation between variables and leveraging isomorphism between the subgraphs resulting from intersecting light cones considerably reduces the number of entries. For instance, for $p=1$, one only needs to run $44$ different QAOA circuits to build and fill the correlation matrices of all the $8,000$ randomly generated problem instances considered in this work, from size $N=32$ to $N=4,096$. This greatly facilitates the collection of statistics required to evaluate the performance of quantum solvers over many problem instances and problem sizes. Moving to $p=2$, one needs to execute $70,425$ circuits to solve $8,000$ randomly generated problem instances from size $N=32$ to $N=4,096$.

\paragraph{Average number of entries versus N---}

\begin{figure*}[!t]
    \centering
    \includegraphics[width=1.0\textwidth]{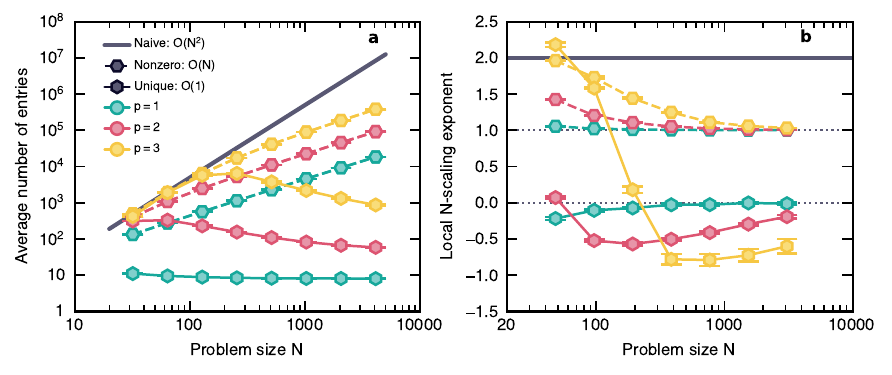}
    \caption{(a) Average number of entries in the correlation matrix $\mathsf{Z}^{(p)}$ defined in Eq.~\eqref{eq:corr_matrix_sm} for $p=1$, $p=2$, and $p=3$ QAOA layers, as a function of the problem size $N$. There are naively $O(N^2)$ entries. Based on the light-cone picture from the QAOA, there are $O(N)$ nonzero entries. Using graph isomorphism, there are $O(1)$ unique entries. (b) Local exponent of the $N$-scaling of the curves in (a), computed using finite differences. The results show the big $O$ scaling with $N$. Error bars indicate one standard deviation.}
    \label{fig:statistics_subgraphs_n_entries}
\end{figure*}

We show in Fig.~\ref{fig:statistics_subgraphs_n_entries}a the average number of entries in the correlation matrix $\mathsf{Z}^{(p)}$ for a typical problem instance, for $p=1$, $p=2$, and $p=3$ QAOA layers, as a function of the problem size $N$. For a given problem size, there are naively $N(N-1)/2$ entries in the correlation matrix, as it is symmetric.

By only considering entries $(i, j)$ where the light cones centered around $i$ and $j$ intersect, one can reduce this $O(N^2)$ scaling to $O(N)$, as shown in Fig.~\ref{fig:statistics_subgraphs_n_entries}a. Indeed, in the limit of large $N$, one expects only $O(N)$ nonzero entries when $p\ll\ln N$, because the average shortest path between two vertices grows as $\ln N$ for $3$-regular graphs (see Fig.~\ref{fig:avg_shortest_path_length}). Each of these nonzero entries corresponds to a subgraph arising from the intersection of the light cones centered around vertices $i$ and $j$.

Among these $O(N)$ subgraphs, many of them turn out to be identical in the large $N$ limit for $p\ll\ln N$. We observe in Fig.~\ref{fig:statistics_subgraphs_n_entries}a that only a constant fraction of the subgraphs are different in a typical problem instance of size $N$. Furthermore, the number of unique subgraphs is independent of the problem size. For instance, for $p=1$ QAOA layer, only about $8$ entries in the correlation matrix are on average nonzero and unique. Hence, in about $8$ QAOA runs one can build the full $N\times N$ correlation matrix for a typical problem instance. The big $O$ scaling is valid in the limit of $N\to+\infty$. For completeness, we plot in Fig.~\ref{fig:statistics_subgraphs_n_entries}b the $N$-dependent local exponent $\alpha(N)$ of $O(N{^\alpha(N)})$.

The aforementioned scaling can be understood as follows. In the large $N$ limit: (i) the majority of the subgraphs saturate the maximum size bound of Eq.~\eqref{eq:n_qubits_with_p} (Fig.~\ref{fig:statistics_subgraphs_size_distribution}); and (ii) the majority of the subgraphs have a treewidth of one (Fig.~\ref{fig:treewidth_subgraphs}), which shows that the majority of the subgraphs are trees. Combining both observations it becomes clear that most of the subgraphs consist of two $3$-Cayley trees connected by a single common vertex. An example is shown in see Fig.~\ref{fig:light_cone}(e). There is only one such graph for a given number of QAOA layers $p$. Therefore the majority of the subgraphs, in the large $N$ limit, are isomorphic.

\subsubsection{Determining (Sub)Graph Isomorphism}

It has recently been shown that the graph isomorphism problem, i.e., the task of determining whether two graphs are isomorphic, can be solved in quasi-polynomial time with an algorithmic complexity $\exp(O(\ln N)^{O(1)})$~\cite{Laszlo2015,Andres2017}. Here, we employ a Boolean satisfiability (SAT) formulation of the problem, for which SAT solvers generally suffer from exponential cost with the input problem size. We summarize the problem formulation as a SAT following Ref.~\cite{MATSUO20232022EAP1159}. We are given two graphs $G(V,E)$ and $G'(V',E')$ with vertices $V$, $V'$, and edges $E$, $E'$, respectively. We introduce literals $x_{i,j}$ with $i=1,2,\ldots |V|$ and $j=1,2,\ldots |V'|$. A literal $x_{i,j}$ is true if a vertex $i\in V$ maps to a vertex $j\in V'$. For the graphs $G$ and $G'$ to be isomorphic, we first need $|V|=|V'|$. Moreover, there should exist a one-to-one mapping between vertices of $V$ and $V'$ fulfilling the following conditions,
\begin{itemize}[label=$\circ$]
    \item \textbf{Condition~\#1:} Each vertex of $G$ is assigned to only one vertex in $G'$
    \begin{equation}
        \bigwedge\nolimits_{i=1}^{|V|}\Bigl(\bigvee\nolimits_{j=1}^{|V'|}x_{i,j}\Bigr)~\land~\bigwedge\nolimits_{i=1}^{|V|}\Bigl(\bigvee\nolimits_{j=2,k=1,j>k}^{|V'|}\neg x_{i,j}\lor\neg x_{i,k}\Bigr).
        \label{eq:sat_cond1_sm}
    \end{equation}
    \item \textbf{Condition~\#2:} Similarly, at most one vertex of $G$ is assigned to a vertex in $G'$
    \begin{equation}
        \bigwedge\nolimits_{k=1}^{|V'|}\Bigl(\bigvee\nolimits_{j=1,i=2,i>j}^{|V|}\neg x_{i,k}\lor\neg x_{j,k}\Bigr).
        \label{eq:sat_cond2_sm}
    \end{equation}
    \item \textbf{Condition~\#3:} Adjacent vertices in $G$ must be adjacent in $G'$
    \begin{equation}
        \bigwedge\nolimits_{(i,j)\in E,k=1}^{|V'|}\Bigl[\neg x_{i,k}\lor\Bigl(\bigvee\nolimits_{(k,k')\in E'}x_{j,k'}\Bigr)\Bigr].
        \label{eq:sat_cond3_sm}
    \end{equation}
\end{itemize}
The conjunction of the three conditions from Eqs.~\eqref{eq:sat_cond1_sm},~\eqref{eq:sat_cond2_sm}, and ~\eqref{eq:sat_cond3_sm} makes a SAT formulation of the graph isomorphism problem. The more general case with $|V|\geq|V'|$ is known as the subgraph isomorphism problem, asking whether $G$ contains a subgraph that is isomorphic to $G'$. The subgraph isomorphism problem is known to be NP-complete.

\paragraph{Practical Implementation---}

In practice, we solve the SAT problem using the SAT solver Glucose~\cite{doi:10.1142/S0218213018400018}, based on Minisat~\cite{10.1007/978-3-540-24605-3_37}, through the Python package PySAT~\cite{imms-sat18}. In addition to returning whether isomorphism exists, the solver returns a satisfying mapping between vertices of $G$ and $G'$, if it exists.

\begin{figure*}[!t]
    \centering
    \includegraphics[width=1\textwidth]{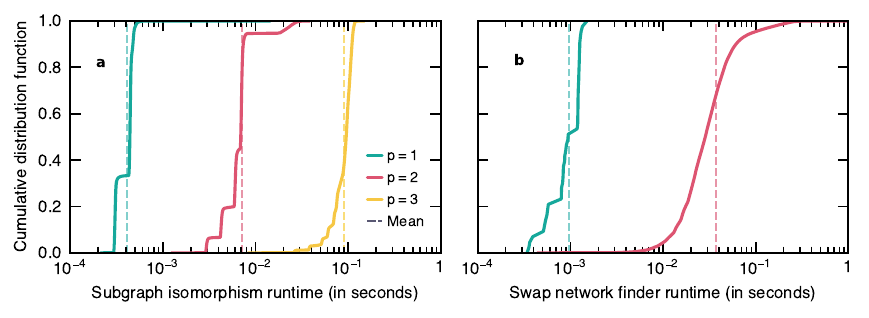}
    \caption{Cumulative distribution functions of the run-time for the SAT solver to determine: (a) subgraph isomorphism; and (b) an efficient swap network, for different number of QAOA layers $p$. All tests were conducted on a $64$Gb MacBook Pro with an Apple M1 Max chip.}
    \label{fig:runtime_sat}
\end{figure*}

We consider a few randomly generated problems with $N=4,096$ variables to estimate the run-time for the SAT solver to determine whether two subgraphs are isomorphic. The run-time includes building the SAT problem from two input subgraphs and running the SAT solver itself. Following Eq.~\eqref{eq:n_qubits_with_p}, the maximum size of a subgraph as a function of the number of QAOA layers $p$ is $1+6(2^p-1)$. The distribution of subgraph sizes was discussed in App.~\ref{sec:implications_qaoa_sm} and plotted in Fig.~\ref{fig:statistics_subgraphs_size_distribution}. Here, we report the run-time to determine isomorphism on a $64$Gb MacBook Pro with an Apple M1 Max chip at fixed $p$, independent of the size of the subgraphs themselves. The results are shown in Fig.~\ref{fig:runtime_sat}a. We observe that the average run-time increases roughly exponentially with $p$, meaning that it is linear in the size of the subgraphs.

As shown in Fig.~\ref{fig:statistics_subgraphs_n_entries}, for a given $N$-variables graph problem, there are $O(N)$ subgraph isomorphism problems to solve if one desires to find the $O(1)$ unique entries of the correlation matrix. Given the cost of running the SAT solver $O(N)$ times for a problem of size $N$, it may be preferable to execute $O(N)$ QAOA circuits instead. In the context of this work, we leverage subgraph isomorphism across different problem instances---of the same and different sizes---as well as within problem instances at low $p$, providing a huge benefit, as shown in Tab.~\ref{tab:isomorphism_db_size}.

\paragraph{Remarks---}

For estimating the run-time of the quantum algorithm in Figs.~2 and~3 of the main text and App.~\ref{sec:executing_circuits_sm}, we assume that we have to execute all of the circuits corresponding to all of the $O(N)$ nonzero entries of the correlation matrix. We do not assume graph isomorphism, given its poor scaling. We only use it as a convenience to get statistics over thousands of problem instances in the context of this work.

\section{Executing Quantum Circuits}
\label{sec:executing_circuits_sm}

\subsection{Quantum Simulations}

\subsubsection{Hardware Topology and Swap Network}
\label{sec:swap_network_sm}

Current superconducting quantum computers have qubit topologies in which there is limited connectivity between the nodes. Thus, to perform quantum logical operations between nonadjacent qubits, one must use swap gates. Naive use of swap gates will introduce an overhead which is at most linear in the number of qubits. However, executing many series of swap gates may still be impractical on noisy devices. Such concerns prompt one to ask: is it possible to find an efficient assignment of qubits and an efficient swap network that minimizes, for example, the total depth of the quantum circuit?

For the QAOA, each qubit encodes a binary variable of the problem of interest. Therefore we seek the mapping between qubits and variables that will minimize the swap overhead needed to encode the phase separator unitary of Eq.~\eqref{eq:qaoa_sm}. For an arbitrary mapping between $N$ variables and $N$ qubits, an arbitrary QAOA phase separator unitary can be encoded in a swap network of depth $O(N)$ on, for example, a linear or square hardware topology~\cite{OGorman2019,Weidenfeller2022scalingofquantum}.

\paragraph{Efficient Swap Network Finder---}
\label{sec:swap_net_finder_sm}

We employ the swap network finder introduced in Ref.~\cite{MATSUO20232022EAP1159} and successfully used in Ref.~\cite{Sack2023}. The hardware topology and phase separator are encoded as graphs $G(V,E)$ and $G'(V',E')$, respectively. Vertices $V'$ in the phase separator graph correspond to qubits, and each edge in $E'$ corresponds to a term $e^{i\gamma\mathsf{W}_{ij}\hat{Z}_i\hat{Z}_j/2}\neq\hat{I}$ from Eq.~\eqref{eq:qaoa_sm}, between two qubits $i$ and $j$. The swap network finder is partly about solving a subgraph isomorphism problem. We perform this task using the approach described in App.~\ref{sec:graph_isomorphism_sm}, based on a Boolean satisfiability (SAT) formulation of the problem. It works as follows: one iteratively updates the connectivity of the hardware topology graph $G$ by applying a set of swap gates. After the swap gates have been applied, one solves a subgraph isomorphism problem on the updated graph. If $G'$ can be embedded onto $G$, then one has found a satisfying mapping of variables onto qubits as well as a satisfying set of swap gates, i.e., a satisfying swap network. If $G'$ cannot be embedded onto $G$, one applies another set of swap gates to $G$ and repeats the operation until a satisfying swap network is found. It is the responsibility of the user to define the set of swap gates since there is no optimization of the swap network per se. Hence, one starts with no or a small swap network and increases its depth until a satisfying solution is found.

For the hardware topology and phase separator graphs considered in this work, we have found that it is relatively easy to find a satisfying swap network after a few different trials.

\begin{figure}[!t]
    \centering
    \includegraphics[width=1\columnwidth]{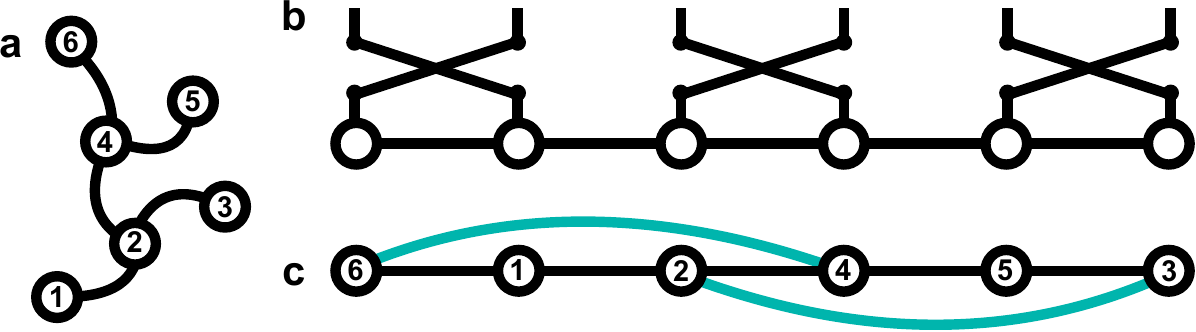}
    \caption{(a) Graph representing a QAOA phase separator made of $N=6$ qubits. (b) Linear chain topology with $6$ qubits and a single layer of swap gates applied on even edges. (c) Mapping of the QAOA phase separator qubits onto the linear chain topology. The extra effective edges introduced by the layer of swap gates are marked in green. A single layer of swap gates enables the embedding of the graph from (a) onto the linear chain topology.}
    \label{fig:swap_network_chain}
\end{figure}

\paragraph{Data---}

We map an $N$-variable QAOA phase separator onto an $N$-variable line topology. At each iteration of the swap network builder, we alternatively add a layer of parallel swap gates on even and odd edges of the line topology. This operation is repeated until a satisfying swap network is found. After this step, the quantum circuit encoding the desired QAOA phase separator for a linear qubit topology is fixed. This strategy is applied to an $N=6$-qubit QAOA phase separator in Fig.~\ref{fig:swap_network_chain}. The swap network builder can find a mapping of the graph nodes onto qubits requiring only a single layer of two-qubit gates. The assignment onto a linear chain is as follows: $[6, 1, 2, 4, 5, 3]$ with two-qubit operations possible on edges $(6,1)$, $(1, 2)$, $(2,4)$, $(4,5)$, and $(5,3)$. From there, a layer of swap gates leads to the following arrangement $[1, 6, 4, 2, 3, 5]$, enabling two-qubit operations on edges $(1,6)$, $(6,4)$, $(4,2)$, $(2,3)$, $(3,5)$. In particular, edges $(6,4)$ and $(2,3)$ are made available through the swap gates, as shown in Fig.~\ref{fig:swap_network_chain}, enabling the encoding of the graph onto the linear topology of the qubits. The strategy is guaranteed to succeed with at most $N$ alternating even and odd layers of swap gates~\cite{OGorman2019}.

The algorithm for finding an efficient swap network requires a SAT solver for establishing graph isomorphism. The run-times required to find such swap networks are shown in Fig.~\ref{fig:runtime_sat}b. Following Eq.~\eqref{eq:n_qubits_with_p}, the maximum size of a subgraph as a function of the number of QAOA layers $p$ is $1+6(2^p-1)$.The distribution of subgraph sizes was discussed in App.~\ref{sec:implications_qaoa_sm} and plotted in Fig.~\ref{fig:statistics_subgraphs_size_distribution}. The run-times in Fig.~\ref{fig:runtime_sat}b were measured on a $64$Gb MacBook Pro with an Apple M1 Max chip at fixed $p$, independent of the size of the subgraphs themselves. We used the SAT solver Glucose~\cite{doi:10.1142/S0218213018400018}, based on Minisat~\cite{10.1007/978-3-540-24605-3_37}, through the Python package PySAT~\cite{imms-sat18}. Given that one has to solve a SAT problem, the method becomes prohibitively expensive as $p$ increases. We refer the reader to Ref.~\cite{MATSUO20232022EAP1159} for further discussion.

A default swap network for encoding an $N$-variable phase separator on a linear chain of $N$ qubits~\cite{OGorman2019} would require $3N(N-1)/2$ two-qubit $\texttt{ISWAP}$ gates. Using an efficient swap network finder can reduce the average number of $\texttt{ISWAP}$ gates for simulating $N$-node subgraphs by about $52\%$ for $p=1$ QAOA layer, $73\%$ for $p=2$ QAOA layers, and $85\%$ for $p=3$ QAOA layers.

\paragraph{Remarks---}

For estimating the run-time of the quantum algorithm in Figs.~2 and~3, we assume that the circuit is mapped onto a hardware-native linear chain of qubits with a default, worst-case-scenario swap network yielding a total number of layers of two-qubit gates proportional to $Np$ for $N$ qubits and $p$ QAOA layers~\cite{OGorman2019}. The quantum algorithm depth (and run-time) does not assume the use of an efficient swap network. In the main text and App.~\ref{sec:executing_circuits_sm}, a default, worst-case-scenario swap network on a linear chain of qubit is assumed.

\subsubsection{Compilation}
\label{sec:compilation_sm}

\begin{figure*}[!t]
    \centering
    \includegraphics[width=0.8\textwidth]{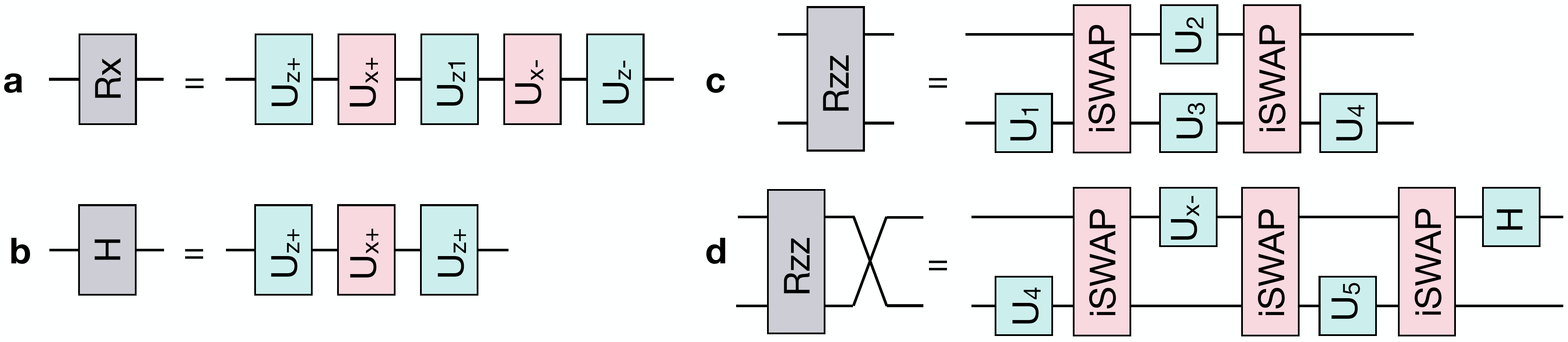}
    \caption{Construction of algorithmic gates in terms of hardware-native gates on Rigetti's Ankaa\textsuperscript{TM}-2 superconducting quantum chip. One-qubit gates are defined in Eq.~\eqref{eq:Un_gates_def_sm}. (a) Implementation of $\texttt{Rx}(\phi)$ for arbitrary angles $\phi$. (b) Implementation of the Hadamard gate. (c) Implementation of the two-qubit $\texttt{Rzz}(\phi)$ gate. (d) Implementation of the two-qubit $\texttt{Rzz}(\phi)\times\texttt{SWAP}$ gate.}
    \label{fig:gate_compilation}
\end{figure*}

The native gates on Rigetti's Ankaa\textsuperscript{TM}-2 are the one-qubit rotation gates about the $x$ axis $\texttt{Rx}(\phi)=\exp(-i\hat{X}\phi/2)$ where $\phi=\pm\pi/2$, and the two-qubit gate $\texttt{ISWAP}=\exp[i\pi(\hat{X}\otimes\hat{X}+\hat{Y}\otimes\hat{Y})/4]$. One-qubit rotations about the $z$ axis, $\texttt{Rz}(\theta\in\mathbb{R})=\exp(-i\hat{Z}\theta/2)$, are implemented virtually. $\hat{X}$, $\hat{Y}$, and $\hat{Z}$ are Pauli operators. Our parameterized circuits have $\texttt{Rx}(\phi)$, $\texttt{Rzz}(\phi)=\exp[-i\phi\hat{Z}\otimes\hat{Z}/2]$, and $\texttt{Rzz}(\phi)\times\texttt{SWAP}$ gates, which are not native to Rigetti's Ankaa\textsuperscript{TM}-2. We implement these gates using the compilation in Fig.~\ref{fig:gate_compilation} where
\begin{align}
    &U_{z\pm} = \texttt{Rz}(\pm\pi/2)\nonumber\\
    &U_{x\pm} = \texttt{Rx}(\pm\pi/2)\nonumber\\
    &U_{z1} = \texttt{Rz}(\phi)\nonumber\\
    &U_1 = \texttt{Rz}(-\pi/2)\texttt{Rx}(\pi/2)\texttt{Rz}(\pi/2)\nonumber\\
    &U_2 = \texttt{Rz}(\pi/2)\texttt{Rx}(\pi/2)\texttt{Rz}(\phi)\texttt{Rx}(-\pi/2)\nonumber\\
    &U_3 = \texttt{Rz}(\pi)\nonumber\\
    &U_4 = \texttt{Rz}(\pi/2)\texttt{Rx}(\pi/2)\nonumber\\
    &U_5 = \texttt{Rx}(-\pi/2)\texttt{Rz}(-\phi).
    \label{eq:Un_gates_def_sm}
\end{align}

\subsubsection{Error Mitigation}
\label{sec:error_mit_sm}

We investigate the effect of error mitigation techniques such as randomized compilation~\cite{PhysRevA.94.052325} and readout error mitigation~\cite{nachman_unfolding_2020} on the two-body expectation values of the quantum relax-and-round (QRR) algorithm.

\paragraph{Randomized Compilation---}

Errors in quantum computers may be divided into coherent and incoherent errors~\cite{Sete2024}. Incoherent errors are the result of decoherence, leakage or other incoherent processes and are generally fundamental to the device. Coherent errors, by contrast, are rotation errors that occur in the computational space. Coherent errors are thus circuit-dependent, making it difficult to predict and mitigate their impact, even if they are known. However, by using the technique of twirling~\cite{Cai2019}, it is possible to mitigate coherent errors and tailor them into more tractable stochastic errors. This principle is widely applied in randomized benchmarking techniques, but it can also be applied to algorithms by using the technique of randomized compiling~\cite{PhysRevA.94.052325,PhysRevX.11.041039}.

To perform randomized compiling, we use the Pauli twirling group~\cite{Cai2019} which is compatible with the hardware-native $\texttt{ISWAP}$ gate. The technique works by arranging the circuit into cycles of ``easy'' one-qubit gates and ``hard'' entangling two-qubit gates. The critical distinction is that the ``easy'' single-qubit gates generally have an order of magnitude lower error rates, and don't typically suffer from significant coherent errors. For every ``hard'' cycle, a pair of Paulis from $\{\hat{I}, \hat{X}, \hat{Y}, \hat{Z}\}$ are selected uniformly at random and inserted before the $\texttt{ISWAP}$ gate. They are compensated by another pair of Paulis inserted after the $\texttt{ISWAP}$ gate, which serves to preserve the logical circuit. The inserted Paulis are then merged with the existing one-qubit gates in the circuit. We repeat the procedure $50$ times to produce a set of logically equivalent, randomized compilations. By executing the ensemble of circuits and aggregating the results, we obtain a set of samples that are explained by a stochastic Pauli noise model rather than a coherent error model. To perform random compilation, we utilize the TrueQ package~\cite{beale_2020_3945250}.

\paragraph{Readout Error Mitigation---}

Readout errors are a significant source of noise in contemporary quantum computers~\cite{Arute2019,PhysRevLett.127.180501,Kim2023}. Measurement errors can occur from either classification errors of the measurement signal, because the qubit decays from the $1$-state to the $0$-state during measurement or more complex interactions. The effect of decay is particularly important, as it results in an asymmetric confusion matrix. Iterative Bayesian Unfolding~\cite{nachman_unfolding_2020} is an effective technique for mitigating measurement error. Like matrix-inversion techniques, the method has calibration and mitigation steps. The calibration step involves preparing the qubits in either the $0$-state or the $1$-state and measuring the outcome. We assume that state preparation is perfect, although, in reality, the state preparation can have errors as well. These errors are typically two to three orders of magnitude lower than readout errors on superconducting platforms, making the assumption reasonable. The calibration is performed with the identical set of measurements as we plan to use in the experiment.

Given the calibration data, we can construct a confusion matrix of prepared states and measured states. The size of the confusion matrix will depend on the size of the subsystems we wish to apply mitigation to. At the individual qubit level, we have $2\times 2$ confusion matrices, while $N$ qubits require learning a $2^N\times 2^N$ confusion matrix. Thus, the largest subsystems that can be fully characterized are on the order of about a dozen qubits.

We then run the algorithm and collect the set of observed bit strings. Since we are interested in the estimation of low-weight observables, we choose to perform mitigation on the marginal distributions. That is, we select the subsystem of the qubit register which is required to estimate the observable, and calculate the bit string distribution of only those qubits. For small subsystems and our experimental value of $10,000$ shots, we are able to generate a statistically accurate bit string distribution. The iterative Bayesian unfolding technique is applied to the subsystem bit string distribution, resulting in a mitigated distribution. The mitigated distribution is then used to calculate the observable value. Because this technique is applied only to low-weight observables in the maximum-cut problem (two-point correlations), it is scalable.

\paragraph{Experimental Data---}

\begin{figure*}[!t]
    \centering
    \includegraphics[width=1\textwidth]{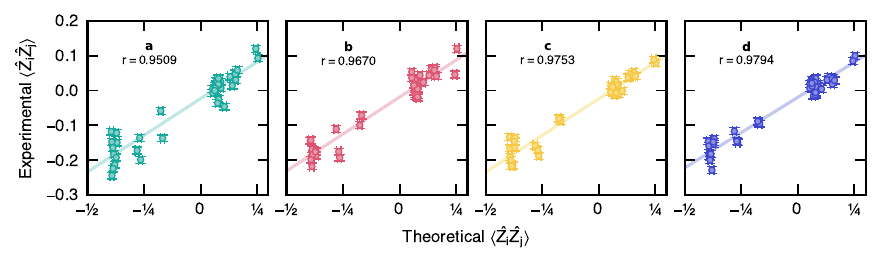}
    \caption{Expectation value of the two-point correlation function $\langle\hat{Z}_i\hat{Z}_j\rangle$ for two qubits $i$ and $j$ relevant for building the correlation matrices of all the graph problems based on the QAOA with $p=1$. Expectation values were estimated through the computation of $10^4$ bit strings. There are $44$ entries corresponding to $44$ unique subgraphs according to Tab.~\ref{tab:isomorphism_db_size}. (a) Raw experimental data from the quantum computer. (b) Readout-error mitigated experimental data. (c) Randomly-compiled experimental data. (d) Readout-error mitigated and randomly compiled experimental data. The Pearson correlation coefficient $r$ is reported. Error bars indicate one standard deviation.}
    \label{fig:experimental_correlation}
\end{figure*}

We consider the $44$ expectation values from noiseless state vector emulations of the quantum circuits and from the experimental runs on the superconducting quantum chip Rigetti Ankaa\textsuperscript{TM}-2. There are four experimental scenarios displayed in Fig.~\ref{fig:experimental_correlation}: (i) raw data, (ii) readout error-mitigated (REM) data, (iii) randomly-compiled (RC) data, and (iv) randomly-compiled and readout error-mitigated data. While differences between the emulation and quantum computer runs are clearly visible, correlations between the expected and experimental data are maintained, as exemplified by the Pearson correlation coefficient $r$ between experimental and emulated expectation values. We observe that error mitigation techniques enhance the correlation between experimental and theoretical data.

While there is a high correlation ($r>0.95$) in all four cases, the magnitudes of the device expectation values are reduced due to hardware noise. This magnitude reduction can be understood through a depolarizing noise model~\cite{Dupont2024}. Indeed, under a depolarizing noise channel, the system is in the mixed state~\cite{Nielsen2011}, $\hat{\rho}_{p,F}=F\vert\Psi\rangle\langle\Psi\vert_p+(1-F)\hat{\mathsf{I}}/2^N$, where $\langle\Psi\vert_p$ is the quantum state resulting from the QAOA with $p$ layers, $F\in[0,1]$ can be interpreted as the overall circuit fidelity and $\hat{\mathsf{I}}$ is the $N$-qubit identity matrix. The expectation value of the two-point correlation for the QRR algorithm reads
\begin{equation}
    \bigl\langle\hat{Z}_i\hat{Z}_j\bigr\rangle_{p,F}=\textrm{tr}\bigl(\hat{\rho}_{p,F}\hat{Z}_i\hat{Z}_j\bigr)=F\bigl\langle\hat{Z}_i\hat{Z}_j\bigr\rangle_{p,F=0}.
\end{equation}
Hence, a depolarizing noise channel simply rescales the correlations by the total fidelity $F$. Under this noise model, and assuming that enough bit strings have been collected to reliably estimate the noisy expectation values, the eigenvectors of the correlation matrix are unaffected by depolarizing noise.

\subsection{Classical State Vector Emulations}

Classical emulations seek to execute a quantum circuit on a classical computer. In the general case, this has an exponential algorithmic scaling with the number of qubits $N$.

\paragraph{Description---}

A state vector emulator enables an exact execution of quantum circuits with an exponential algorithmic cost in the number of qubits $N$. Indeed, it stores the full quantum state, i.e., $\vert\Psi\rangle\in\mathbb{C}^{2^N}$, requiring an exponentially large amount of memory and operations to execute quantum logical operations on $\vert\Psi\rangle$.

\paragraph{Numerical Implementation---}

We have implemented a state vector emulator tailored for the QAOA circuit with $p$ layers of Eq.~\eqref{eq:qaoa_sm} using the Python packages NumPy~\cite{harris2020array} and Numba~\cite{lam2015numba}. Once the normalized quantum state $\vert\Psi\rangle_p$ is obtained, we sample bit strings $\boldsymbol{z}$ according to the probability distribution $\vert\langle\boldsymbol{z}\vert\Psi\rangle_p\vert^2$.

\section{Additional Data and Analyses}

\subsection{Relax-and-Round versus Goemans-Williamson Algorithms}
\label{sec:rr_comparisons_sm}

\begin{figure}[!t]
    \centering
    \includegraphics[width=0.5\textwidth]{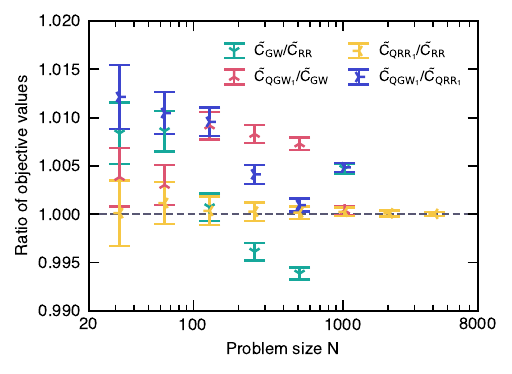}
    \caption{Ratio of the average of various objective values from different solvers as a function of the problem size $N$: $\tilde{C}_\textrm{GW}$ is returned by the classical Goemans-Williamson algorithm of App.~\ref{sec:cl_gw_sm}, $\tilde{C}_\textrm{RR}$ is returned by the classical relax-and-round algorithm of App.~\ref{sec:cl_relax_and_round_sm}, $\tilde{C}_{\textrm{QGW}_1}$ is returned by the quantum Goemans-Williamson algorithm with one layer of App.~\ref{sec:qgw_sm}, $\tilde{C}_{\textrm{QRR}_1}$ is returned by the quantum relax-and-round algorithm with one layer of App.~\ref{sec:qrr_sm}. Each data point is averaged over $1,000$ problem instances. Error bars indicate one standard deviation.}
    \label{fig:compare_rr_gw_q_cl}
\end{figure}

We compare the empirical performance of classical and quantum solvers based on a relaxation of the problem: classical relax-and-round (App.~\ref{sec:cl_relax_and_round_sm}), classical Goemans-Williamson (App.~\ref{sec:cl_gw_sm}), quantum relax-and-round (App.~\ref{sec:qrr_sm}), and quantum Goemans-Williamson (App.~\ref{sec:qgw_sm}) algorithms. The main difference between the algorithms is that Goemans-Williamson introduces extra degrees of freedom, such that the objective value of the relaxed version of the problem is tighter to the original binary one. However, whether this translates into a better sign-rounded solution is not guaranteed. We show in Fig.~\ref{fig:compare_rr_gw_q_cl} that the two algorithms have similar performances, yielding solutions within $1\%$ of each other. We also observe that for the two algorithms, the classical and quantum counterparts with $p=1$ layer have similar performances. This is explained and discussed further in App.~\ref{app:eq_classical_quantum_rr}.

For these reasons, we focus on the classical (App.~\ref{sec:cl_relax_and_round_sm}) and quantum relax-and-round (App.~\ref{sec:qrr_sm}) algorithms.

\subsection{Linear Versus Geometric Temperature Interpolation in Simulated Annealing}
\label{sec:sa_lin_geo_sm}

\begin{figure}[!t]
    \centering
    \includegraphics[width=0.5\textwidth]{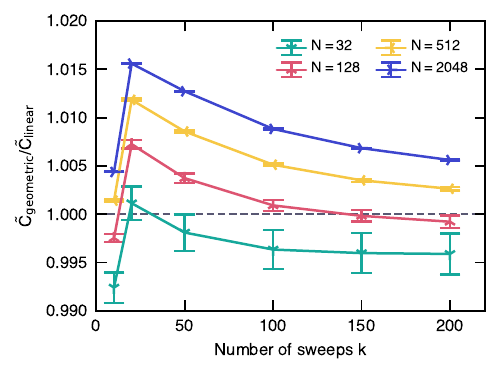}
    \caption{Ratio of the average objective values using geometric ($\tilde{C}_{\text{geometric}}$) and linear temperature ($\tilde{C}_{\text{linear}}$) schedules, as a function of the number of simulated annealing sweeps $k$. Each data point is averaged over $1,000$ randomly generated problem instances. Error bars indicate one standard deviation.}
    \label{fig:sa_linear_vs_geometric}
\end{figure}

In this section we compare the performance of linear and geometric temperature interpolation schemes for interpolating between the hot and cold temperatures used in simulated annealing (See App.~\ref{sec:sa_sm} and  Eqs.~\eqref{eq:lin_temp_sched_sa_sm} and~\eqref{eq:geo_temp_sched_sa_sm}). In Fig.~\ref{fig:sa_linear_vs_geometric}, we show the ratio of the average objective values using geometric ($\tilde{C}_{\text{geometric}}$) and linear temperature ($\tilde{C}_{\text{linear}}$) schedules, as a function of the number of simulated annealing sweeps $k$. We find that as the problem size $N$ increases ($N\gtrsim 128$), for a number of sweeps ranging from a few tens to a few hundreds, the geometric temperature schedule of Eq.~\eqref{eq:geo_temp_sched_sa_sm} yields better results. As the number of sweeps $k$ goes to infinity, both temperature schedules should return the same optimal solution, making the ratio converge to one. Throughout this work, we focus on the geometric temperature schedule, unless explicitly specified otherwise.

\subsection{Effect of the Number of Bit Strings for Expectation-Based Quantum Relax-and-Round Algorithm}
\label{app:number_of_bitstrings}

\begin{figure*}[!t]
    \centering
    \includegraphics[width=1.0\textwidth]{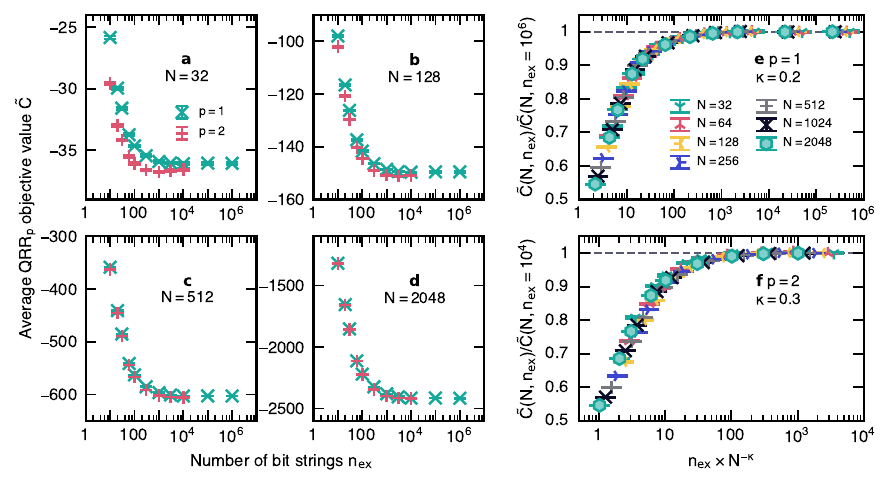}
    \caption{(a-d) Average objective value computed using the QRR algorithm based on the QAOA with $p=1$ and $p=2$ layers for different problem sizes $N$, as a function of the number of circuit executions $n_\textrm{ex}$ (and, thus, number of bit strings).  Correlation matrix elements were computed using Eq.~\eqref{eq:estimate_expectation_value_sm}. (a) $N=32$, (b) $N=128$, (c) $N=512$, and (d) $N=2048$. (e-f) Empirical scaling relation for the two-variable average objective value $\tilde{C}(N, n_\textrm{ex})$, relating $N$ to $n_\textrm{ex}$ following Eq.~\eqref{eq:scaling_relation_N_nex_sm}. (e) $p=1$ with $\kappa=0.2$. (f) $p=2$ with $\kappa=0.3$.  Each data point is averaged over $1,000$ problem instances. Error bars indicate one standard deviation.}
    \label{fig:bitstrings_effect_qrr}
\end{figure*}

We investigate the effect of the number of circuit executions (shots), $n_\textrm{ex}$, on the performance of the quantum relax-and-round (QRR) algorithm. Each circuit execution yields a bit string $\boldsymbol{z}\in{\pm 1}^N$ that is used to estimate the expectation value of two-point correlations
\begin{equation}
    \bigl\langle\hat{Z}_i\hat{Z}_j\bigr\rangle\approx\frac{1}{n_\textrm{ex}}\sum\nolimits_{k=1}^{n_\textrm{ex}}z^{(k)}_iz^{(k)}_j,
    \label{eq:estimate_expectation_value_sm}
\end{equation}
where $z^{(k)}_i$ is the value of bit $i$ in the $k$th bit string $\boldsymbol{z}^{(k)}$ from the $k$th execution of the circuit. The effect of a finite number of executions $n_\textrm{ex}$ can be modeled by a random component on the expectation values
\begin{equation}
    \bigl\langle\hat{Z}_i\hat{Z}_j\bigr\rangle(n_\textrm{ex})\sim\bigl\langle\hat{Z}_i\hat{Z}_j\bigr\rangle(n_\textrm{ex}\to+\infty) + \frac{\mathcal{N}(0,1)}{\sqrt{n_\textrm{ex}}},
    \label{eq:expectation_convergence_sm}
\end{equation}
where $\mathcal{N}(0,1)$ is a random normal variable of mean zero and unit variance. As per the central limit theorem, the expectation value converges asymptotically as $O(n_\textrm{ex}^{-1/2})$. In the context of the QRR algorithm, the effect of a finite $n_\textrm{ex}$ is that of performing eigendecomposition on a noisy symmetric matrix. Parameters such as the size $N$ of the correlation matrix, the level spacing between eigenvalues, the condition number of the matrix, and the sign-rounding step play a role in the perturbation strength of its eigenvectors~\cite{Chen2021}.

\subsubsection{Numerical Data}

In the absence of an analytical framework, we perform numerical state vector simulations with $p=1$ and $p=2$ QAOA layers for various problem sizes $N$ and number of circuit executions $n_\textrm{ex}$. In Figs.~\ref{fig:bitstrings_effect_qrr}a-d,  we show the average objective value for different problem sizes $N$, obtained using the QRR algorithm where the correlation matrix elements were estimated using Eq.~\eqref{eq:estimate_expectation_value_sm}. We observe that the objective value converges asymptotically with the number of bit strings. We find empirically that the two parameters $N$ and $n_\textrm{ex}$ in the average objective value $\tilde{C}(N, n_\textrm{ex})$ are connected by a scaling relation
\begin{equation}
    \tilde{C}\bigl(N, n_\textrm{ex}\bigr)\bigr/\tilde{C}\bigl(N, n_\textrm{ex}\to\infty\bigr)=\mathcal{F}\bigl(n_\textrm{ex}\times N^{-\kappa}\bigr),
    \label{eq:scaling_relation_N_nex_sm}
\end{equation}
where $\mathcal{F}$ is a scaling function. Using $\kappa=0.2$ for $p=1$ and $\kappa=0.3$ for $p=2$, respectively, leads to a satisfactory data collapse (Figs.~\ref{fig:bitstrings_effect_qrr}e and~\ref{fig:bitstrings_effect_qrr}f), supporting the scaling relation of Eq.~\eqref{eq:estimate_expectation_value_sm}. Such a scaling relation was also observed in the context of solving Sherrington-Kirkpatrick spin glasses~\cite{PhysRevLett.35.1792} with the QRR algorithm~\cite{Dupont2024}; albeit with a larger exponent $\kappa\approx 1.5$. Thus, at small $p$, the QRR algorithm is much more robust to the number of circuit executions $n_\textrm{ex}$ for unit-weight random $3$-regular graphs, which we attribute to the sparser nature of the correlation matrices, compared to Sherrington-Kirkpatrick problem instances. This is consistent with the exponent $\kappa$ increasing with $p$ in Figs.~\ref{fig:bitstrings_effect_qrr}e and~\ref{fig:bitstrings_effect_qrr}f.

\begin{figure*}[!t]
    \centering
    \includegraphics[width=1\textwidth]{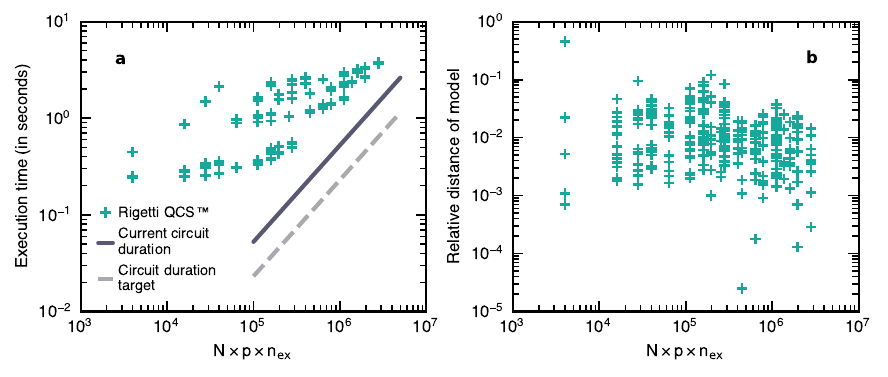}
    \caption{(a) Time in seconds for executing and collecting $n_\textrm{ex}$ bit strings from QAOA circuits involving $N$ qubits and $p$ layers, compiled to a linear chain topology using a full swap network~\cite{OGorman2019}. The plot shows the time to execute on the Ankaa\textsuperscript{TM}-2 Quantum Processing Unit accessed through Rigetti Quantum Cloud Services\textsuperscript{TM} (QCS), the current circuit duration and the basis for quantum timing in this paper (solid line), and a target execution time based on the best superconducting gate times in the literature. (b) Residuals between the model of Eq.~\ref{eq:qpu_run-time_model} and the experimental data in panel (a).}
    \label{fig:runtime_qpu}
\end{figure*}

\subsubsection{Implications}

\paragraph{Practical meaning---}

For the QRR algorithm (based on a one-layer QAOA circuit) to maintain a fixed performance on unit-weight random $3$-regular graphs, the number of quantum circuit executions should scale with the problem size as $n_\textrm{ex}\sim O(N^{0.2})$. Eq.~\eqref{eq:scaling_relation_N_nex_sm} and Fig.~\ref{fig:bitstrings_effect_qrr}e suggest that $\mathcal{F}(x)\approx 1$ for $x\approx 10^3$, i.e., in practice one wants to perform $n_\textrm{ex}\gtrsim 10^3N^{0.2}$ circuit executions. For the maximum problem size considered $N=4,096$, this means $n_\textrm{ex}\gtrsim 5,000$. A similar analysis can be performed at $p=2$ where $n_\textrm{ex}\gtrsim 10^2N^{0.3}$, yielding $n_\textrm{ex}\gtrsim 1,000$ for $N=4,096$. Here, we employ $n_\textrm{ex}=10^4$ independent of the problem size $N$, ensuring near-optimality of the QRR algorithm.

\paragraph{Algorithmic Complexity of the QRR---}

Finally, we note that in the absence of a step for finding an efficient swap network on a linear chain topology, encoding $p$ $N$-qubit phase separators takes at most $O(pN)$ layers of parallel two-qubit gates~\cite{OGorman2019}. Because circuit run-time is linear in the number of layers, maintaining a fixed performance for the QRR algorithm requires a run-time of $O(pN^{1+\kappa})$ with $\kappa\ll 1$ for the unit-weight random $3$-regular graphs considered. As such, collecting bit strings might be a limiting factor for the QRR algorithm, at least in terms of algorithmic complexity. In contrast, the eigendecomposition of the correlation matrix has an algorithmic complexity of $O(N)$. Nevertheless, for larger depth $p$ and intermediate sizes $N$, the eigendecomposition should tend toward an $O(N^2)$ scaling given that the correlation matrix becomes denser with increasing $p$. The number of bit strings required for asymptotic convergence of the QRR algorithm is unknown in this regime, but one expects that $n_\textrm{ex}=1$ bit string be sufficient in the $p\to+\infty$ limit given that it would correspond to the optimal solution. In this situation $\kappa=0$. Therefore, at fixed $N$, we expect a nontrivial relationship between the number of bit strings $n_\textrm{ex}$ required to achieve asymptotic convergence, and the depth $p$.

\subsection{Equivalence Between Classical and Quantum Relax-and-Round at \texorpdfstring{$p=1$}{p=1}}
\label{app:eq_classical_quantum_rr}

\subsubsection{Analytical Derivation Based on the QAOA at \texorpdfstring{$p=1$}{p=1}}

\begin{figure*}[!t]
    \centering
    \includegraphics[width=0.95\textwidth]{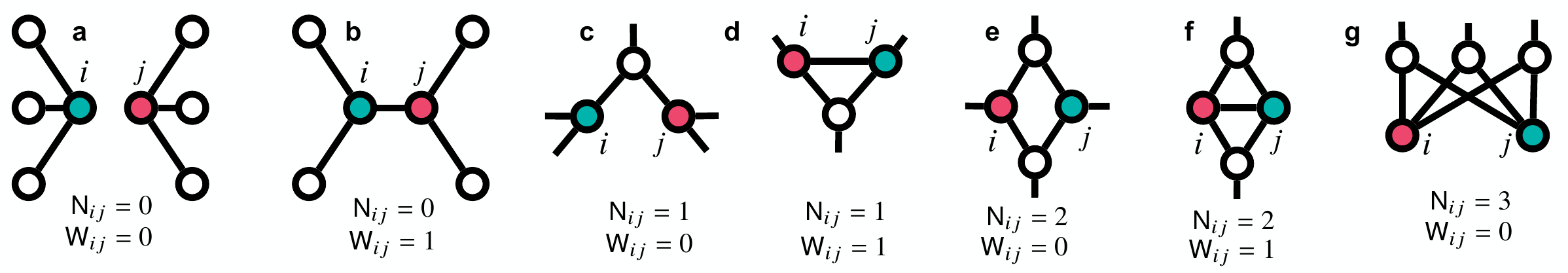}
    \caption{Local structures appearing in $3$-regular graphs with respect to two nodes $i$ and $j$. (a) $\mathsf{N}_{ij}=\mathsf{W}_{ij}=0$ (b) $\mathsf{N}_{ij}=0$ and $\mathsf{W}_{ij}=1$ (c) $\mathsf{N}_{ij}=1$ and $\mathsf{W}_{ij}=0$ (d) $\mathsf{N}_{ij}=1$ and $\mathsf{W}_{ij}=1$ (e) $\mathsf{N}_{ij}=2$ and $\mathsf{W}_{ij}=0$ (f) $\mathsf{N}_{ij}=2$ and $\mathsf{W}_{ij}=1$ (g) $\mathsf{N}_{ij}=3$ and $\mathsf{W}_{ij}=0$.}
    \label{fig:3regular_graph_structures}
\end{figure*}

At $p=1$, it is possible to obtain an analytic expression for the expectation value of the two-point correlation $\langle\hat{Z}_i\hat{Z}_j\rangle$ between qubits $i$ and $j$, as a function of the parameters $\gamma_1\equiv\gamma$ and $\beta_1\equiv\beta$. The expression was derived in App. B of Ref.~\cite{Dupont2024} and we simply state the result:
\begin{align}
    \bigl\langle\hat{Z}_i\hat{Z}_j\bigr\rangle=& -\sin\bigl(2\beta\bigr)\cos\bigl(2\beta\bigr)\sin\bigl(\gamma \mathsf{W}_{ij}\bigr)\nonumber\\
    &\times\Biggl[\prod\nolimits_{k\neq i,j}\cos\bigl(\gamma \mathsf{W}_{ik}\bigr)+\prod\nolimits_{k\neq i,j}\cos\bigl(\gamma \mathsf{W}_{jk}\bigr)\Biggr]\nonumber\\
    &-\frac{\sin^2\bigl(2\beta\bigr)}{2}\Biggl[\prod\nolimits_{k\neq i,j}\cos\gamma \bigl(\mathsf{W}_{ik}+\mathsf{W}_{jk}\bigr)\nonumber\\
    &-\prod\nolimits_{k\neq i,j}\cos\gamma\bigl(\mathsf{W}_{jk}-\mathsf{W}_{ik}\bigr)\Biggr].
    \label{eq:zz_formula_qaoa_p1}
\end{align}
For unit-weight random $3$-regular graphs, $\mathsf{W}_{ij}\in\{0,1\}$, so when $i\neq j$~\cite{Dupont2024} the above expression can be simplified to 
\begin{align}
    \bigl\langle\hat{Z}_i\hat{Z}_j\bigr\rangle_{i \ne j}=&-2\sin(2\beta)\cos(2\beta)\sin(\gamma)\cos^2(\gamma)\nonumber\\
    &\times\Biggl\{\mathsf{W}_{ij} + \frac{\tan\bigl(2\beta\bigr)\cos^2\bigl(\gamma\bigr)\sin\bigl(\gamma\bigr)}{2}\nonumber\\
    &\times \Bigl[\mathsf{N}_{ij} + \delta_{\mathsf{N}_{ij}3}\tan^4\bigl(\gamma\bigr) + \mathsf{W}_{ij}\mathsf{N}_{ij}\tan^2\bigl(\gamma\bigr)\Bigr]\Biggr\},
    \label{eq:zizj_3regular_graphs}
\end{align}
where $\mathsf{N}_{ij}=\sum_k\mathsf{W}_{ik}\mathsf{W}_{kj}$ and $\delta_{\mathsf{N}_{ij}3}$ is the Kronecker delta. The quantity $0\leq\mathsf{N}_{ij}\leq 3$ counts the number of times nodes $i$ and $j$ share a common nearest-neighbor $k$, independently of whether $i$ and $j$ are nearest-neighbors themselves. In Fig.~\ref{fig:3regular_graph_structures} we show the local structures around nodes $i$ and $j$ of $3$-regular graphs and the corresponding values of $\mathsf{N}_{ij}$ and $\mathsf{W}_{ij}$.

For $i\neq j$, the correlation matrix entries are simply the signed correlations $\langle\hat{Z}_i\hat{Z}_j\rangle_{i \ne j}$ in Eq.~\eqref{eq:zizj_3regular_graphs} and $\mathsf{Z}_{i=j}=0$ otherwise. In the following, we show that for unit-weight $3$-regular graphs, the correlation and adjacency matrices commute in the large $N$ limit. As a result, the two matrices share the same eigenvectors, assuming that either matrix has distinct eigenvalues. As such, a relax-and-round approach on either of the two matrices will provide the same solution. The matrix resulting from the commutator $[\mathsf{W},\mathsf{Z}]$ has entries
\begin{equation}
    \bigl[\mathsf{W}\mathsf{Z}\bigr]_{ij} - \bigl[\mathsf{Z}\mathsf{W}\bigr]_{ij}=\sum\nolimits_k\mathsf{W}_{ik}\mathsf{Z}_{kj}-\sum\nolimits_k\mathsf{Z}_{ik}\mathsf{W}_{kj}.
\end{equation}
There are four terms in $\mathsf{Z}_{ij}$ (see Eq.~\eqref{eq:zizj_3regular_graphs}) which we consider in order:
\begin{itemize}[label=$\circ$,leftmargin=12pt]

    \item The first one is $\propto\mathsf{W}_{ij}$: The adjacency matrix commutes with itself and the commutator corresponding to this term is zero for all angles $\beta$ and $\gamma$.

    \item The second term is $\propto\mathsf{N}_{ij}=\sum_k\mathsf{W}_{ik}\mathsf{W}_{kj}$: It is straightforward to show that the commutator entries are zero for all angles $\beta$ and $\gamma$ since $[\mathsf{W},\mathsf{N}]_{ij}=\sum_{kq}\mathsf{W}_{ik}\mathsf{W}_{kq}\mathsf{W}_{qj}-\sum_{kq}\mathsf{W}_{ik}\mathsf{W}_{kq}\mathsf{W}_{qj}=[\mathsf{W}^3]_{ij}-[\mathsf{W}^3]_{ij}=0$.

    \item The third term is $\propto\delta_{\mathsf{N}_{ij}3}$. Up to a prefactor depending on $\gamma$ and $\beta$, the commutator entries read $[\mathsf{W},\delta_{\mathsf{N}=3}]_{ij}=\sum_k\mathsf{W}_{ik}\delta_{\mathsf{N}_{kj}=3}-\sum_k\delta_{\mathsf{N}_{ik}=3}\mathsf{W}_{kj}$. According to Fig.~\ref{fig:3regular_graph_structures}g, if $\delta_{\mathsf{N}_{kj}=3}=1$ for two nodes $k$ and $j$, then, necessarily $\mathsf{W}_{ik}=1$ if and only if $\mathsf{W}_{jk}=\mathsf{W}_{kj}=0$ and $\mathsf{W}_{ij}=\mathsf{W}_{ji}=1$, and zero otherwise. As such, $[\mathsf{W},\delta_{\mathsf{N}=3}]_{ij}=0$ for all angles $\gamma$ and $\beta$.

    \item The last term is $\propto\mathsf{W}_{ij}\mathsf{N}_{ij}$. Unlike the three other cases, it is not straightforward to show whether or not the adjacency matrix commutes with this term for arbitrary unit-weight $3$-regular graphs and angles. According to Fig.~\ref{fig:3regular_graph_structures}, this term is nonzero for local structures involving a triangle, where $\mathsf{W}_{ij}=1$ and $\mathsf{N}_{ij}\geq 1$. Up to a constant which depends on angles $\beta$ and $\gamma$, the elements $(i,j)$ of the commutator read $\sum_k\mathsf{W}_{ik}\mathsf{W}_{kj}\mathsf{N}_{kj}-\sum_k\mathsf{W}_{ik}\mathsf{N}_{ik}\mathsf{W}_{kj}$ where $\mathsf{N}_{ij}=\sum_k\mathsf{W}_{ik}\mathsf{W}_{kj}$. Given that asymptotically with the problem size $N$, the average number of triangles in a random $3$-regular graph is $4/3$  independently of $N$ (see App.~\ref{sec:problem_instances_sm}), this term's contribution to the overall commutator will be vanishingly small in the limit of large $N$.
\end{itemize}
The adjacency $\mathsf{W}$ and correlation $\mathsf{Z}$ matrices therefore commute, up to a vanishingly small contribution as $N$ increases, arising from local $3$-node triangular structures in the graph. While this is strictly an approximation, its accuracy has been verified by simulations (see the next section). As such, $\mathsf{W}$ and $\mathsf{Z}$ share the same eigenvectors and a relax-and-round strategy on either matrix will lead to the same solution. We note that while the eigenvectors are the same, they may not necessarily be ordered in the same fashion. However, as the two relaxation problems on $\mathsf{W}$ and $\mathsf{Z}$ attempt to achieve the same goal, it is reasonable to assume a similar ordering.

\subsubsection{Numerical Data}

\begin{figure}[!t]
    \centering
    \includegraphics[width=0.5\textwidth]{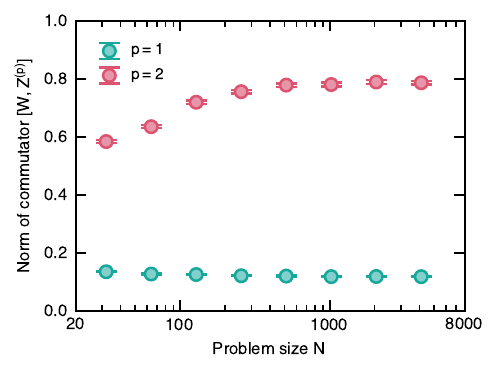}
    \caption{Norm of the commutator between the adjacency $\mathsf{W}$ and correlation $\mathsf{Z}^{(p)}$ matrices, as defined in Eq.~\eqref{eq:commutator_norm}, as a function of the problem size $N$. Each data point is averaged over $1,000$ problem instances. Error bars indicate one standard deviation.}
    \label{fig:commutator_norm_adj_corr_matrices}
\end{figure}

To evaluate the accuracy of the commutation relation between the adjacency and correlation matrices, we consider the operator norm $\|\cdot\|_2$ of the commutator $[\mathsf{W}, \mathsf{Z}^{(p)}]$
\begin{equation}
    \Bigl\|\bigl[\mathsf{W}, \mathsf{Z}^{(p)}\bigr]\Bigr\|_2 = \Bigl\|\mathsf{W}\mathsf{Z}^{(p)} - \mathsf{Z}^{(p)}\mathsf{W}\Bigr\|_2 = \sigma_\textrm{max},
    \label{eq:commutator_norm}
\end{equation}
where $\sigma_\textrm{max}$ is the largest singular value of the matrix resulting from the commutator. It is shown in Fig.~\ref{fig:commutator_norm_adj_corr_matrices} as a function of the problem size $N$ for $p=1$ and $p=2$. At $p=1$, we find that the norm of the commutator slowly decreases with $N$, in line with theoretical expectations. Although the norm of the commutator is not strictly zero at finite $N$, we verify that the solutions returned by the classical and quantum $(p=1)$ relax-and-round methods are the same (see App.~\ref{app:eq_classical_quantum_rr}). At $p=2$, the norm increases with $N$, in stark contrast to the behavior for $p=1$. Indeed, for $p>1$, the correlation matrix is not expected to commute with the adjacency matrix.

\subsection{Evaluating the Optimal Solution for Computing the Approximation Ratio}

In order to compute the approximation ratio $\alpha$ (Eq.~\eqref{eq:app_ratio}) of a solution $\boldsymbol{z}$, one needs to know the problem's optimal solution, $\boldsymbol{z}_\textrm{opt}$. Here, we employ the best-in-class Burer-Monteiro classical heuristic~\cite{doi:10.1137/S1052623400382467}, implemented in the library MQLib~\cite{DunningEtAl2018}, and introduced in App.~\ref{sec:burer_monteiro_sm}.

While the Burer-Monteiro algorithm cannot guarantee that it will find the optimal solution, the likelihood of improving the average quality solution increases the longer the algorithm runs. Moreover, by initializing the solver with different seeds and examining the distribution of solutions, the existence of a peak at a minimum objective value $\tilde{C}$ is a good indication of the optimality of the solution.

For $N\leq 512$, we limited the total run-time to one second and found that each problem instance was solved about $100$ times, with a peak in the distribution of solutions at a minimum objective value $\tilde{C}$. Thus, we are quite confident in the optimality of the solution used for estimating the approximation ratio.

For $N\geq 1,024$, we ran the algorithm for a few hundred seconds for each problem instance. The limited statistics were insufficient to estimate the distribution of solutions, lowering our confidence in the genuine optimality of the solution used for computing the approximation ratio.

\subsection{Performance Gain from the Greedy-Enhanced Quantum-Relax-and-Round Solver}
\label{sec:grdy_enh_qrr_sm}

We study the performance gain of the greedy-enhanced quantum relax-and-round solver (App.~\ref{sec:grdy_enh_qrr_alg_sm}) over its non-enhanced version (App.~\ref{sec:qrr_sm}). We implement two versions, where sign-flips are attempted on variables that are: (i) randomly selected; and (ii) guided by the non-rounded eigenvector entries. We define the numbers of variables visited as $fN$ for an $N$-variable problem with $f\in\mathbb{R}$ a control parameter.

\begin{figure*}[!t]
    \centering
    \includegraphics[width=1.0\textwidth]{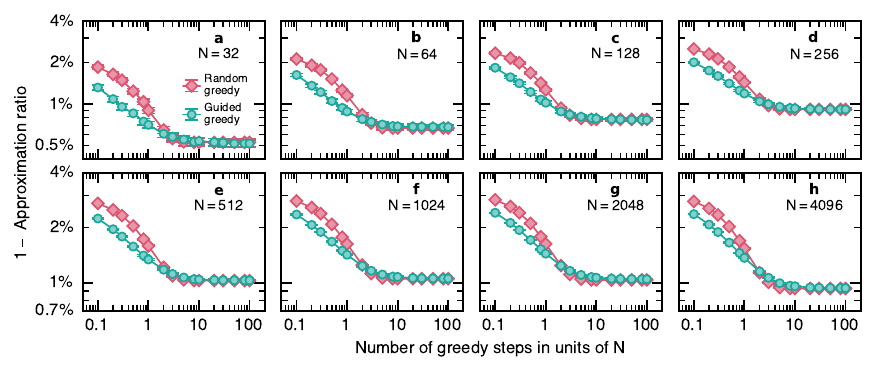}
    \caption{Average distance to the perfect approximation ratio, as a function of the number of greedy steps. Results are presented for both the random and guided sign-flip strategies with problem sizes: (a) $N=32$, (b) $N=64$, (c) $N=128$, (d) $N=256$, (e) $N=512$, (f) $N=1024$, (g) $N=2048$, and (h) $N=4096$. Each data point is averaged over $1,000$ randomly generated problem instances. Error bars indicate one standard deviation.  Number of steps are in units of $N$ i.e., $f$.}
    \label{fig:performance_vs_ngreedy_steps}
\end{figure*}

We show in Fig.~\ref{fig:performance_vs_ngreedy_steps} the average distance to the perfect approximation ratio as a function of $f$ for both sign-flipping strategies and various problem sizes $N$. We find that they both converge toward the same, higher, average approximation ratio as compared to the nonenhanced standard version. We observe that the guided version converges faster, but the random version asymptotically yields the same average approximation ratio. The convergence occurs with $f\sim O(1)$, independently of $N$. This means the random version visits all variables $O(1)$ times on average, thus enabling it to visit variables that the guided version identifies as being favorable, albeit with somewhat slower convergence.

As currently implemented, the greedy sign-flip strategy acts like a local gradient descent method in the sense that it may get stuck in local minima. Allowing sign-flips that increase the objective value may enable the algorithm to find solutions with better approximation ratios in the large $f$ limit, similar to a Markov-Chain Monte Carlo algorithm initialized with the solution of the original quantum-relax-and-round algorithm.

\subsection{Comparison of Various Combinatorial Optimization Solvers}
\label{sec:comparison_solvers}

\begin{figure*}[!ht]
    \centering
    \includegraphics[width=1\textwidth]{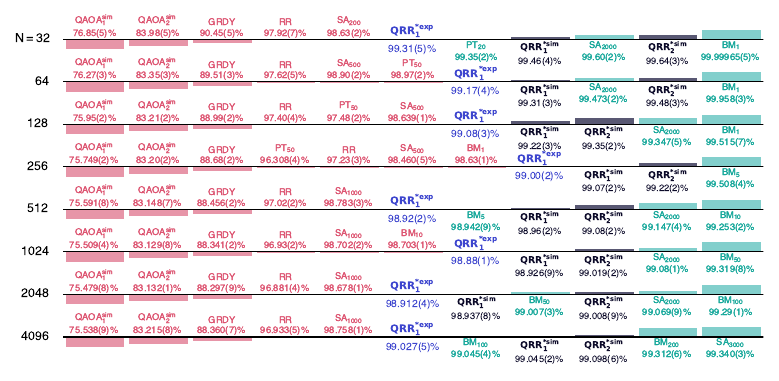}
    \caption{Approximation ratio averaged over $1,000$ randomly generated $3$-regular graph problem instances, computed using a range of solvers for each problem size $N$. Solvers are: ``QAOA$_p$'' (QAOA with $p$ layers); ``GRDY'' (a classical greedy solver); ``RR'' (a classical relax-and-round solver); ``SA$_k$'' (simulated annealing with $k$ sweeps); ``PT$_k$'' (parallel tempering with $k$ sweeps); ``BM$_t$'' (the Burer-Monteiro solver with maximum run-time $t$ in milliseconds; and  ``QRR*$_p$ (greedy-enhanced quantum relax-and-round-solver based on the QAOA with $p$ layers). For the quantum solvers, ``sim'' and ``exp'' indicate emulated and experimental results, respectively. The approximation ratio is presented as a percentage below each solver, with the standard error in parentheses.}
    \label{fig:average_performance_comparison}
\end{figure*}

In this section, we benchmark the performance of QRR* against a suite of methods. The results are shown in Fig.~\ref{fig:average_performance_comparison}. The comparison includes the QAOA (App. ~\ref{sec:qaoa_sm}), the most competitive classical solvers (simulated annealing and the Burer-Monteiro algorithm; see Appendices~\ref{sec:sa_sm} and~\ref{sec:burer_monteiro_sm}, respectively), and the following:
\begin{description}
    \item[Classical Greedy Solver] The classical greedy solver (App.~\ref{sec:greedy_solver_sm}) serves as a comparison point against quantum iterative solvers bearing similarities~\cite{PhysRevLett.125.260505,Dupont2023,Brady2023}. Fig.~\ref{fig:average_performance_comparison} shows the greedy solver is not competitive against QRR*. However, it should be noted that it performs better than the QAOA at $p=2$, highlighting the need to embed the QAOA into hybrid quantum-classical workflows (such as the QRR algorithm) to get the most out of it.

    \item[Classical Relax-and-Round Solver] The classical relax-and-round solver was compared against the classical and quantum Goemans-Williamson methods in App.~\ref{sec:rr_comparisons_sm}. In the main text we also discussed how the QRR algorithm at $p=1$ is equivalent to the classical relax-and-round solver. Fig.~\ref{fig:average_performance_comparison} shows that the relax-and-round solver yields an average approximation ratio of about $97\%$, much higher than both the classical greedy algorithm and the QAOA.

    \item[Parallel Tempering] Parallel tempering was not considered in the performance comparison presented in the main text because it was not competitive. We found that at fixed resources (number of sweeps for simulated annealing and number of sweeps multiplied by the number of replicas for parallel tempering), simulated annealing performed better for the largest system sizes. It should be noted that all the classical solvers were tested in a single-core context, to enable a fair comparison against the quantum solver. Parallel tempering is particularly well-suited to parallel computation using multiple computing cores. Indeed, each replica is independent with communication needed only for exchanging nearby replicas.
\end{description}

\subsection{Measuring the Run-time of Various Combinatorial Optimization Solvers}

\subsubsection{Run-time of the Greedy-Enhanced Quantum Relax-and-Round Solver}
\label{app:run-time_qrr_solver}

The greedy-enhanced quantum relax-and-round solver (QRR*) is the reference quantum solver used throughout this work. Its run-time can be decomposed into four blocks: (i) run-time on the quantum processor for collecting bit strings; (ii) building the correlation matrix from the bit strings; (iii) computing the leading eigenvectors of the correlation matrix and sign-rounding them; and (iv) performing a greedy local search on the solution.

The run-time on the quantum processor assumes that the circuit is built using a worst-case-scenario swap network on a hardware-native linear chain topology of qubits, which requires $Np$ layers of two-qubit gates~\cite{OGorman2019}. An efficient compilation technique, such as the one described in App.~\ref{sec:swap_network_sm}, is not used. Moreover, the run-time on the quantum processor does not rely on the graph isomorphism described in App.~\ref{sec:graph_isomorphism_sm}, which was just a convenience for the purpose of this work to obtain statistics on thousands of problem instances. The run-time assumes that one has to execute a circuit for each of the nonzero entries of the correlation matrix.

\begin{figure*}[!t]
    \centering
    \includegraphics[width=1.0\textwidth]{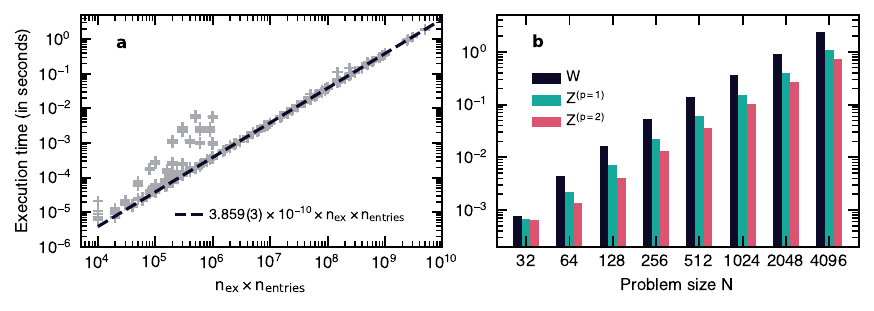}
    \caption{(a) Run-time for computing the correlation matrix made of $n_\textrm{entries}$ nonzero entries from $n_\textrm{ex}$ bit strings. The run-time is linear per entry and bit string, with the prefactor approximately $3.859(3)\times 10^{-10}$ per least-squares fitting. (b) Run-time for the eigendecomposition of the $N \times N$ adjacency ($\mathsf{W}$) and correlation matrices ($\mathsf{Z}$) at $p=1$ and $p=2$ using the Lanczos algorithm. Each data point is averaged over $20$ matrices. Units in both panels: seconds.}
    \label{fig:runtime_qrr}
\end{figure*}

\paragraph{Run-time on the Quantum Processor---}

We can measure the time it takes to execute a QAOA circuit compiled to a linear chain topology using a full swap network~\cite{OGorman2019} on Rigetti Quantum Cloud Services\textsuperscript{TM} (QCS) by instrumenting calls to the cloud service API as exposed by the Pyquil open-source software package. Using the Ankaa\textsuperscript{TM}-2 Quantum Processing Unit (QPU) with $84$ qubits, we compiled $300$ test circuits across a range of $N$ (number of qubits), $p$ (number of QAOA layers), and $n_\textrm{ex}$ (number of bit strings collected). The test circuits implemented the QAOA, compiled to a linear chain topology using a full swap network~\cite{OGorman2019}. The results are shown in Fig.~\ref{fig:runtime_qpu}a.

The QCS run-time can be modeled as
\begin{align}
    t_\textrm{ex} =& c_0 + c_N N + c_{n_\textrm{ex}} n_\textrm{ex} + c_{N.p} Np\nonumber\\
    &+ c_{N.n_\textrm{ex}} Nn_\textrm{ex} + c_{N.p.n_\textrm{ex}} Np n_\textrm{ex},
    \label{eq:qpu_run-time_model}
\end{align}
where each $c_i$ is a prefactor for a specific computational expense that might be improved. When we fit this model to the measured data using least-squares fitting over a linear residual, we obtain the coefficients of Tab.~\ref{tab:run-time_qpu_fit} such that $t_\textrm{ex}$ is in seconds. The residual of the fit is illustrated in Fig.~\ref{fig:runtime_qpu}b.

\begin{table*}[t!]
    \centering
    \begin{tabular}{c@{\hskip 20pt}c@{\hskip 20pt}l}
        \hline\hline\\[-0.8em]
        Prefactor & Value & Description \\
        \hline\\[-0.8em]
        $c_0$            & $2.77 \times 10^{-2}$ & Constant overheads include network latency and QCS\textsuperscript{TM} job management\\
        $c_N$            & $1.07 \times 10^{-3}$ & The number of modular control system cards utilized grows $\propto N$\\
        $c_{n_\textrm{ex}}$     & $2.04 \times 10^{-4}$ & Per-sample cost ascribed to reset delay and qubit measurement\\
        $c_{N.p}$        & $3.83 \times 10^{-4}$ & The size of the program on each control system card grows $\propto Np$\\
        $c_{N.n_\textrm{ex}}$   & $8.37 \times 10^{-7}$ & The number of measured bits returned to user memory grows $\propto N n_\textrm{ex}$\\
        $c_{N.p.n_\textrm{ex}}$ & $4.46 \times 10^{-7}$ & Per-gate-layer cost within a linear-chain QAOA ansatz grows $\propto N p n_\textrm{ex}$\\
        \hline\hline\\[-0.8em]
    \end{tabular}
    \caption{Prefactor values for the run-time model of Eq.~\eqref{eq:qpu_run-time_model} (in seconds) on the Rigetti QCS\textsuperscript{TM} data of Fig.~\ref{fig:runtime_qpu}.}
    \label{tab:run-time_qpu_fit}
\end{table*}

We anticipate $c_0$ could be minimized by reducing network and service latency, or by in-service integration of the QRR* algorithm. We suspect $c_N$ might be eliminated by horizontally scaling classical compute resources so all modular parts of the control system are initialized in $O(1)$ time. $c_{N.p}$ could be improved by increasing the rate at which individual cards in the control system can be programmed, while $c_{N.n_\textrm{ex}}$ could be improved with faster measurement classification and higher bandwidth in returning bit strings to the QRR* algorithm. $c_{n_\textrm{ex}}$ resolves to the expected value accounting for a default passive reset delay ($200~\mu\text{s}$) and measurement operations, and could be improved by an active reset protocol~\cite{Karalekas_2020}. Finally, $c_{N.p.n_\textrm{ex}}$ establishes the per-gate-layer cost, given the total number of layers in a linear-chain QAOA ansatz of depth $p$ is $Np$~\cite{OGorman2019}; the fitted value of $446$ nanoseconds is generally consistent with the timing of three $\texttt{ISWAP}$ plus one-qubit gates (see App.~\ref{sec:compilation_sm}), noting that this measurement was conducted across a different set of qubits and with the possibility of parallel one-qubit and two-qubit operations.

Assuming adoption of an active reset protocol~\cite{Karalekas_2020} and elimination of all other overheads tabulated above, we establish the quantum computer run-time in our analyses of Fig.~\ref{fig:runtime_to_match_qrr} of the main text as the current circuit duration for the performance data of Tab.~\ref{tab:qubits_characteristics}. Precisely
\begin{equation}
    \tilde{t}_\textrm{ex} = n_\textrm{ex}\Bigl[t_\textrm{init} + p\bigl(t_\textrm{mx} + Nt_\textrm{ps}\bigr) + t_\textrm{mes+res}\Bigr],
    \label{eq:circuit_exec_run-time}
\end{equation}
where $t_\textrm{init}$ accounts for the initial Hadamard gate, $t_\textrm{mx}$ the mixer time, $t_\textrm{ps}$ the phase separator time based on a linear architecture topology requiring a swap network~\cite{OGorman2019}, and $t_\textrm{mes+res}$ the measurement time and active reset of the qubit before the next run. The compilation overhead of the QAOA circuit into hardware-native gates (see App.~\ref{sec:compilation_sm}) leads to
\begin{equation}
    t_\textrm{init} = t_\textrm{1Q},\quad t_\textrm{mx} = 2t_\textrm{1Q},\quad t_\textrm{ps} = 4t_\textrm{1Q} + 3t_\textrm{2Q},
\end{equation}
where $t_\textrm{1Q}\equiv t_\texttt{Rx}$ and $t_\textrm{2Q}\equiv t_\texttt{ISWAP}$ are the one- and two-qubit gate durations, respectively. The measurement time includes reading the qubit ($\simeq 1\mu\text{s}$) and performing an active reset based on its measured value, which involves applying a one-qubit gate and a classical feedback-loop limited by network latency of $\simeq 1~\mu\text{s}$. This reading/active reset strategy is repeated three times in a row, resulting in $t_\textrm{mes+res}\simeq 6~\mu\text{s}$~\cite{Karalekas_2020}. Current average values for the one- and two-qubit gates operations are reported in Tab.~\ref{tab:qubits_characteristics} where $t_\texttt{Rx}=40$ ns and $t_\texttt{ISWAP}=122$ ns. This model, shown as the solid line in Fig.~\ref{fig:runtime_qpu}, is approximately three times faster than the asymptotic performance measured on Rigetti QCS\textsuperscript{TM}.

A further threefold reduction in quantum computer run-time could be achieved by improving gate speeds toward the state-of-the-art for transmon-based superconducting quantum processors~\cite{Arute2019}; this is shown as the dashed line in Fig.~\ref{fig:runtime_qpu} where $t_\texttt{Rx}=20$ ns and $t_\texttt{ISWAP}=50$ ns are used. This model was used in Fig.~\ref{fig:runtime_vs_performance} of the main text.

\paragraph{The Light Cone Technique on the Quantum Processor---}

The light cone technique trades the simulation of a unique large QAOA circuit for many smaller ones. On a chip of $M$ qubits, one may be able to fit several such subcircuits, thus reducing the overall run-time burden of the light cone decomposition. From the data of App.~\ref{sec:light_cone_sm}, we can estimate the average size of the subcircuits for different values of $(N, p)$ and the average number of such subcircuits for a given problem as a function of $(N, p)$. Without relying on graph isomorphism the average number of such subcircuits is $O(N)$. We do not rely on graph isomorphism for estimating run-times, given its poor algorithmic scaling.

To establish the run-time of the quantum relax-and-round algorithm, we consider the number of times the $M$-qubit quantum computer needs to be used to compute all nonzero elements of the correlation matrix, given an average number of subcircuits of average size. The depth of those average-size subcircuits is related to using a complete swap network embedding of the phase separator component of the QAOA circuit onto a linear hardware-native topology (see App.~\ref{sec:swap_network_sm}) where no optimal embedding is sought~\cite{OGorman2019}. For example, if one has an $M=84$-qubit quantum computer and needs to execute $180$ different subcircuits involving on average $7$ qubits, $12$ such subcircuits can be executed in parallel across the full quantum computer. This operation will need to be repeated $15$ times to obtain results for all subcircuits. 

\paragraph{Run-time on the Classical Processor---}

At this stage, we assume we have $n_\textrm{ex}$ bit strings $\boldsymbol{b}=\{0,1\}^N$, each of length $N$. First, we loop over all nontrivially zero entries $n_\textrm{entries}$ of the upper triangle of the $N\times N$ correlation matrix $\mathsf{Z}$. For each such entry $\mathsf{Z}_{ij}=(\delta_{ij}-1)\langle\hat{Z}_i\hat{Z}_j\rangle$, we estimate the expectation value 
\begin{equation}
    \bigl\langle\hat{Z}_i\hat{Z}_j\bigr\rangle\approx\frac{4}{n_\textrm{ex}}\sum\nolimits_{k=1}^{n_\textrm{ex}}\left(b^{(k)}_i-\frac{1}{2}\right)\left(b^{(k)}_j-\frac{1}{2}\right),
    \label{eq:correlation_estimate_value_sm}
\end{equation}
where $b^{(k)}_i$ is the value of bit $i$ in the $k$th bit string $\boldsymbol{b}^{(k)}$ from the $k$th quantum circuit execution. We report in Fig.~\ref{fig:runtime_qrr}a the run-time of a C++ implementation of the above strategy on a $64$Gb MacBook Pro with an Apple M1 Max chip. We find that the run-time per bit string and entry is linear and can be fitted to $3.859(3)\times 10^{-10}\times n_\textrm{entries} \times n_\textrm{ex}$.

Once the correlation matrix $\mathsf{Z}$ has been constructed, the next step is to compute its leading eigenvectors and sign-round them. We employ a C++ implementation of the Lanczos method to find the $k=8$ leading eigenvectors of the correlation matrix. Each of the eigenvectors is then sign-rounded and their corresponding objective values computed. The one with the best objective value is returned as the solution of the quantum relax-and-round algorithm. The run-time on a $64$Gb MacBook Pro with an Apple M1 Max chip is reported in Fig.~\ref{fig:runtime_qrr}b for the adjacency ($\mathsf{W}$) and correlation matrices ($\mathsf{Z}$) at $p=1$ and $p=2$ with different sizes $N$. There is no clear dependence on the number of entries in the matrix ($\mathsf{W}$ is sparser than $\mathsf{Z}^{(p=1)}$, itself sparser than $\mathsf{Z}^{(p=2)}$). We attribute that to the faster convergence of Lanczos with better-conditioned matrices. When needed, we will assume that the eigendecomposition for the correlation matrix at $p=3$ has the same run-time as for $p=2$.

\paragraph{Tabulated Run-time Summary for Quantum Solver---}

The average run-time of the greedy-enhanced quantum relax-and-round solver (QRR*) for different scenarios and number of variables $N$ are in Tab.~\ref{tab:qrr_run-time}. We report $t_\textrm{Q}$ and $t_\textrm{C}$, which are the run-times of the quantum and the classical components, respectively. These values were used for generating Figs. 2 and 3 in the main text. The quantum run-time is based on
\begin{align}
    t_\textrm{Q}=&\mathbb{E}\left[{\tilde{t}_\textrm{ex}(N,p)}\right]\nonumber\\
    &\times\left\lceil\mathbb{E}\left[{n_\textrm{subgraphs}(N,p)}\right] \frac{\mathbb{E}\left[{n_\textrm{qubits}(N,p)}\right]}{M}\right\rceil,
    \label{eq:quantum_run-time}
\end{align}
where $\mathbb{E}[\cdot]$ denotes the ensemble average over the subgraphs (light-cone induced subcircuits), $\tilde{t}_\textrm{ex}$ the circuit duration based on Eq.~\eqref{eq:circuit_exec_run-time}, $n_\textrm{subgraphs}$ the number of subgraphs (App.~\ref{sec:graph_isomorphism_sm}), $n_\textrm{qubits}$ the number of qubits for encoding a subgraph. These numbers depend on the number of variables $N$ of the problem and the number of QAOA layers $p$. $M$ denotes the number of qubits available on the quantum computer.

\begin{table*}[!ht]
    \centering
    \begin{tabular}{l@{\hskip 16pt}c@{\hskip 22pt}c@{\hskip 22pt}c@{\hskip 22pt}c@{\hskip 22pt}c@{\hskip 22pt}}
        \hline\hline\\[-0.8em]
        \makecell[l]{} & \multicolumn{3}{c}{\makecell[c]{$p=1$, $n_\textrm{ex}=5\times 10^3$}} & \makecell[c]{$p=2$, $n_\textrm{ex}=10^3$} & \makecell[c]{$p=3$, $n_\textrm{ex}=5\times 10^2$}\\
        \hline\\[-0.8em]
        \makecell[l]{\textbf{Number}\\\textbf{of}\\\textbf{variables}} & \makecell[c]{$M=10^2$,\\$t_{\texttt{Rx}}=40$ ns,\\$t_{\texttt{ISWAP}}=122$ ns} & \makecell[c]{$M=10^3$,\\$t_{\texttt{Rx}}=40$ ns,\\$t_{\texttt{ISWAP}}=122$ ns} & \makecell[c]{$M=10^3$,\\$t_{\texttt{Rx}}=20$ ns,\\$t_{\texttt{ISWAP}}=50$ ns} & \makecell[c]{$M=10^3$,\\$t_{\texttt{Rx}}=20$ ns,\\$t_{\texttt{ISWAP}}=50$ ns} & \makecell[c]{$M=10^3$,\\$t_{\texttt{Rx}}=20$ ns,\\$t_{\texttt{ISWAP}}=50$ ns}\\
        \hline\\[-0.8em]
        \makecell[l]{$N=32$} & \makecell[c]{$t_\textrm{Q}=425.29$~ms\\$t_\textrm{C}=0.94$~ms\\$\textbf{=426.23}$~\textbf{ms}} & \makecell[c]{$t_\textrm{Q}=47.25$~ms\\$t_\textrm{C}=0.94$~ms\\$\textbf{=48.20}$~\textbf{ms}} & \makecell[c]{$t_\textrm{Q}=37.65$~ms\\$t_\textrm{C}=0.94$~ms\\$\textbf{=38.60}$~\textbf{ms}} & \makecell[c]{$t_\textrm{Q}=92.48$~ms\\$t_\textrm{C}=0.82$~ms\\$\textbf{=93.29}$~\textbf{ms}} & \makecell[c]{$t_\textrm{Q}=153.19$~ms\\$t_\textrm{C}=0.76$~ms\\$\textbf{=153.95}$~\textbf{ms}}\\
        \hline\\[-0.8em]
        \makecell[l]{$N=64$} & \makecell[c]{$t_\textrm{Q}=900.51$~ms\\$t_\textrm{C}=2.78$~ms\\$\textbf{=903.28}$~\textbf{ms}} & \makecell[c]{$t_\textrm{Q}=94.79$~ms\\$t_\textrm{C}=2.78$~ms\\$\textbf{=97.57}$~\textbf{ms}} & \makecell[c]{$t_\textrm{Q}=75.43$~ms\\$t_\textrm{C}=2.78$~ms\\$\textbf{=97.57}$~\textbf{ms}} & \makecell[c]{$t_\textrm{Q}=261.96$~ms\\$t_\textrm{C}=1.79$~ms\\$\textbf{=263.75}$~\textbf{ms}} & \makecell[c]{$t_\textrm{Q}=945.35$~ms\\$t_\textrm{C}=1.75$~ms\\$\textbf{=947.10}$~\textbf{ms}}\\
        \hline\\[-0.8em]
        \makecell[l]{$N=128$} & \makecell[c]{$t_\textrm{Q}=1,803.23$~ms\\$t_\textrm{C}=8.32$~ms\\$\textbf{=1,811.54}$~\textbf{ms}} & \makecell[c]{$t_\textrm{Q}=189.81$~ms\\$t_\textrm{C}=8.32$~ms\\$\textbf{=198.13}$~\textbf{ms}} & \makecell[c]{$t_\textrm{Q}=150.96$~ms\\$t_\textrm{C}=8.32$~ms\\$\textbf{=159.28}$~\textbf{ms}} & \makecell[c]{$t_\textrm{Q}=616.94$~ms\\$t_\textrm{C}=5.03$~ms\\$\textbf{=621.97}$~\textbf{ms}} & \makecell[c]{$t_\textrm{Q}=3,820.99$~ms\\$t_\textrm{C}=5.30$~ms\\$\textbf{=3,826.29}$~\textbf{ms}}\\
        \hline\\[-0.8em]
        \makecell[l]{$N=256$} & \makecell[c]{$t_\textrm{Q}=3,656.26$~ms\\$t_\textrm{C}=24.60$~ms\\$\textbf{=3,680.86}$~\textbf{ms}} & \makecell[c]{$t_\textrm{Q}=379.87$~ms\\$t_\textrm{C}=24.60$~ms\\$\textbf{=404.47}$~\textbf{ms}} & \makecell[c]{$t_\textrm{Q}=302.03$~ms\\$t_\textrm{C}=24.60$~ms\\$\textbf{=326.63}$~\textbf{ms}} & \makecell[c]{$t_\textrm{Q}=1,356.93$~ms\\$t_\textrm{C}=15.18$~ms\\$\textbf{=1,372.11}$~\textbf{ms}} & \makecell[c]{$t_\textrm{Q}=11,482.77$~ms\\$t_\textrm{C}=16.45$~ms\\$\textbf{=11,499.22}$~\textbf{ms}}\\
        \hline\\[-0.8em]
        \makecell[l]{$N=512$} & \makecell[c]{$t_\textrm{Q}=7,267.29$~ms\\$t_\textrm{C}=64.04$~ms\\$\textbf{=7,331.33}$~\textbf{ms}} & \makecell[c]{$t_\textrm{Q}=759.98$~ms\\$t_\textrm{C}=64.04$~ms\\$\textbf{=824.02}$~\textbf{ms}} & \makecell[c]{$t_\textrm{Q}=604.16$~ms\\$t_\textrm{C}=64.04$~ms\\$\textbf{=668.21}$~\textbf{ms}} & \makecell[c]{$t_\textrm{Q}=2,823.41$~ms\\$t_\textrm{C}=40.48$~ms\\$\textbf{=2,863.88}$~\textbf{ms}} & \makecell[c]{$t_\textrm{Q}=28,547.53$~ms\\$t_\textrm{C}=44.13$~ms\\$\textbf{=28,591.66}$~\textbf{ms}}\\
        \hline\\[-0.8em]
        \makecell[l]{$N=1,024$} & \makecell[c]{$t_\textrm{Q}=14,584.41$~ms\\$t_\textrm{C}=163.10$~ms\\$\textbf{=14,747.51}$~\textbf{ms}} & \makecell[c]{$t_\textrm{Q}=1,472.69$~ms\\$t_\textrm{C}=163.10$~ms\\$\textbf{=1,635.80}$~\textbf{ms}} & \makecell[c]{$t_\textrm{Q}=1,170.67$~ms\\$t_\textrm{C}=163.10$~ms\\$\textbf{=1,333.77}$~\textbf{ms}} & \makecell[c]{$t_\textrm{Q}=5,785.37$~ms\\$t_\textrm{C}=109.32$~ms\\$\textbf{=5,894.68}$~\textbf{ms}} & \makecell[c]{$t_\textrm{Q}=62,757.71$~ms\\$t_\textrm{C}=117.74$~ms\\$\textbf{=62,875.45}$~\textbf{ms}}\\
        \hline\\[-0.8em]
        \makecell[l]{$N=2,048$} & \makecell[c]{$t_\textrm{Q}=29,170.96$~ms\\$t_\textrm{C}=411.97$~ms\\$\textbf{=29,582.93}$~\textbf{ms}} & \makecell[c]{$t_\textrm{Q}=2,445.60$~ms\\$t_\textrm{C}=411.97$~ms\\$\textbf{=3,357.57}$~\textbf{ms}} & \makecell[c]{$t_\textrm{Q}=2,341.43$~ms\\$t_\textrm{C}=411.97$~ms\\$\textbf{=2,753.40}$~\textbf{ms}} & \makecell[c]{$t_\textrm{Q}=11,680.59$~ms\\$t_\textrm{C}=280.52$~ms\\$\textbf{=11,961.11}$~\textbf{ms}} & \makecell[c]{$t_\textrm{Q}=132,025.62$~ms\\$t_\textrm{C}=298.68$~ms\\$\textbf{=132,324.30}$~\textbf{ms}}\\
        \hline\\[-0.8em]
        \makecell[l]{$N=4,096$} & \makecell[c]{$t_\textrm{Q}=58,391.69$~ms\\$t_\textrm{C}=1,107.77$~ms\\$\textbf{=59,499.46}$~\textbf{ms}} & \makecell[c]{$t_\textrm{Q}=5,843.92$~ms\\$t_\textrm{C}=1,107.77$~ms\\$\textbf{=6,951.69}$~\textbf{ms}} & \makecell[c]{$t_\textrm{Q}=4,645.20$~ms\\$t_\textrm{C}=1,107.77$~ms\\$\textbf{=5,752.98}$~\textbf{ms}} & \makecell[c]{$t_\textrm{Q}=23,499.65$~ms\\$t_\textrm{C}=756.55$~ms\\$\textbf{=24,256.20}$~\textbf{ms}} & \makecell[c]{$t_\textrm{Q}=271,015.06$~ms\\$t_\textrm{C}=794.26$~ms\\$\textbf{=271,809.32}$~\textbf{ms}}\\
        \hline\hline\\[-0.8em]
    \end{tabular}
    \caption{Average run-time in milliseconds (ms) of the greedy-enhanced quantum relax-and-round solver (QRR*) for different numbers of variables $N$ on random $3$-regular graphs. The different columns correspond to the various scenarios explored. $t_\textrm{Q}$ refers to the quantum computing run-time (based on Eq.~\eqref{eq:quantum_run-time}) on an $M$-qubit quantum computer. $t_\textrm{C}$ refers to the classical component of the run-time, consisting of the time to compute expectation values, perform the eigendecomposition and $10$ local greedy steps.}
    \label{tab:qrr_run-time}
\end{table*}

\subsubsection{Measuring the Run-time of Classical Solvers}

\begin{figure}[!t]
    \centering
    \includegraphics[width=0.5\textwidth]{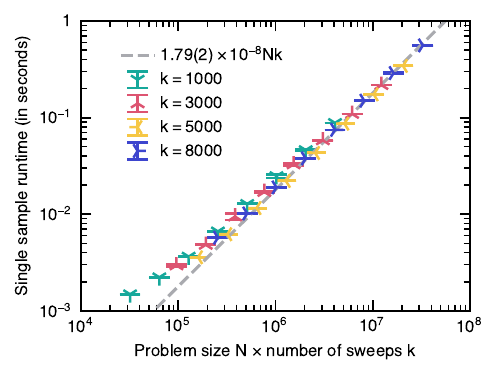}
    \caption{Run-time in seconds on a $64$Gb MacBook Pro with an Apple M1 Max chip for collecting one sample using simulated annealing with $k$ sweeps on problems of size $N$. The dashed line shows an asymptotic linear relationship yields an average run-time of $1.79(2)\times 10^{-8}$ seconds per sweep and per variable.}
    \label{fig:runtime_sa}
\end{figure}

We focus on classical solvers which are most competitive against the low-depth quantum relax-and-round algorithm considered in this work: Simulated annealing (App.~\ref{sec:sa_sm}) and the heuristic by Burer and Monteiro (App.~\ref{sec:burer_monteiro_sm}). The time required for constructing the problem or loading into the solver is excluded from the run-time. From there, each of these algorithms has a run-time dictated by a control parameter provided as input: the number of sweeps for simulated annealing, and the run-time limit for the MQLib/Burer2002 implementation of the Burer-Monteiro heuristic.

For simulated annealing, we control the number of sweeps $k$, which is proportional to the total run-time. As described in App.~\ref{sec:sa_sm}, each sweep involves an update over $N$ variables, each having only $O(1)$ nearest neighbors. Thus, the run-time is also proportional to $N$. In Fig.~\ref{fig:runtime_sa}, we show the run-time (measured on a $64$Gb MacBook Pro with an Apple M1 Max chip) for collecting one sample as a function of the problem size $N$ multiplied by the number of sweeps $k$. Using a linear asymptotic fit we estimate the average run-time per sweep and per variable to be $1.79(2)\times 10^{-8}$ seconds.

\subsection{Optimal Run-time to Match the Performance of the Greedy-Enhanced Quantum Relax-and-Round Solver}
\label{sec:opt_run-time_cl_sm}

We investigate the time $t^*$ for a classical solver to match the performance of the QRR* algorithm. It is defined as
\begin{equation}
    t^* = t\bigr/P\Bigl(C(\boldsymbol{z}_{\textrm{cl}})\geq C(\boldsymbol{z}_{\textrm{QRR*}})\Bigr),
    \label{eq:t_to_match_sm}
\end{equation}
where $t$ is the actual run-time of the classical solver and $P$ the probability that the classical solver returns a solution $\boldsymbol{z}_{\textrm{cl}}$ at least matching that of the quantum algorithm $\boldsymbol{z}_{\textrm{QRR*}}$. For the solvers considered here, $P$ is intrinsically related to $t$ given that a longer run-time is more likely to return a better solution, and conversely, a smaller run-time $t$ is more likely to result in a lower value of $P$. Therefore, we seek the optimal time $t^*_\textrm{opt}$ for the classical solver to find a solution matching the quality of the QRR* solution, which concurrently minimizes $t$ and maximizes $P$. We consider three classical solvers: Gurobi, simulated annealing, and the Burer-Monteiro approach.

\subsubsection{Simulated Annealing and the Burer-Monteiro Solvers}

The run-time $t$ of simulated annealing in Eq.~\eqref{eq:t_to_match_sm} is proportional to the number of variables in the problem and the number of sweeps performed, as shown in Fig.~\ref{fig:runtime_sa}. The Burer-Monteiro implementation in MQLib/Burer2002, which was used in this work, takes $t$ as an input parameter.

With knowledge of the solution $\boldsymbol{z}_{\textrm{QRR*}}$ for a given problem instance, we run both solvers between $10^2$ and $10^4$ times with random initial seeds for a fixed set of parameters (number of sweeps for simulated annealing and maximum run-time for MQLib/Burer2002) and collect a solution from each of these runs to estimate $P$. This operation is repeated over $1,000$ randomly generated problem instances for each size $N$. We then compute an average time $\mathbb{E}[t^*]$ over all the problem instances, and subsets of the hardest instances. The time $t^*$ is used to identify the hardest instances.

\begin{figure*}[!t]
    \centering
    \includegraphics[width=1\textwidth]{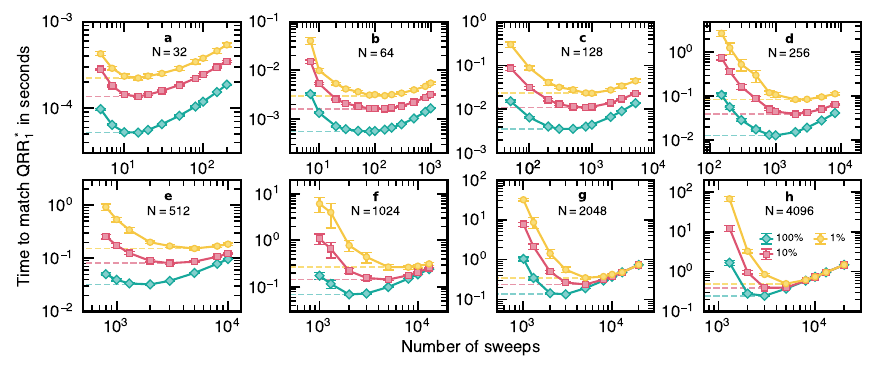}
    \caption{Run-time required for simulated annealing to match the performance of the greedy-enhanced quantum relax-and-round solver as a function of the number of sweeps. We consider all problem instances ($100\%$), as well as the hardest $10\%$ and $1\%$. Each panel corresponds to different problem sizes: (a) $N=32$, (b) $N=64$, (c) $N=128$, (d) $N=256$, (e) $N=512$, (f) $N=1024$, (g) $N=2048$, (h) $N=4096$. Error bars indicate one standard deviation. Run-times were measured on a $64$Gb MacBook Pro with an Apple M1 Max chip. Units: seconds.}
    \label{fig:optimal_n_sweeps}
\end{figure*}

\begin{figure*}[!t]
    \centering
    \includegraphics[width=1\textwidth]{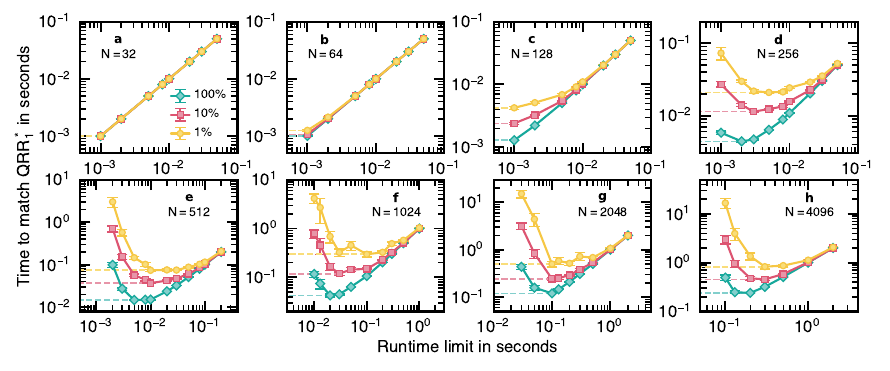}
    \caption{Run-time required for the MQLib/Burer2002 implementation of the Burer-Monteiro heuristic to match the performance of the greedy-enhanced quantum relax-and-round solver as a function of the run-time input parameter. We consider all problem instances ($100\%$), as well as the hardest $10\%$ and $1\%$. Each panel corresponds to different problem sizes: (a) $N=32$, (b) $N=64$, (c) $N=128$, (d) $N=256$, (e) $N=512$, (f) $N=1024$, (g) $N=2048$, (h) $N=4096$. Error bars indicate one standard deviation. Run-times were measured on a $64$Gb MacBook Pro with an Apple M1 Max chip. Units: seconds.}
    \label{fig:optimal_burer2002}
\end{figure*}

The results for simulated annealing and MQLib/Burer2002 are shown in Figs.~\ref{fig:optimal_n_sweeps} and~\ref{fig:optimal_burer2002}, respectively. The optimal time $t^*_\textrm{opt}$ depends on $N$, ranging from a fraction of milliseconds $(N=32)$ to a fraction of seconds ($N=4,096$). In the main text, we compare this optimal run-time with the run-time of the QRR* algorithm.

\subsubsection{Gurobi Solver}

\begin{figure*}[!t]
    \centering
    \includegraphics[width=1\textwidth]{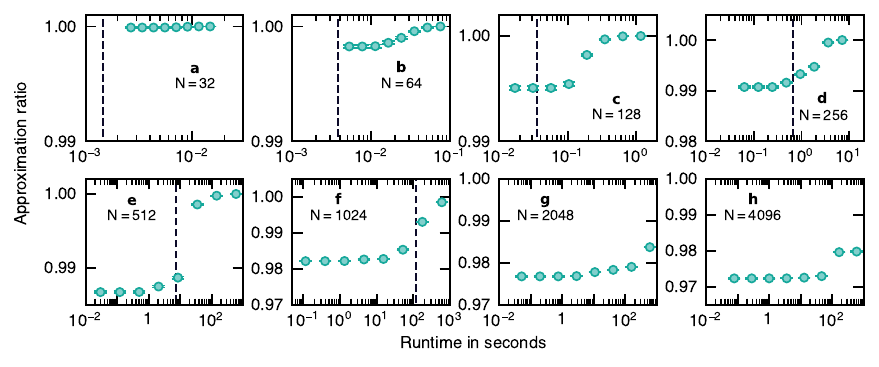}
    \caption{Average approximation ratio returned by the classical solver Gurobi, as a function of the run-time. The left-most data point corresponds to the minimum time across $1,000$ problem instances for Gurobi to return any solution at all. The right-most data point is either the run-time limit of $600$ seconds or the maximum time taken by the solver across $1,000$ problem instances to return the optimal solution. Each panel corresponds to different problem sizes: (a) $N=32$, (b) $N=64$, (c) $N=128$, (d) $N=256$, (e) $N=512$, (f) $N=1024$, (g) $N=2048$, (h) $N=4096$. Error bars indicate one standard deviation. Units: seconds.}
    \label{fig:gurobi_ar_vs_time}
\end{figure*}

The run-time required for Gurobi to find a solution matching that of QRR* for $N\geq 256$ is orders of magnitude larger than the other solvers considered. Hence, it is computationally expensive to estimate $P$ in Eq.~\eqref{eq:t_to_match_sm} using multiple runs with random initial seeds, similar to what was done for simulated annealing and MQLib/Burer2002. Instead, we run Gurobi once for each of the problem instances for a maximum run-time of $600$ seconds. Over time, increasingly better solutions are found and they are tracked.

In Fig.~\ref{fig:gurobi_ar_vs_time}, we show the approximation ratio averaged over $1,000$ randomly generated problem instances for each size $N$, as a function of Gurobi's run-time. We display the average run-time (vertical dashed line) for Gurobi to return a solution matching that of QRR* on a problem instance by problem instance basis. This run-time is equivalent to $t^*_\textrm{opt}$ in Eq.~\eqref{eq:t_to_match_sm}. For $N\geq 2,048$, Gurobi failed to return a solution matching that of QRR* for all of the individual $1,000$ problem instances. Specifically, at $N=2,048$, Gurobi found matching solutions for only $6\%$ of the problem instances, while at $N=4,096$ it found matching solutions for $0.4\%$ of the problem instances. Therefore, we can only provide a lower bound on Gurobi's run-time to match QRR*. We take $t^*_\textrm{opt}=600$ seconds when Gurobi fails to provide a solution matching that of QRR* within the time limit for a given problem instance. This is a lower bound that assumes that it is possible for Gurobi to yield a solution matching that of QRR* for $t^*_\textrm{opt}+\varepsilon$ with $\varepsilon\to 0$. The average $\mathbb{E}[t^*_\textrm{opt}]$ over the $1,000$ problem instances was used for Fig.~2 of the main text.

\subsubsection{Tabulated Optimal Run-time Summary}

In Tab.~\ref{tab:cl_solvers_opt_run-time}, we summarize the optimal run-time for three classical methods---simulated annealing, the Burer-Monteiro heuristic, and Gurobi---to match the experimental performance of the QRR* algorithm at $p=1$ on $1,000$ randomly generated problem instances. The data were extracted from Figs.~\ref{fig:optimal_n_sweeps}, ~\ref{fig:optimal_burer2002} and ~\ref{fig:gurobi_ar_vs_time}. For simulated annealing, we assumed knowledge of the optimal number of sweeps for estimating the optimal run-time, which would otherwise need to be searched. As such, while simulated annealing and the Burer-Monteiro algorithm display similar optimal run-times as a function of $N$, the Burer-Monteiro solver may be considered superior.

\begin{table*}[!ht]
    \centering
    \begin{tabular}{l@{\hskip 16pt}c@{\hskip 22pt}c@{\hskip 22pt}c@{\hskip 22pt}}
        \hline\hline\\[-0.8em]
        \makecell[l]{\textbf{Number of}\\\textbf{variables}} & \makecell[c]{\textbf{Gurobi}} & \makecell[c]{\textbf{Simulated annealing}} & \makecell[c]{\textbf{Burer-Monteiro}}\\
        \hline\\[-0.8em]
        \makecell[l]{$N=32$} & \makecell[c]{$1.44$~ms} & \makecell[c]{$0.05$~ms} & \makecell[c]{$1.00$~ms}\\
        \hline\\[-0.8em]
        \makecell[l]{$N=64$} & \makecell[c]{$2.98$~ms} & \makecell[c]{$0.56$~ms} & \makecell[c]{$1.01$~ms}\\
        \hline\\[-0.8em]
        \makecell[l]{$N=128$} & \makecell[c]{$17.38$~ms} & \makecell[c]{$3.62$~ms} & \makecell[c]{$1.29$~ms}\\
        \hline\\[-0.8em]
        \makecell[l]{$N=256$} & \makecell[c]{$266.66$~ms} & \makecell[c]{$12.83$~ms} & \makecell[c]{$4.40$~ms}\\
        \hline\\[-0.8em]
        \makecell[l]{$N=512$} & \makecell[c]{$4,679.07$~ms} & \makecell[c]{$32.30$~ms} & \makecell[c]{$14.89$~ms}\\
        \hline\\[-0.8em]
        \makecell[l]{$N=1,024$} & \makecell[c]{$98,420.74$~ms} & \makecell[c]{$69.52$~ms} & \makecell[c]{$42.58$~ms}\\
        \hline\\[-0.8em]
        \makecell[l]{$N=2,048$} & \makecell[c]{$571,179.90$~ms$^{\star}$} & \makecell[c]{$134.91$~ms} & \makecell[c]{$120.63$~ms}\\
        \hline\\[-0.8em]
        \makecell[l]{$N=4,096$} & \makecell[c]{$597,340.04$~ms$^{\star}$} & \makecell[c]{$246.71$~ms} & \makecell[c]{$240.60$~ms}\\
        \hline\hline\\[-0.8em]
    \end{tabular}
    \caption{Average optimal run-time in milliseconds (ms) for selected solvers to match the experimental performance of the QRR* algorithm at $p=1$ on $1,000$ randomly generated problem instances. $^{\star}$The Gurobi run-time for $N\geq 2,048$ is a lower bound given that on most problem instances, the solver exceeded its time limit of $600$ seconds before finding a solution matching that of QRR*.}
    \label{tab:cl_solvers_opt_run-time}
\end{table*}

\let\oldaddcontentsline\addcontentsline
\renewcommand{\addcontentsline}[3]{}
\bibliography{references}

\begin{thebibliography}{110}%
\makeatletter
\providecommand \@ifxundefined [1]{%
 \@ifx{#1\undefined}
}%
\providecommand \@ifnum [1]{%
 \ifnum #1\expandafter \@firstoftwo
 \else \expandafter \@secondoftwo
 \fi
}%
\providecommand \@ifx [1]{%
 \ifx #1\expandafter \@firstoftwo
 \else \expandafter \@secondoftwo
 \fi
}%
\providecommand \natexlab [1]{#1}%
\providecommand \enquote  [1]{``#1''}%
\providecommand \bibnamefont  [1]{#1}%
\providecommand \bibfnamefont [1]{#1}%
\providecommand \citenamefont [1]{#1}%
\providecommand \href@noop [0]{\@secondoftwo}%
\providecommand \href [0]{\begingroup \@sanitize@url \@href}%
\providecommand \@href[1]{\@@startlink{#1}\@@href}%
\providecommand \@@href[1]{\endgroup#1\@@endlink}%
\providecommand \@sanitize@url [0]{\catcode `\\12\catcode `\$12\catcode `\&12\catcode `\#12\catcode `\^12\catcode `\_12\catcode `\%12\relax}%
\providecommand \@@startlink[1]{}%
\providecommand \@@endlink[0]{}%
\providecommand \url  [0]{\begingroup\@sanitize@url \@url }%
\providecommand \@url [1]{\endgroup\@href {#1}{\urlprefix }}%
\providecommand \urlprefix  [0]{URL }%
\providecommand \Eprint [0]{\href }%
\providecommand \doibase [0]{https://doi.org/}%
\providecommand \selectlanguage [0]{\@gobble}%
\providecommand \bibinfo  [0]{\@secondoftwo}%
\providecommand \bibfield  [0]{\@secondoftwo}%
\providecommand \translation [1]{[#1]}%
\providecommand \BibitemOpen [0]{}%
\providecommand \bibitemStop [0]{}%
\providecommand \bibitemNoStop [0]{.\EOS\space}%
\providecommand \EOS [0]{\spacefactor3000\relax}%
\providecommand \BibitemShut  [1]{\csname bibitem#1\endcsname}%
\let\auto@bib@innerbib\@empty
\bibitem [{\citenamefont {Kochenberger}\ \emph {et~al.}(2014)\citenamefont {Kochenberger}, \citenamefont {Hao}, \citenamefont {Glover}, \citenamefont {Lewis}, \citenamefont {L{\"u}}, \citenamefont {Wang},\ and\ \citenamefont {Wang}}]{Kochenberger2014}%
  \BibitemOpen
  \bibfield  {author} {\bibinfo {author} {\bibfnamefont {G.}~\bibnamefont {Kochenberger}}, \bibinfo {author} {\bibfnamefont {J.-K.}\ \bibnamefont {Hao}}, \bibinfo {author} {\bibfnamefont {F.}~\bibnamefont {Glover}}, \bibinfo {author} {\bibfnamefont {M.}~\bibnamefont {Lewis}}, \bibinfo {author} {\bibfnamefont {Z.}~\bibnamefont {L{\"u}}}, \bibinfo {author} {\bibfnamefont {H.}~\bibnamefont {Wang}},\ and\ \bibinfo {author} {\bibfnamefont {Y.}~\bibnamefont {Wang}},\ }\bibfield  {title} {\bibinfo {title} {The unconstrained binary quadratic programming problem: a survey},\ }\href {https://doi.org/10.1007/s10878-014-9734-0} {\bibfield  {journal} {\bibinfo  {journal} {J. Comb. Optim.}\ }\textbf {\bibinfo {volume} {28}},\ \bibinfo {pages} {58} (\bibinfo {year} {2014})}\BibitemShut {NoStop}%
\bibitem [{\citenamefont {Farhi}\ \emph {et~al.}(2001)\citenamefont {Farhi}, \citenamefont {Goldstone}, \citenamefont {Gutmann}, \citenamefont {Lapan}, \citenamefont {Lundgren},\ and\ \citenamefont {Preda}}]{Farhi2001}%
  \BibitemOpen
  \bibfield  {author} {\bibinfo {author} {\bibfnamefont {E.}~\bibnamefont {Farhi}}, \bibinfo {author} {\bibfnamefont {J.}~\bibnamefont {Goldstone}}, \bibinfo {author} {\bibfnamefont {S.}~\bibnamefont {Gutmann}}, \bibinfo {author} {\bibfnamefont {J.}~\bibnamefont {Lapan}}, \bibinfo {author} {\bibfnamefont {A.}~\bibnamefont {Lundgren}},\ and\ \bibinfo {author} {\bibfnamefont {D.}~\bibnamefont {Preda}},\ }\bibfield  {title} {\bibinfo {title} {A {Q}uantum {A}diabatic {E}volution {A}lgorithm {A}pplied to {R}andom {I}nstances of an {NP}-{C}omplete {P}roblem},\ }\href {https://doi.org/10.1126/science.1057726} {\bibfield  {journal} {\bibinfo  {journal} {Science}\ }\textbf {\bibinfo {volume} {292}},\ \bibinfo {pages} {472} (\bibinfo {year} {2001})}\BibitemShut {NoStop}%
\bibitem [{\citenamefont {Farhi}\ \emph {et~al.}(2014{\natexlab{a}})\citenamefont {Farhi}, \citenamefont {Goldstone},\ and\ \citenamefont {Gutmann}}]{Farhi2014}%
  \BibitemOpen
  \bibfield  {author} {\bibinfo {author} {\bibfnamefont {E.}~\bibnamefont {Farhi}}, \bibinfo {author} {\bibfnamefont {J.}~\bibnamefont {Goldstone}},\ and\ \bibinfo {author} {\bibfnamefont {S.}~\bibnamefont {Gutmann}},\ }\bibfield  {title} {\bibinfo {title} {{A} {Q}uantum {A}pproximate {O}ptimization {A}lgorithm},\ }\href {https://arxiv.org/abs/1411.4028} {\bibfield  {journal} {\bibinfo  {journal} {arXiv:1411.4028}\ } (\bibinfo {year} {2014}{\natexlab{a}})}\BibitemShut {NoStop}%
\bibitem [{\citenamefont {Albash}\ and\ \citenamefont {Lidar}(2018)}]{RevModPhys.90.015002}%
  \BibitemOpen
  \bibfield  {author} {\bibinfo {author} {\bibfnamefont {T.}~\bibnamefont {Albash}}\ and\ \bibinfo {author} {\bibfnamefont {D.~A.}\ \bibnamefont {Lidar}},\ }\bibfield  {title} {\bibinfo {title} {Adiabatic quantum computation},\ }\href {https://doi.org/10.1103/RevModPhys.90.015002} {\bibfield  {journal} {\bibinfo  {journal} {Rev. Mod. Phys.}\ }\textbf {\bibinfo {volume} {90}},\ \bibinfo {pages} {015002} (\bibinfo {year} {2018})}\BibitemShut {NoStop}%
\bibitem [{\citenamefont {Blekos}\ \emph {et~al.}(2024)\citenamefont {Blekos}, \citenamefont {Brand}, \citenamefont {Ceschini}, \citenamefont {Chou}, \citenamefont {Li}, \citenamefont {Pandya},\ and\ \citenamefont {Summer}}]{BLEKOS20241}%
  \BibitemOpen
  \bibfield  {author} {\bibinfo {author} {\bibfnamefont {K.}~\bibnamefont {Blekos}}, \bibinfo {author} {\bibfnamefont {D.}~\bibnamefont {Brand}}, \bibinfo {author} {\bibfnamefont {A.}~\bibnamefont {Ceschini}}, \bibinfo {author} {\bibfnamefont {C.-H.}\ \bibnamefont {Chou}}, \bibinfo {author} {\bibfnamefont {R.-H.}\ \bibnamefont {Li}}, \bibinfo {author} {\bibfnamefont {K.}~\bibnamefont {Pandya}},\ and\ \bibinfo {author} {\bibfnamefont {A.}~\bibnamefont {Summer}},\ }\bibfield  {title} {\bibinfo {title} {A review on quantum approximate optimization algorithm and its variants},\ }\href {https://doi.org/https://doi.org/10.1016/j.physrep.2024.03.002} {\bibfield  {journal} {\bibinfo  {journal} {Phys. Rep.}\ }\textbf {\bibinfo {volume} {1068}},\ \bibinfo {pages} {1} (\bibinfo {year} {2024})}\BibitemShut {NoStop}%
\bibitem [{\citenamefont {Ebadi}\ \emph {et~al.}(2022)\citenamefont {Ebadi}, \citenamefont {Keesling}, \citenamefont {Cain}, \citenamefont {Wang}, \citenamefont {Levine}, \citenamefont {Bluvstein}, \citenamefont {Semeghini}, \citenamefont {Omran}, \citenamefont {Liu}, \citenamefont {Samajdar}, \citenamefont {Luo}, \citenamefont {Nash}, \citenamefont {Gao}, \citenamefont {Barak}, \citenamefont {Farhi}, \citenamefont {Sachdev}, \citenamefont {Gemelke}, \citenamefont {Zhou}, \citenamefont {Choi}, \citenamefont {Pichler}, \citenamefont {Wang}, \citenamefont {Greiner}, \citenamefont {Vuletić},\ and\ \citenamefont {Lukin}}]{Ebadi2022}%
  \BibitemOpen
  \bibfield  {author} {\bibinfo {author} {\bibfnamefont {S.}~\bibnamefont {Ebadi}}, \bibinfo {author} {\bibfnamefont {A.}~\bibnamefont {Keesling}}, \bibinfo {author} {\bibfnamefont {M.}~\bibnamefont {Cain}}, \bibinfo {author} {\bibfnamefont {T.~T.}\ \bibnamefont {Wang}}, \bibinfo {author} {\bibfnamefont {H.}~\bibnamefont {Levine}}, \bibinfo {author} {\bibfnamefont {D.}~\bibnamefont {Bluvstein}}, \bibinfo {author} {\bibfnamefont {G.}~\bibnamefont {Semeghini}}, \bibinfo {author} {\bibfnamefont {A.}~\bibnamefont {Omran}}, \bibinfo {author} {\bibfnamefont {J.-G.}\ \bibnamefont {Liu}}, \bibinfo {author} {\bibfnamefont {R.}~\bibnamefont {Samajdar}}, \bibinfo {author} {\bibfnamefont {X.-Z.}\ \bibnamefont {Luo}}, \bibinfo {author} {\bibfnamefont {B.}~\bibnamefont {Nash}}, \bibinfo {author} {\bibfnamefont {X.}~\bibnamefont {Gao}}, \bibinfo {author} {\bibfnamefont {B.}~\bibnamefont {Barak}}, \bibinfo {author} {\bibfnamefont {E.}~\bibnamefont {Farhi}}, \bibinfo {author} {\bibfnamefont {S.}~\bibnamefont {Sachdev}},
  \bibinfo {author} {\bibfnamefont {N.}~\bibnamefont {Gemelke}}, \bibinfo {author} {\bibfnamefont {L.}~\bibnamefont {Zhou}}, \bibinfo {author} {\bibfnamefont {S.}~\bibnamefont {Choi}}, \bibinfo {author} {\bibfnamefont {H.}~\bibnamefont {Pichler}}, \bibinfo {author} {\bibfnamefont {S.-T.}\ \bibnamefont {Wang}}, \bibinfo {author} {\bibfnamefont {M.}~\bibnamefont {Greiner}}, \bibinfo {author} {\bibfnamefont {V.}~\bibnamefont {Vuletić}},\ and\ \bibinfo {author} {\bibfnamefont {M.~D.}\ \bibnamefont {Lukin}},\ }\bibfield  {title} {\bibinfo {title} {Quantum optimization of maximum independent set using rydberg atom arrays},\ }\href {https://doi.org/10.1126/science.abo6587} {\bibfield  {journal} {\bibinfo  {journal} {Science}\ }\textbf {\bibinfo {volume} {376}},\ \bibinfo {pages} {1209} (\bibinfo {year} {2022})}\BibitemShut {NoStop}%
\bibitem [{\citenamefont {Andrist}\ \emph {et~al.}(2023)\citenamefont {Andrist}, \citenamefont {Schuetz}, \citenamefont {Minssen}, \citenamefont {Yalovetzky}, \citenamefont {Chakrabarti}, \citenamefont {Herman}, \citenamefont {Kumar}, \citenamefont {Salton}, \citenamefont {Shaydulin}, \citenamefont {Sun}, \citenamefont {Pistoia},\ and\ \citenamefont {Katzgraber}}]{PhysRevResearch.5.043277}%
  \BibitemOpen
  \bibfield  {author} {\bibinfo {author} {\bibfnamefont {R.~S.}\ \bibnamefont {Andrist}}, \bibinfo {author} {\bibfnamefont {M.~J.~A.}\ \bibnamefont {Schuetz}}, \bibinfo {author} {\bibfnamefont {P.}~\bibnamefont {Minssen}}, \bibinfo {author} {\bibfnamefont {R.}~\bibnamefont {Yalovetzky}}, \bibinfo {author} {\bibfnamefont {S.}~\bibnamefont {Chakrabarti}}, \bibinfo {author} {\bibfnamefont {D.}~\bibnamefont {Herman}}, \bibinfo {author} {\bibfnamefont {N.}~\bibnamefont {Kumar}}, \bibinfo {author} {\bibfnamefont {G.}~\bibnamefont {Salton}}, \bibinfo {author} {\bibfnamefont {R.}~\bibnamefont {Shaydulin}}, \bibinfo {author} {\bibfnamefont {Y.}~\bibnamefont {Sun}}, \bibinfo {author} {\bibfnamefont {M.}~\bibnamefont {Pistoia}},\ and\ \bibinfo {author} {\bibfnamefont {H.~G.}\ \bibnamefont {Katzgraber}},\ }\bibfield  {title} {\bibinfo {title} {Hardness of the maximum-independent-set problem on unit-disk graphs and prospects for quantum speedups},\ }\href {https://doi.org/10.1103/PhysRevResearch.5.043277} {\bibfield
  {journal} {\bibinfo  {journal} {Phys. Rev. Res.}\ }\textbf {\bibinfo {volume} {5}},\ \bibinfo {pages} {043277} (\bibinfo {year} {2023})}\BibitemShut {NoStop}%
\bibitem [{\citenamefont {King}\ \emph {et~al.}(2023)\citenamefont {King}, \citenamefont {Raymond}, \citenamefont {Lanting}, \citenamefont {Harris}, \citenamefont {Zucca}, \citenamefont {Altomare}, \citenamefont {Berkley}, \citenamefont {Boothby}, \citenamefont {Ejtemaee}, \citenamefont {Enderud}, \citenamefont {Hoskinson}, \citenamefont {Huang}, \citenamefont {Ladizinsky}, \citenamefont {MacDonald}, \citenamefont {Marsden}, \citenamefont {Molavi}, \citenamefont {Oh}, \citenamefont {Poulin-Lamarre}, \citenamefont {Reis}, \citenamefont {Rich}, \citenamefont {Sato}, \citenamefont {Tsai}, \citenamefont {Volkmann}, \citenamefont {Whittaker}, \citenamefont {Yao}, \citenamefont {Sandvik},\ and\ \citenamefont {Amin}}]{King2023}%
  \BibitemOpen
  \bibfield  {author} {\bibinfo {author} {\bibfnamefont {A.~D.}\ \bibnamefont {King}}, \bibinfo {author} {\bibfnamefont {J.}~\bibnamefont {Raymond}}, \bibinfo {author} {\bibfnamefont {T.}~\bibnamefont {Lanting}}, \bibinfo {author} {\bibfnamefont {R.}~\bibnamefont {Harris}}, \bibinfo {author} {\bibfnamefont {A.}~\bibnamefont {Zucca}}, \bibinfo {author} {\bibfnamefont {F.}~\bibnamefont {Altomare}}, \bibinfo {author} {\bibfnamefont {A.~J.}\ \bibnamefont {Berkley}}, \bibinfo {author} {\bibfnamefont {K.}~\bibnamefont {Boothby}}, \bibinfo {author} {\bibfnamefont {S.}~\bibnamefont {Ejtemaee}}, \bibinfo {author} {\bibfnamefont {C.}~\bibnamefont {Enderud}}, \bibinfo {author} {\bibfnamefont {E.}~\bibnamefont {Hoskinson}}, \bibinfo {author} {\bibfnamefont {S.}~\bibnamefont {Huang}}, \bibinfo {author} {\bibfnamefont {E.}~\bibnamefont {Ladizinsky}}, \bibinfo {author} {\bibfnamefont {A.~J.~R.}\ \bibnamefont {MacDonald}}, \bibinfo {author} {\bibfnamefont {G.}~\bibnamefont {Marsden}}, \bibinfo {author} {\bibfnamefont
  {R.}~\bibnamefont {Molavi}}, \bibinfo {author} {\bibfnamefont {T.}~\bibnamefont {Oh}}, \bibinfo {author} {\bibfnamefont {G.}~\bibnamefont {Poulin-Lamarre}}, \bibinfo {author} {\bibfnamefont {M.}~\bibnamefont {Reis}}, \bibinfo {author} {\bibfnamefont {C.}~\bibnamefont {Rich}}, \bibinfo {author} {\bibfnamefont {Y.}~\bibnamefont {Sato}}, \bibinfo {author} {\bibfnamefont {N.}~\bibnamefont {Tsai}}, \bibinfo {author} {\bibfnamefont {M.}~\bibnamefont {Volkmann}}, \bibinfo {author} {\bibfnamefont {J.~D.}\ \bibnamefont {Whittaker}}, \bibinfo {author} {\bibfnamefont {J.}~\bibnamefont {Yao}}, \bibinfo {author} {\bibfnamefont {A.~W.}\ \bibnamefont {Sandvik}},\ and\ \bibinfo {author} {\bibfnamefont {M.~H.}\ \bibnamefont {Amin}},\ }\bibfield  {title} {\bibinfo {title} {Quantum critical dynamics in a 5,000-qubit programmable spin glass},\ }\href {https://doi.org/10.1038/s41586-023-05867-2} {\bibfield  {journal} {\bibinfo  {journal} {Nature}\ }\textbf {\bibinfo {volume} {617}},\ \bibinfo {pages} {61} (\bibinfo {year}
  {2023})}\BibitemShut {NoStop}%
\bibitem [{\citenamefont {Koch}\ \emph {et~al.}(2007)\citenamefont {Koch}, \citenamefont {Yu}, \citenamefont {Gambetta}, \citenamefont {Houck}, \citenamefont {Schuster}, \citenamefont {Majer}, \citenamefont {Blais}, \citenamefont {Devoret}, \citenamefont {Girvin},\ and\ \citenamefont {Schoelkopf}}]{PhysRevA.76.042319}%
  \BibitemOpen
  \bibfield  {author} {\bibinfo {author} {\bibfnamefont {J.}~\bibnamefont {Koch}}, \bibinfo {author} {\bibfnamefont {T.~M.}\ \bibnamefont {Yu}}, \bibinfo {author} {\bibfnamefont {J.}~\bibnamefont {Gambetta}}, \bibinfo {author} {\bibfnamefont {A.~A.}\ \bibnamefont {Houck}}, \bibinfo {author} {\bibfnamefont {D.~I.}\ \bibnamefont {Schuster}}, \bibinfo {author} {\bibfnamefont {J.}~\bibnamefont {Majer}}, \bibinfo {author} {\bibfnamefont {A.}~\bibnamefont {Blais}}, \bibinfo {author} {\bibfnamefont {M.~H.}\ \bibnamefont {Devoret}}, \bibinfo {author} {\bibfnamefont {S.~M.}\ \bibnamefont {Girvin}},\ and\ \bibinfo {author} {\bibfnamefont {R.~J.}\ \bibnamefont {Schoelkopf}},\ }\bibfield  {title} {\bibinfo {title} {Charge-insensitive qubit design derived from the cooper pair box},\ }\href {https://doi.org/10.1103/PhysRevA.76.042319} {\bibfield  {journal} {\bibinfo  {journal} {Phys. Rev. A}\ }\textbf {\bibinfo {volume} {76}},\ \bibinfo {pages} {042319} (\bibinfo {year} {2007})}\BibitemShut {NoStop}%
\bibitem [{\citenamefont {Barends}\ \emph {et~al.}(2013)\citenamefont {Barends}, \citenamefont {Kelly}, \citenamefont {Megrant}, \citenamefont {Sank}, \citenamefont {Jeffrey}, \citenamefont {Chen}, \citenamefont {Yin}, \citenamefont {Chiaro}, \citenamefont {Mutus}, \citenamefont {Neill}, \citenamefont {O'Malley}, \citenamefont {Roushan}, \citenamefont {Wenner}, \citenamefont {White}, \citenamefont {Cleland},\ and\ \citenamefont {Martinis}}]{PhysRevLett.111.080502}%
  \BibitemOpen
  \bibfield  {author} {\bibinfo {author} {\bibfnamefont {R.}~\bibnamefont {Barends}}, \bibinfo {author} {\bibfnamefont {J.}~\bibnamefont {Kelly}}, \bibinfo {author} {\bibfnamefont {A.}~\bibnamefont {Megrant}}, \bibinfo {author} {\bibfnamefont {D.}~\bibnamefont {Sank}}, \bibinfo {author} {\bibfnamefont {E.}~\bibnamefont {Jeffrey}}, \bibinfo {author} {\bibfnamefont {Y.}~\bibnamefont {Chen}}, \bibinfo {author} {\bibfnamefont {Y.}~\bibnamefont {Yin}}, \bibinfo {author} {\bibfnamefont {B.}~\bibnamefont {Chiaro}}, \bibinfo {author} {\bibfnamefont {J.}~\bibnamefont {Mutus}}, \bibinfo {author} {\bibfnamefont {C.}~\bibnamefont {Neill}}, \bibinfo {author} {\bibfnamefont {P.}~\bibnamefont {O'Malley}}, \bibinfo {author} {\bibfnamefont {P.}~\bibnamefont {Roushan}}, \bibinfo {author} {\bibfnamefont {J.}~\bibnamefont {Wenner}}, \bibinfo {author} {\bibfnamefont {T.~C.}\ \bibnamefont {White}}, \bibinfo {author} {\bibfnamefont {A.~N.}\ \bibnamefont {Cleland}},\ and\ \bibinfo {author} {\bibfnamefont {J.~M.}\ \bibnamefont
  {Martinis}},\ }\bibfield  {title} {\bibinfo {title} {Coherent josephson qubit suitable for scalable quantum integrated circuits},\ }\href {https://doi.org/10.1103/PhysRevLett.111.080502} {\bibfield  {journal} {\bibinfo  {journal} {Phys. Rev. Lett.}\ }\textbf {\bibinfo {volume} {111}},\ \bibinfo {pages} {080502} (\bibinfo {year} {2013})}\BibitemShut {NoStop}%
\bibitem [{\citenamefont {Sipser}(2013)}]{sipser13}%
  \BibitemOpen
  \bibfield  {author} {\bibinfo {author} {\bibfnamefont {M.}~\bibnamefont {Sipser}},\ }\href@noop {} {\emph {\bibinfo {title} {Introduction to the {T}heory of {C}omputation}}},\ \bibinfo {edition} {3rd}\ ed.\ (\bibinfo  {publisher} {Course Technology},\ \bibinfo {address} {Boston, MA},\ \bibinfo {year} {2013})\BibitemShut {NoStop}%
\bibitem [{\citenamefont {H\r{a}stad}(2001)}]{10.1145/502090.502098}%
  \BibitemOpen
  \bibfield  {author} {\bibinfo {author} {\bibfnamefont {J.}~\bibnamefont {H\r{a}stad}},\ }\bibfield  {title} {\bibinfo {title} {Some optimal inapproximability results},\ }\href {https://doi.org/10.1145/502090.502098} {\bibfield  {journal} {\bibinfo  {journal} {J. ACM}\ }\textbf {\bibinfo {volume} {48}},\ \bibinfo {pages} {798–859} (\bibinfo {year} {2001})}\BibitemShut {NoStop}%
\bibitem [{\citenamefont {Goemans}\ and\ \citenamefont {Williamson}(1995)}]{Goemans1995}%
  \BibitemOpen
  \bibfield  {author} {\bibinfo {author} {\bibfnamefont {M.~X.}\ \bibnamefont {Goemans}}\ and\ \bibinfo {author} {\bibfnamefont {D.~P.}\ \bibnamefont {Williamson}},\ }\bibfield  {title} {\bibinfo {title} {Improved {A}pproximation {A}lgorithms for {M}aximum {C}ut and {S}atisfiability {P}roblems {U}sing {S}emidefinite {P}rogramming},\ }\href {https://doi.org/10.1145/227683.227684} {\bibfield  {journal} {\bibinfo  {journal} {J. ACM}\ }\textbf {\bibinfo {volume} {42}},\ \bibinfo {pages} {1115–1145} (\bibinfo {year} {1995})}\BibitemShut {NoStop}%
\bibitem [{\citenamefont {{Gurobi Optimization, LLC}}(2024)}]{gurobi}%
  \BibitemOpen
  \bibfield  {author} {\bibinfo {author} {\bibnamefont {{Gurobi Optimization, LLC}}},\ }\href {https://www.gurobi.com} {\bibinfo {title} {{Gurobi Optimizer Reference Manual}}} (\bibinfo {year} {2024})\BibitemShut {NoStop}%
\bibitem [{\citenamefont {Kirkpatrick}\ \emph {et~al.}(1983)\citenamefont {Kirkpatrick}, \citenamefont {Gelatt},\ and\ \citenamefont {Vecchi}}]{doi:10.1126/science.220.4598.671}%
  \BibitemOpen
  \bibfield  {author} {\bibinfo {author} {\bibfnamefont {S.}~\bibnamefont {Kirkpatrick}}, \bibinfo {author} {\bibfnamefont {C.~D.}\ \bibnamefont {Gelatt}},\ and\ \bibinfo {author} {\bibfnamefont {M.~P.}\ \bibnamefont {Vecchi}},\ }\bibfield  {title} {\bibinfo {title} {Optimization by {S}imulated {A}nnealing},\ }\href {https://doi.org/10.1126/science.220.4598.671} {\bibfield  {journal} {\bibinfo  {journal} {Science}\ }\textbf {\bibinfo {volume} {220}},\ \bibinfo {pages} {671} (\bibinfo {year} {1983})}\BibitemShut {NoStop}%
\bibitem [{\citenamefont {Burer}\ \emph {et~al.}(2002)\citenamefont {Burer}, \citenamefont {Monteiro},\ and\ \citenamefont {Zhang}}]{doi:10.1137/S1052623400382467}%
  \BibitemOpen
  \bibfield  {author} {\bibinfo {author} {\bibfnamefont {S.}~\bibnamefont {Burer}}, \bibinfo {author} {\bibfnamefont {R.~D.~C.}\ \bibnamefont {Monteiro}},\ and\ \bibinfo {author} {\bibfnamefont {Y.}~\bibnamefont {Zhang}},\ }\bibfield  {title} {\bibinfo {title} {Rank-two relaxation heuristics for max-cut and other binary quadratic programs},\ }\href {https://doi.org/10.1137/S1052623400382467} {\bibfield  {journal} {\bibinfo  {journal} {SIAM J. Optim.}\ }\textbf {\bibinfo {volume} {12}},\ \bibinfo {pages} {503} (\bibinfo {year} {2002})}\BibitemShut {NoStop}%
\bibitem [{\citenamefont {Burer}\ and\ \citenamefont {Monteiro}(2003)}]{Burer2003}%
  \BibitemOpen
  \bibfield  {author} {\bibinfo {author} {\bibfnamefont {S.}~\bibnamefont {Burer}}\ and\ \bibinfo {author} {\bibfnamefont {R.~D.~C.}\ \bibnamefont {Monteiro}},\ }\bibfield  {title} {\bibinfo {title} {A nonlinear programming algorithm for solving semidefinite programs via low-rank factorization},\ }\href {https://doi.org/10.1007/s10107-002-0352-8} {\bibfield  {journal} {\bibinfo  {journal} {Math. Program.}\ }\textbf {\bibinfo {volume} {95}},\ \bibinfo {pages} {329} (\bibinfo {year} {2003})}\BibitemShut {NoStop}%
\bibitem [{\citenamefont {Harrigan}\ \emph {et~al.}(2021)\citenamefont {Harrigan}, \citenamefont {Sung}, \citenamefont {Neeley}, \citenamefont {Satzinger}, \citenamefont {Arute}, \citenamefont {Arya}, \citenamefont {Atalaya}, \citenamefont {Bardin}, \citenamefont {Barends}, \citenamefont {Boixo}, \citenamefont {Broughton}, \citenamefont {Buckley}, \citenamefont {Buell}, \citenamefont {Burkett}, \citenamefont {Bushnell}, \citenamefont {Chen}, \citenamefont {Chen}, \citenamefont {Chiaro}, \citenamefont {Collins}, \citenamefont {Courtney}, \citenamefont {Demura}, \citenamefont {Dunsworth}, \citenamefont {Eppens}, \citenamefont {Fowler}, \citenamefont {Foxen}, \citenamefont {Gidney}, \citenamefont {Giustina}, \citenamefont {Graff}, \citenamefont {Habegger}, \citenamefont {Ho}, \citenamefont {Hong}, \citenamefont {Huang}, \citenamefont {Ioffe}, \citenamefont {Isakov}, \citenamefont {Jeffrey}, \citenamefont {Jiang}, \citenamefont {Jones}, \citenamefont {Kafri}, \citenamefont {Kechedzhi}, \citenamefont {Kelly},
  \citenamefont {Kim}, \citenamefont {Klimov}, \citenamefont {Korotkov}, \citenamefont {Kostritsa}, \citenamefont {Landhuis}, \citenamefont {Laptev}, \citenamefont {Lindmark}, \citenamefont {Leib}, \citenamefont {Martin}, \citenamefont {Martinis}, \citenamefont {McClean}, \citenamefont {McEwen}, \citenamefont {Megrant}, \citenamefont {Mi}, \citenamefont {Mohseni}, \citenamefont {Mruczkiewicz}, \citenamefont {Mutus}, \citenamefont {Naaman}, \citenamefont {Neill}, \citenamefont {Neukart}, \citenamefont {Niu}, \citenamefont {O'Brien}, \citenamefont {O'Gorman}, \citenamefont {Ostby}, \citenamefont {Petukhov}, \citenamefont {Putterman}, \citenamefont {Quintana}, \citenamefont {Roushan}, \citenamefont {Rubin}, \citenamefont {Sank}, \citenamefont {Skolik}, \citenamefont {Smelyanskiy}, \citenamefont {Strain}, \citenamefont {Streif}, \citenamefont {Szalay}, \citenamefont {Vainsencher}, \citenamefont {White}, \citenamefont {Yao}, \citenamefont {Yeh}, \citenamefont {Zalcman}, \citenamefont {Zhou}, \citenamefont {Neven},
  \citenamefont {Bacon}, \citenamefont {Lucero}, \citenamefont {Farhi},\ and\ \citenamefont {Babbush}}]{Harrigan2021}%
  \BibitemOpen
  \bibfield  {author} {\bibinfo {author} {\bibfnamefont {M.~P.}\ \bibnamefont {Harrigan}}, \bibinfo {author} {\bibfnamefont {K.~J.}\ \bibnamefont {Sung}}, \bibinfo {author} {\bibfnamefont {M.}~\bibnamefont {Neeley}}, \bibinfo {author} {\bibfnamefont {K.~J.}\ \bibnamefont {Satzinger}}, \bibinfo {author} {\bibfnamefont {F.}~\bibnamefont {Arute}}, \bibinfo {author} {\bibfnamefont {K.}~\bibnamefont {Arya}}, \bibinfo {author} {\bibfnamefont {J.}~\bibnamefont {Atalaya}}, \bibinfo {author} {\bibfnamefont {J.~C.}\ \bibnamefont {Bardin}}, \bibinfo {author} {\bibfnamefont {R.}~\bibnamefont {Barends}}, \bibinfo {author} {\bibfnamefont {S.}~\bibnamefont {Boixo}}, \bibinfo {author} {\bibfnamefont {M.}~\bibnamefont {Broughton}}, \bibinfo {author} {\bibfnamefont {B.~B.}\ \bibnamefont {Buckley}}, \bibinfo {author} {\bibfnamefont {D.~A.}\ \bibnamefont {Buell}}, \bibinfo {author} {\bibfnamefont {B.}~\bibnamefont {Burkett}}, \bibinfo {author} {\bibfnamefont {N.}~\bibnamefont {Bushnell}}, \bibinfo {author} {\bibfnamefont
  {Y.}~\bibnamefont {Chen}}, \bibinfo {author} {\bibfnamefont {Z.}~\bibnamefont {Chen}}, \bibinfo {author} {\bibfnamefont {B.}~\bibnamefont {Chiaro}}, \bibinfo {author} {\bibfnamefont {R.}~\bibnamefont {Collins}}, \bibinfo {author} {\bibfnamefont {W.}~\bibnamefont {Courtney}}, \bibinfo {author} {\bibfnamefont {S.}~\bibnamefont {Demura}}, \bibinfo {author} {\bibfnamefont {A.}~\bibnamefont {Dunsworth}}, \bibinfo {author} {\bibfnamefont {D.}~\bibnamefont {Eppens}}, \bibinfo {author} {\bibfnamefont {A.}~\bibnamefont {Fowler}}, \bibinfo {author} {\bibfnamefont {B.}~\bibnamefont {Foxen}}, \bibinfo {author} {\bibfnamefont {C.}~\bibnamefont {Gidney}}, \bibinfo {author} {\bibfnamefont {M.}~\bibnamefont {Giustina}}, \bibinfo {author} {\bibfnamefont {R.}~\bibnamefont {Graff}}, \bibinfo {author} {\bibfnamefont {S.}~\bibnamefont {Habegger}}, \bibinfo {author} {\bibfnamefont {A.}~\bibnamefont {Ho}}, \bibinfo {author} {\bibfnamefont {S.}~\bibnamefont {Hong}}, \bibinfo {author} {\bibfnamefont {T.}~\bibnamefont {Huang}},
  \bibinfo {author} {\bibfnamefont {L.~B.}\ \bibnamefont {Ioffe}}, \bibinfo {author} {\bibfnamefont {S.~V.}\ \bibnamefont {Isakov}}, \bibinfo {author} {\bibfnamefont {E.}~\bibnamefont {Jeffrey}}, \bibinfo {author} {\bibfnamefont {Z.}~\bibnamefont {Jiang}}, \bibinfo {author} {\bibfnamefont {C.}~\bibnamefont {Jones}}, \bibinfo {author} {\bibfnamefont {D.}~\bibnamefont {Kafri}}, \bibinfo {author} {\bibfnamefont {K.}~\bibnamefont {Kechedzhi}}, \bibinfo {author} {\bibfnamefont {J.}~\bibnamefont {Kelly}}, \bibinfo {author} {\bibfnamefont {S.}~\bibnamefont {Kim}}, \bibinfo {author} {\bibfnamefont {P.~V.}\ \bibnamefont {Klimov}}, \bibinfo {author} {\bibfnamefont {A.~N.}\ \bibnamefont {Korotkov}}, \bibinfo {author} {\bibfnamefont {F.}~\bibnamefont {Kostritsa}}, \bibinfo {author} {\bibfnamefont {D.}~\bibnamefont {Landhuis}}, \bibinfo {author} {\bibfnamefont {P.}~\bibnamefont {Laptev}}, \bibinfo {author} {\bibfnamefont {M.}~\bibnamefont {Lindmark}}, \bibinfo {author} {\bibfnamefont {M.}~\bibnamefont {Leib}}, \bibinfo
  {author} {\bibfnamefont {O.}~\bibnamefont {Martin}}, \bibinfo {author} {\bibfnamefont {J.~M.}\ \bibnamefont {Martinis}}, \bibinfo {author} {\bibfnamefont {J.~R.}\ \bibnamefont {McClean}}, \bibinfo {author} {\bibfnamefont {M.}~\bibnamefont {McEwen}}, \bibinfo {author} {\bibfnamefont {A.}~\bibnamefont {Megrant}}, \bibinfo {author} {\bibfnamefont {X.}~\bibnamefont {Mi}}, \bibinfo {author} {\bibfnamefont {M.}~\bibnamefont {Mohseni}}, \bibinfo {author} {\bibfnamefont {W.}~\bibnamefont {Mruczkiewicz}}, \bibinfo {author} {\bibfnamefont {J.}~\bibnamefont {Mutus}}, \bibinfo {author} {\bibfnamefont {O.}~\bibnamefont {Naaman}}, \bibinfo {author} {\bibfnamefont {C.}~\bibnamefont {Neill}}, \bibinfo {author} {\bibfnamefont {F.}~\bibnamefont {Neukart}}, \bibinfo {author} {\bibfnamefont {M.~Y.}\ \bibnamefont {Niu}}, \bibinfo {author} {\bibfnamefont {T.~E.}\ \bibnamefont {O'Brien}}, \bibinfo {author} {\bibfnamefont {B.}~\bibnamefont {O'Gorman}}, \bibinfo {author} {\bibfnamefont {E.}~\bibnamefont {Ostby}}, \bibinfo {author}
  {\bibfnamefont {A.}~\bibnamefont {Petukhov}}, \bibinfo {author} {\bibfnamefont {H.}~\bibnamefont {Putterman}}, \bibinfo {author} {\bibfnamefont {C.}~\bibnamefont {Quintana}}, \bibinfo {author} {\bibfnamefont {P.}~\bibnamefont {Roushan}}, \bibinfo {author} {\bibfnamefont {N.~C.}\ \bibnamefont {Rubin}}, \bibinfo {author} {\bibfnamefont {D.}~\bibnamefont {Sank}}, \bibinfo {author} {\bibfnamefont {A.}~\bibnamefont {Skolik}}, \bibinfo {author} {\bibfnamefont {V.}~\bibnamefont {Smelyanskiy}}, \bibinfo {author} {\bibfnamefont {D.}~\bibnamefont {Strain}}, \bibinfo {author} {\bibfnamefont {M.}~\bibnamefont {Streif}}, \bibinfo {author} {\bibfnamefont {M.}~\bibnamefont {Szalay}}, \bibinfo {author} {\bibfnamefont {A.}~\bibnamefont {Vainsencher}}, \bibinfo {author} {\bibfnamefont {T.}~\bibnamefont {White}}, \bibinfo {author} {\bibfnamefont {Z.~J.}\ \bibnamefont {Yao}}, \bibinfo {author} {\bibfnamefont {P.}~\bibnamefont {Yeh}}, \bibinfo {author} {\bibfnamefont {A.}~\bibnamefont {Zalcman}}, \bibinfo {author}
  {\bibfnamefont {L.}~\bibnamefont {Zhou}}, \bibinfo {author} {\bibfnamefont {H.}~\bibnamefont {Neven}}, \bibinfo {author} {\bibfnamefont {D.}~\bibnamefont {Bacon}}, \bibinfo {author} {\bibfnamefont {E.}~\bibnamefont {Lucero}}, \bibinfo {author} {\bibfnamefont {E.}~\bibnamefont {Farhi}},\ and\ \bibinfo {author} {\bibfnamefont {R.}~\bibnamefont {Babbush}},\ }\bibfield  {title} {\bibinfo {title} {Quantum approximate optimization of non-planar graph problems on a planar superconducting processor},\ }\href {https://doi.org/10.1038/s41567-020-01105-y} {\bibfield  {journal} {\bibinfo  {journal} {Nat. Phys.}\ }\textbf {\bibinfo {volume} {17}},\ \bibinfo {pages} {332} (\bibinfo {year} {2021})}\BibitemShut {NoStop}%
\bibitem [{\citenamefont {DeCross}\ \emph {et~al.}(2023)\citenamefont {DeCross}, \citenamefont {Chertkov}, \citenamefont {Kohagen},\ and\ \citenamefont {Foss-Feig}}]{PhysRevX.13.041057}%
  \BibitemOpen
  \bibfield  {author} {\bibinfo {author} {\bibfnamefont {M.}~\bibnamefont {DeCross}}, \bibinfo {author} {\bibfnamefont {E.}~\bibnamefont {Chertkov}}, \bibinfo {author} {\bibfnamefont {M.}~\bibnamefont {Kohagen}},\ and\ \bibinfo {author} {\bibfnamefont {M.}~\bibnamefont {Foss-Feig}},\ }\bibfield  {title} {\bibinfo {title} {Qubit-reuse compilation with mid-circuit measurement and reset},\ }\href {https://doi.org/10.1103/PhysRevX.13.041057} {\bibfield  {journal} {\bibinfo  {journal} {Phys. Rev. X}\ }\textbf {\bibinfo {volume} {13}},\ \bibinfo {pages} {041057} (\bibinfo {year} {2023})}\BibitemShut {NoStop}%
\bibitem [{\citenamefont {Sack}\ and\ \citenamefont {Egger}(2024{\natexlab{a}})}]{PhysRevResearch.6.013223}%
  \BibitemOpen
  \bibfield  {author} {\bibinfo {author} {\bibfnamefont {S.~H.}\ \bibnamefont {Sack}}\ and\ \bibinfo {author} {\bibfnamefont {D.~J.}\ \bibnamefont {Egger}},\ }\bibfield  {title} {\bibinfo {title} {Large-scale quantum approximate optimization on nonplanar graphs with machine learning noise mitigation},\ }\href {https://doi.org/10.1103/PhysRevResearch.6.013223} {\bibfield  {journal} {\bibinfo  {journal} {Phys. Rev. Res.}\ }\textbf {\bibinfo {volume} {6}},\ \bibinfo {pages} {013223} (\bibinfo {year} {2024}{\natexlab{a}})}\BibitemShut {NoStop}%
\bibitem [{\citenamefont {Sciorilli}\ \emph {et~al.}(2025)\citenamefont {Sciorilli}, \citenamefont {Borges}, \citenamefont {Patti}, \citenamefont {García-Mart\`in}, \citenamefont {Camilo}, \citenamefont {Anandkumar},\ and\ \citenamefont {Aolita}}]{Sciorilli2024}%
  \BibitemOpen
  \bibfield  {author} {\bibinfo {author} {\bibfnamefont {M.}~\bibnamefont {Sciorilli}}, \bibinfo {author} {\bibfnamefont {L.}~\bibnamefont {Borges}}, \bibinfo {author} {\bibfnamefont {T.~L.}\ \bibnamefont {Patti}}, \bibinfo {author} {\bibfnamefont {D.}~\bibnamefont {García-Mart\`in}}, \bibinfo {author} {\bibfnamefont {G.}~\bibnamefont {Camilo}}, \bibinfo {author} {\bibfnamefont {A.}~\bibnamefont {Anandkumar}},\ and\ \bibinfo {author} {\bibfnamefont {L.}~\bibnamefont {Aolita}},\ }\bibfield  {title} {\bibinfo {title} {Towards large-scale quantum optimization solvers with few qubits},\ }\href {https://doi.org/10.1038/s41467-024-55346-z} {\bibfield  {journal} {\bibinfo  {journal} {Nat. Commun.}\ }\textbf {\bibinfo {volume} {16}},\ \bibinfo {pages} {476} (\bibinfo {year} {2025})}\BibitemShut {NoStop}%
\bibitem [{\citenamefont {Lucas}(2014)}]{lucas2014}%
  \BibitemOpen
  \bibfield  {author} {\bibinfo {author} {\bibfnamefont {A.}~\bibnamefont {Lucas}},\ }\bibfield  {title} {\bibinfo {title} {Ising formulations of many {NP} problems},\ }\href {https://doi.org/10.3389/fphy.2014.00005} {\bibfield  {journal} {\bibinfo  {journal} {Front. Phys.}\ }\textbf {\bibinfo {volume} {2}},\ \bibinfo {pages} {5} (\bibinfo {year} {2014})}\BibitemShut {NoStop}%
\bibitem [{\citenamefont {Otterbach}\ \emph {et~al.}(2017)\citenamefont {Otterbach}, \citenamefont {Manenti}, \citenamefont {Alidoust}, \citenamefont {Bestwick}, \citenamefont {Block}, \citenamefont {Bloom}, \citenamefont {Caldwell}, \citenamefont {Didier}, \citenamefont {Fried}, \citenamefont {Hong}, \citenamefont {Karalekas}, \citenamefont {Osborn}, \citenamefont {Papageorge}, \citenamefont {Peterson}, \citenamefont {Prawiroatmodjo}, \citenamefont {Rubin}, \citenamefont {Ryan}, \citenamefont {Scarabelli}, \citenamefont {Scheer}, \citenamefont {Sete}, \citenamefont {Sivarajah}, \citenamefont {Smith}, \citenamefont {Staley}, \citenamefont {Tezak}, \citenamefont {Zeng}, \citenamefont {Hudson}, \citenamefont {Johnson}, \citenamefont {Reagor}, \citenamefont {da~Silva},\ and\ \citenamefont {Rigetti}}]{Otterbach2017}%
  \BibitemOpen
  \bibfield  {author} {\bibinfo {author} {\bibfnamefont {J.~S.}\ \bibnamefont {Otterbach}}, \bibinfo {author} {\bibfnamefont {R.}~\bibnamefont {Manenti}}, \bibinfo {author} {\bibfnamefont {N.}~\bibnamefont {Alidoust}}, \bibinfo {author} {\bibfnamefont {A.}~\bibnamefont {Bestwick}}, \bibinfo {author} {\bibfnamefont {M.}~\bibnamefont {Block}}, \bibinfo {author} {\bibfnamefont {B.}~\bibnamefont {Bloom}}, \bibinfo {author} {\bibfnamefont {S.}~\bibnamefont {Caldwell}}, \bibinfo {author} {\bibfnamefont {N.}~\bibnamefont {Didier}}, \bibinfo {author} {\bibfnamefont {E.~S.}\ \bibnamefont {Fried}}, \bibinfo {author} {\bibfnamefont {S.}~\bibnamefont {Hong}}, \bibinfo {author} {\bibfnamefont {P.}~\bibnamefont {Karalekas}}, \bibinfo {author} {\bibfnamefont {C.~B.}\ \bibnamefont {Osborn}}, \bibinfo {author} {\bibfnamefont {A.}~\bibnamefont {Papageorge}}, \bibinfo {author} {\bibfnamefont {E.~C.}\ \bibnamefont {Peterson}}, \bibinfo {author} {\bibfnamefont {G.}~\bibnamefont {Prawiroatmodjo}}, \bibinfo {author} {\bibfnamefont
  {N.}~\bibnamefont {Rubin}}, \bibinfo {author} {\bibfnamefont {C.~A.}\ \bibnamefont {Ryan}}, \bibinfo {author} {\bibfnamefont {D.}~\bibnamefont {Scarabelli}}, \bibinfo {author} {\bibfnamefont {M.}~\bibnamefont {Scheer}}, \bibinfo {author} {\bibfnamefont {E.~A.}\ \bibnamefont {Sete}}, \bibinfo {author} {\bibfnamefont {P.}~\bibnamefont {Sivarajah}}, \bibinfo {author} {\bibfnamefont {R.~S.}\ \bibnamefont {Smith}}, \bibinfo {author} {\bibfnamefont {A.}~\bibnamefont {Staley}}, \bibinfo {author} {\bibfnamefont {N.}~\bibnamefont {Tezak}}, \bibinfo {author} {\bibfnamefont {W.~J.}\ \bibnamefont {Zeng}}, \bibinfo {author} {\bibfnamefont {A.}~\bibnamefont {Hudson}}, \bibinfo {author} {\bibfnamefont {B.~R.}\ \bibnamefont {Johnson}}, \bibinfo {author} {\bibfnamefont {M.}~\bibnamefont {Reagor}}, \bibinfo {author} {\bibfnamefont {M.~P.}\ \bibnamefont {da~Silva}},\ and\ \bibinfo {author} {\bibfnamefont {C.}~\bibnamefont {Rigetti}},\ }\bibfield  {title} {\bibinfo {title} {Unsupervised {M}achine {L}earning on a {H}ybrid
  {Q}uantum {C}omputer},\ }\href {https://arxiv.org/abs/1712.05771} {\bibfield  {journal} {\bibinfo  {journal} {arXiv:1712.05771}\ } (\bibinfo {year} {2017})}\BibitemShut {NoStop}%
\bibitem [{\citenamefont {Pagano}\ \emph {et~al.}(2020)\citenamefont {Pagano}, \citenamefont {Bapat}, \citenamefont {Becker}, \citenamefont {Collins}, \citenamefont {De}, \citenamefont {Hess}, \citenamefont {Kaplan}, \citenamefont {Kyprianidis}, \citenamefont {Tan}, \citenamefont {Baldwin}, \citenamefont {Brady}, \citenamefont {Deshpande}, \citenamefont {Liu}, \citenamefont {Jordan}, \citenamefont {Gorshkov},\ and\ \citenamefont {Monroe}}]{Pagano2020}%
  \BibitemOpen
  \bibfield  {author} {\bibinfo {author} {\bibfnamefont {G.}~\bibnamefont {Pagano}}, \bibinfo {author} {\bibfnamefont {A.}~\bibnamefont {Bapat}}, \bibinfo {author} {\bibfnamefont {P.}~\bibnamefont {Becker}}, \bibinfo {author} {\bibfnamefont {K.~S.}\ \bibnamefont {Collins}}, \bibinfo {author} {\bibfnamefont {A.}~\bibnamefont {De}}, \bibinfo {author} {\bibfnamefont {P.~W.}\ \bibnamefont {Hess}}, \bibinfo {author} {\bibfnamefont {H.~B.}\ \bibnamefont {Kaplan}}, \bibinfo {author} {\bibfnamefont {A.}~\bibnamefont {Kyprianidis}}, \bibinfo {author} {\bibfnamefont {W.~L.}\ \bibnamefont {Tan}}, \bibinfo {author} {\bibfnamefont {C.}~\bibnamefont {Baldwin}}, \bibinfo {author} {\bibfnamefont {L.~T.}\ \bibnamefont {Brady}}, \bibinfo {author} {\bibfnamefont {A.}~\bibnamefont {Deshpande}}, \bibinfo {author} {\bibfnamefont {F.}~\bibnamefont {Liu}}, \bibinfo {author} {\bibfnamefont {S.}~\bibnamefont {Jordan}}, \bibinfo {author} {\bibfnamefont {A.~V.}\ \bibnamefont {Gorshkov}},\ and\ \bibinfo {author} {\bibfnamefont
  {C.}~\bibnamefont {Monroe}},\ }\bibfield  {title} {\bibinfo {title} {Quantum approximate optimization of the long-range ising model with a trapped-ion quantum simulator},\ }\href {https://doi.org/10.1073/pnas.2006373117} {\bibfield  {journal} {\bibinfo  {journal} {Proc. Natl. Acad. Sci. U.S.A.}\ }\textbf {\bibinfo {volume} {117}},\ \bibinfo {pages} {25396} (\bibinfo {year} {2020})}\BibitemShut {NoStop}%
\bibitem [{\citenamefont {Wurtz}\ and\ \citenamefont {Lykov}(2021)}]{Wurtz2021}%
  \BibitemOpen
  \bibfield  {author} {\bibinfo {author} {\bibfnamefont {J.}~\bibnamefont {Wurtz}}\ and\ \bibinfo {author} {\bibfnamefont {D.}~\bibnamefont {Lykov}},\ }\bibfield  {title} {\bibinfo {title} {Fixed-angle conjectures for the quantum approximate optimization algorithm on regular maxcut graphs},\ }\href {https://doi.org/10.1103/PhysRevA.104.052419} {\bibfield  {journal} {\bibinfo  {journal} {Phys. Rev. A}\ }\textbf {\bibinfo {volume} {104}},\ \bibinfo {pages} {052419} (\bibinfo {year} {2021})}\BibitemShut {NoStop}%
\bibitem [{\citenamefont {Basso}\ \emph {et~al.}(2022)\citenamefont {Basso}, \citenamefont {Farhi}, \citenamefont {Marwaha}, \citenamefont {Villalonga},\ and\ \citenamefont {Zhou}}]{Basso2022}%
  \BibitemOpen
  \bibfield  {author} {\bibinfo {author} {\bibfnamefont {J.}~\bibnamefont {Basso}}, \bibinfo {author} {\bibfnamefont {E.}~\bibnamefont {Farhi}}, \bibinfo {author} {\bibfnamefont {K.}~\bibnamefont {Marwaha}}, \bibinfo {author} {\bibfnamefont {B.}~\bibnamefont {Villalonga}},\ and\ \bibinfo {author} {\bibfnamefont {L.}~\bibnamefont {Zhou}},\ }\bibfield  {title} {\bibinfo {title} {{The Quantum Approximate Optimization Algorithm at High Depth for MaxCut on Large-Girth Regular Graphs and the Sherrington-Kirkpatrick Model}},\ }in\ \href {https://doi.org/10.4230/LIPIcs.TQC.2022.7} {\emph {\bibinfo {booktitle} {17th Conference on the Theory of Quantum Computation, Communication and Cryptography (TQC 2022)}}},\ \bibinfo {series} {Leibniz International Proceedings in Informatics (LIPIcs)}, Vol.\ \bibinfo {volume} {232},\ \bibinfo {editor} {edited by\ \bibinfo {editor} {\bibfnamefont {F.}~\bibnamefont {Le~Gall}}\ and\ \bibinfo {editor} {\bibfnamefont {T.}~\bibnamefont {Morimae}}}\ (\bibinfo  {publisher} {Schloss Dagstuhl
  -- Leibniz-Zentrum f{\"u}r Informatik},\ \bibinfo {address} {Dagstuhl, Germany},\ \bibinfo {year} {2022})\ pp.\ \bibinfo {pages} {7:1--7:21}\BibitemShut {NoStop}%
\bibitem [{\citenamefont {Maciejewski}\ \emph {et~al.}(2024)\citenamefont {Maciejewski}, \citenamefont {Hadfield}, \citenamefont {Hall}, \citenamefont {Hodson}, \citenamefont {Dupont}, \citenamefont {Evert}, \citenamefont {Sud}, \citenamefont {Alam}, \citenamefont {Wang}, \citenamefont {Jeffrey}, \citenamefont {Sundar}, \citenamefont {Lott}, \citenamefont {Grabbe}, \citenamefont {Rieffel}, \citenamefont {Reagor},\ and\ \citenamefont {Venturelli}}]{Maciejewski2023}%
  \BibitemOpen
  \bibfield  {author} {\bibinfo {author} {\bibfnamefont {F.~B.}\ \bibnamefont {Maciejewski}}, \bibinfo {author} {\bibfnamefont {S.}~\bibnamefont {Hadfield}}, \bibinfo {author} {\bibfnamefont {B.}~\bibnamefont {Hall}}, \bibinfo {author} {\bibfnamefont {M.}~\bibnamefont {Hodson}}, \bibinfo {author} {\bibfnamefont {M.}~\bibnamefont {Dupont}}, \bibinfo {author} {\bibfnamefont {B.}~\bibnamefont {Evert}}, \bibinfo {author} {\bibfnamefont {J.}~\bibnamefont {Sud}}, \bibinfo {author} {\bibfnamefont {M.~S.}\ \bibnamefont {Alam}}, \bibinfo {author} {\bibfnamefont {Z.}~\bibnamefont {Wang}}, \bibinfo {author} {\bibfnamefont {S.}~\bibnamefont {Jeffrey}}, \bibinfo {author} {\bibfnamefont {B.}~\bibnamefont {Sundar}}, \bibinfo {author} {\bibfnamefont {P.~A.}\ \bibnamefont {Lott}}, \bibinfo {author} {\bibfnamefont {S.}~\bibnamefont {Grabbe}}, \bibinfo {author} {\bibfnamefont {E.~G.}\ \bibnamefont {Rieffel}}, \bibinfo {author} {\bibfnamefont {M.~J.}\ \bibnamefont {Reagor}},\ and\ \bibinfo {author} {\bibfnamefont
  {D.}~\bibnamefont {Venturelli}},\ }\bibfield  {title} {\bibinfo {title} {Design and execution of quantum circuits using tens of superconducting qubits and thousands of gates for dense ising optimization problems},\ }\href {https://doi.org/10.1103/PhysRevApplied.22.044074} {\bibfield  {journal} {\bibinfo  {journal} {Phys. Rev. Appl.}\ }\textbf {\bibinfo {volume} {22}},\ \bibinfo {pages} {044074} (\bibinfo {year} {2024})}\BibitemShut {NoStop}%
\bibitem [{\citenamefont {Pelofske}\ \emph {et~al.}(2024)\citenamefont {Pelofske}, \citenamefont {B{\"a}rtschi},\ and\ \citenamefont {Eidenbenz}}]{Pelofske2024}%
  \BibitemOpen
  \bibfield  {author} {\bibinfo {author} {\bibfnamefont {E.}~\bibnamefont {Pelofske}}, \bibinfo {author} {\bibfnamefont {A.}~\bibnamefont {B{\"a}rtschi}},\ and\ \bibinfo {author} {\bibfnamefont {S.}~\bibnamefont {Eidenbenz}},\ }\bibfield  {title} {\bibinfo {title} {Short-depth qaoa circuits and quantum annealing on higher-order ising models},\ }\href {https://doi.org/10.1038/s41534-024-00825-w} {\bibfield  {journal} {\bibinfo  {journal} {Npj Quantum Inf.}\ }\textbf {\bibinfo {volume} {10}},\ \bibinfo {pages} {30} (\bibinfo {year} {2024})}\BibitemShut {NoStop}%
\bibitem [{\citenamefont {Dupont}\ and\ \citenamefont {Sundar}(2024)}]{Dupont2024}%
  \BibitemOpen
  \bibfield  {author} {\bibinfo {author} {\bibfnamefont {M.}~\bibnamefont {Dupont}}\ and\ \bibinfo {author} {\bibfnamefont {B.}~\bibnamefont {Sundar}},\ }\bibfield  {title} {\bibinfo {title} {Extending relax-and-round combinatorial optimization solvers with quantum correlations},\ }\href {https://doi.org/10.1103/PhysRevA.109.012429} {\bibfield  {journal} {\bibinfo  {journal} {Phys. Rev. A}\ }\textbf {\bibinfo {volume} {109}},\ \bibinfo {pages} {012429} (\bibinfo {year} {2024})}\BibitemShut {NoStop}%
\bibitem [{3re()}]{3reggraphs}%
  \BibitemOpen
  \href@noop {} {}\bibinfo {note} {Devising an algorithm for the maximum cut problem on random $3$-regular graphs guaranteeing an approximation ratio of at least $99.7\%$~\cite{10.1007/3-540-48523-6_17} is NP-hard, with the current record sitting at $\alpha\simeq 93.3\%$~\cite{HALPERIN2004169} from a modified version of the Goemans-Williamson algorithm based on semidefinite programming.}\BibitemShut {Stop}%
\bibitem [{\citenamefont {Wurtz}\ and\ \citenamefont {Love}(2021)}]{PhysRevA.103.042612}%
  \BibitemOpen
  \bibfield  {author} {\bibinfo {author} {\bibfnamefont {J.}~\bibnamefont {Wurtz}}\ and\ \bibinfo {author} {\bibfnamefont {P.}~\bibnamefont {Love}},\ }\bibfield  {title} {\bibinfo {title} {Maxcut quantum approximate optimization algorithm performance guarantees for p>1},\ }\href {https://doi.org/10.1103/PhysRevA.103.042612} {\bibfield  {journal} {\bibinfo  {journal} {Phys. Rev. A}\ }\textbf {\bibinfo {volume} {103}},\ \bibinfo {pages} {042612} (\bibinfo {year} {2021})}\BibitemShut {NoStop}%
\bibitem [{\citenamefont {Cai}\ \emph {et~al.}(2023)\citenamefont {Cai}, \citenamefont {Babbush}, \citenamefont {Benjamin}, \citenamefont {Endo}, \citenamefont {Huggins}, \citenamefont {Li}, \citenamefont {McClean},\ and\ \citenamefont {O'Brien}}]{RevModPhys.95.045005}%
  \BibitemOpen
  \bibfield  {author} {\bibinfo {author} {\bibfnamefont {Z.}~\bibnamefont {Cai}}, \bibinfo {author} {\bibfnamefont {R.}~\bibnamefont {Babbush}}, \bibinfo {author} {\bibfnamefont {S.~C.}\ \bibnamefont {Benjamin}}, \bibinfo {author} {\bibfnamefont {S.}~\bibnamefont {Endo}}, \bibinfo {author} {\bibfnamefont {W.~J.}\ \bibnamefont {Huggins}}, \bibinfo {author} {\bibfnamefont {Y.}~\bibnamefont {Li}}, \bibinfo {author} {\bibfnamefont {J.~R.}\ \bibnamefont {McClean}},\ and\ \bibinfo {author} {\bibfnamefont {T.~E.}\ \bibnamefont {O'Brien}},\ }\bibfield  {title} {\bibinfo {title} {Quantum error mitigation},\ }\href {https://doi.org/10.1103/RevModPhys.95.045005} {\bibfield  {journal} {\bibinfo  {journal} {Rev. Mod. Phys.}\ }\textbf {\bibinfo {volume} {95}},\ \bibinfo {pages} {045005} (\bibinfo {year} {2023})}\BibitemShut {NoStop}%
\bibitem [{\citenamefont {Sete}\ \emph {et~al.}(2021{\natexlab{a}})\citenamefont {Sete}, \citenamefont {Chen}, \citenamefont {Manenti}, \citenamefont {Kulshreshtha},\ and\ \citenamefont {Poletto}}]{PhysRevApplied.15.064063}%
  \BibitemOpen
  \bibfield  {author} {\bibinfo {author} {\bibfnamefont {E.~A.}\ \bibnamefont {Sete}}, \bibinfo {author} {\bibfnamefont {A.~Q.}\ \bibnamefont {Chen}}, \bibinfo {author} {\bibfnamefont {R.}~\bibnamefont {Manenti}}, \bibinfo {author} {\bibfnamefont {S.}~\bibnamefont {Kulshreshtha}},\ and\ \bibinfo {author} {\bibfnamefont {S.}~\bibnamefont {Poletto}},\ }\bibfield  {title} {\bibinfo {title} {Floating tunable coupler for scalable quantum computing architectures},\ }\href {https://doi.org/10.1103/PhysRevApplied.15.064063} {\bibfield  {journal} {\bibinfo  {journal} {Phys. Rev. Appl.}\ }\textbf {\bibinfo {volume} {15}},\ \bibinfo {pages} {064063} (\bibinfo {year} {2021}{\natexlab{a}})}\BibitemShut {NoStop}%
\bibitem [{\citenamefont {Sete}\ \emph {et~al.}(2021{\natexlab{b}})\citenamefont {Sete}, \citenamefont {Didier}, \citenamefont {Chen}, \citenamefont {Kulshreshtha}, \citenamefont {Manenti},\ and\ \citenamefont {Poletto}}]{PhysRevApplied.16.024050}%
  \BibitemOpen
  \bibfield  {author} {\bibinfo {author} {\bibfnamefont {E.~A.}\ \bibnamefont {Sete}}, \bibinfo {author} {\bibfnamefont {N.}~\bibnamefont {Didier}}, \bibinfo {author} {\bibfnamefont {A.~Q.}\ \bibnamefont {Chen}}, \bibinfo {author} {\bibfnamefont {S.}~\bibnamefont {Kulshreshtha}}, \bibinfo {author} {\bibfnamefont {R.}~\bibnamefont {Manenti}},\ and\ \bibinfo {author} {\bibfnamefont {S.}~\bibnamefont {Poletto}},\ }\bibfield  {title} {\bibinfo {title} {Parametric-resonance entangling gates with a tunable coupler},\ }\href {https://doi.org/10.1103/PhysRevApplied.16.024050} {\bibfield  {journal} {\bibinfo  {journal} {Phys. Rev. Appl.}\ }\textbf {\bibinfo {volume} {16}},\ \bibinfo {pages} {024050} (\bibinfo {year} {2021}{\natexlab{b}})}\BibitemShut {NoStop}%
\bibitem [{\citenamefont {O'Gorman}\ \emph {et~al.}(2019)\citenamefont {O'Gorman}, \citenamefont {Huggins}, \citenamefont {Rieffel},\ and\ \citenamefont {Whaley}}]{OGorman2019}%
  \BibitemOpen
  \bibfield  {author} {\bibinfo {author} {\bibfnamefont {B.}~\bibnamefont {O'Gorman}}, \bibinfo {author} {\bibfnamefont {W.~J.}\ \bibnamefont {Huggins}}, \bibinfo {author} {\bibfnamefont {E.~G.}\ \bibnamefont {Rieffel}},\ and\ \bibinfo {author} {\bibfnamefont {K.~B.}\ \bibnamefont {Whaley}},\ }\bibfield  {title} {\bibinfo {title} {Generalized swap networks for near-term quantum computing},\ }\href {https://arxiv.org/abs/1905.05118} {\bibfield  {journal} {\bibinfo  {journal} {arXiv:1905.05118}\ } (\bibinfo {year} {2019})}\BibitemShut {NoStop}%
\bibitem [{\citenamefont {Nachman}\ \emph {et~al.}(2020)\citenamefont {Nachman}, \citenamefont {Urbanek}, \citenamefont {de~Jong},\ and\ \citenamefont {Bauer}}]{nachman_unfolding_2020}%
  \BibitemOpen
  \bibfield  {author} {\bibinfo {author} {\bibfnamefont {B.}~\bibnamefont {Nachman}}, \bibinfo {author} {\bibfnamefont {M.}~\bibnamefont {Urbanek}}, \bibinfo {author} {\bibfnamefont {W.~A.}\ \bibnamefont {de~Jong}},\ and\ \bibinfo {author} {\bibfnamefont {C.~W.}\ \bibnamefont {Bauer}},\ }\bibfield  {title} {\bibinfo {title} {Unfolding quantum computer readout noise},\ }\href {https://doi.org/10.1038/s41534-020-00309-7} {\bibfield  {journal} {\bibinfo  {journal} {Npj Quantum Inf.}\ }\textbf {\bibinfo {volume} {6}},\ \bibinfo {pages} {84} (\bibinfo {year} {2020})}\BibitemShut {NoStop}%
\bibitem [{\citenamefont {Neira}\ \emph {et~al.}(2024)\citenamefont {Neira}, \citenamefont {Brown}, \citenamefont {Sathe}, \citenamefont {Wudarski}, \citenamefont {Pavone}, \citenamefont {Rieffel},\ and\ \citenamefont {Venturelli}}]{Bernal2024}%
  \BibitemOpen
  \bibfield  {author} {\bibinfo {author} {\bibfnamefont {D.~E.~B.}\ \bibnamefont {Neira}}, \bibinfo {author} {\bibfnamefont {R.}~\bibnamefont {Brown}}, \bibinfo {author} {\bibfnamefont {P.}~\bibnamefont {Sathe}}, \bibinfo {author} {\bibfnamefont {F.}~\bibnamefont {Wudarski}}, \bibinfo {author} {\bibfnamefont {M.}~\bibnamefont {Pavone}}, \bibinfo {author} {\bibfnamefont {E.~G.}\ \bibnamefont {Rieffel}},\ and\ \bibinfo {author} {\bibfnamefont {D.}~\bibnamefont {Venturelli}},\ }\bibfield  {title} {\bibinfo {title} {Benchmarking the {O}peration of {Q}uantum {H}euristics and {I}sing {M}achines: {S}coring {P}arameter {S}etting {S}trategies on {O}ptimization {A}pplications},\ }\href {https://arxiv.org/abs/2402.10255} {\bibfield  {journal} {\bibinfo  {journal} {arXiv:2402.10255}\ } (\bibinfo {year} {2024})}\BibitemShut {NoStop}%
\bibitem [{\citenamefont {Lykov}\ \emph {et~al.}(2023)\citenamefont {Lykov}, \citenamefont {Wurtz}, \citenamefont {Poole}, \citenamefont {Saffman}, \citenamefont {Noel},\ and\ \citenamefont {Alexeev}}]{Lykov2023}%
  \BibitemOpen
  \bibfield  {author} {\bibinfo {author} {\bibfnamefont {D.}~\bibnamefont {Lykov}}, \bibinfo {author} {\bibfnamefont {J.}~\bibnamefont {Wurtz}}, \bibinfo {author} {\bibfnamefont {C.}~\bibnamefont {Poole}}, \bibinfo {author} {\bibfnamefont {M.}~\bibnamefont {Saffman}}, \bibinfo {author} {\bibfnamefont {T.}~\bibnamefont {Noel}},\ and\ \bibinfo {author} {\bibfnamefont {Y.}~\bibnamefont {Alexeev}},\ }\bibfield  {title} {\bibinfo {title} {Sampling frequency thresholds for the quantum advantage of the quantum approximate optimization algorithm},\ }\href {https://doi.org/10.1038/s41534-023-00718-4} {\bibfield  {journal} {\bibinfo  {journal} {Npj Quantum Inf.}\ }\textbf {\bibinfo {volume} {9}},\ \bibinfo {pages} {73} (\bibinfo {year} {2023})}\BibitemShut {NoStop}%
\bibitem [{\citenamefont {Kibble}(1976)}]{Kibble1976}%
  \BibitemOpen
  \bibfield  {author} {\bibinfo {author} {\bibfnamefont {T.~W.~B.}\ \bibnamefont {Kibble}},\ }\bibfield  {title} {\bibinfo {title} {Topology of cosmic domains and strings},\ }\href {https://doi.org/10.1088/0305-4470/9/8/029} {\bibfield  {journal} {\bibinfo  {journal} {J. Phys. A Math.}\ }\textbf {\bibinfo {volume} {9}},\ \bibinfo {pages} {1387} (\bibinfo {year} {1976})}\BibitemShut {NoStop}%
\bibitem [{\citenamefont {Zurek}(1985)}]{Zurek1985}%
  \BibitemOpen
  \bibfield  {author} {\bibinfo {author} {\bibfnamefont {W.~H.}\ \bibnamefont {Zurek}},\ }\bibfield  {title} {\bibinfo {title} {Cosmological experiments in superfluid helium?},\ }\href {https://doi.org/10.1038/317505a0} {\bibfield  {journal} {\bibinfo  {journal} {Nature}\ }\textbf {\bibinfo {volume} {317}},\ \bibinfo {pages} {505} (\bibinfo {year} {1985})}\BibitemShut {NoStop}%
\bibitem [{\citenamefont {Liu}\ \emph {et~al.}(2014)\citenamefont {Liu}, \citenamefont {Polkovnikov},\ and\ \citenamefont {Sandvik}}]{PhysRevB.89.054307}%
  \BibitemOpen
  \bibfield  {author} {\bibinfo {author} {\bibfnamefont {C.-W.}\ \bibnamefont {Liu}}, \bibinfo {author} {\bibfnamefont {A.}~\bibnamefont {Polkovnikov}},\ and\ \bibinfo {author} {\bibfnamefont {A.~W.}\ \bibnamefont {Sandvik}},\ }\bibfield  {title} {\bibinfo {title} {Dynamic scaling at classical phase transitions approached through nonequilibrium quenching},\ }\href {https://doi.org/10.1103/PhysRevB.89.054307} {\bibfield  {journal} {\bibinfo  {journal} {Phys. Rev. B}\ }\textbf {\bibinfo {volume} {89}},\ \bibinfo {pages} {054307} (\bibinfo {year} {2014})}\BibitemShut {NoStop}%
\bibitem [{\citenamefont {Liu}\ \emph {et~al.}(2015)\citenamefont {Liu}, \citenamefont {Polkovnikov},\ and\ \citenamefont {Sandvik}}]{PhysRevLett.114.147203}%
  \BibitemOpen
  \bibfield  {author} {\bibinfo {author} {\bibfnamefont {C.-W.}\ \bibnamefont {Liu}}, \bibinfo {author} {\bibfnamefont {A.}~\bibnamefont {Polkovnikov}},\ and\ \bibinfo {author} {\bibfnamefont {A.~W.}\ \bibnamefont {Sandvik}},\ }\bibfield  {title} {\bibinfo {title} {Quantum versus classical annealing: Insights from scaling theory and results for spin glasses on 3-regular graphs},\ }\href {https://doi.org/10.1103/PhysRevLett.114.147203} {\bibfield  {journal} {\bibinfo  {journal} {Phys. Rev. Lett.}\ }\textbf {\bibinfo {volume} {114}},\ \bibinfo {pages} {147203} (\bibinfo {year} {2015})}\BibitemShut {NoStop}%
\bibitem [{\citenamefont {Farhi}\ \emph {et~al.}(2012)\citenamefont {Farhi}, \citenamefont {Gosset}, \citenamefont {Hen}, \citenamefont {Sandvik}, \citenamefont {Shor}, \citenamefont {Young},\ and\ \citenamefont {Zamponi}}]{PhysRevA.86.052334}%
  \BibitemOpen
  \bibfield  {author} {\bibinfo {author} {\bibfnamefont {E.}~\bibnamefont {Farhi}}, \bibinfo {author} {\bibfnamefont {D.}~\bibnamefont {Gosset}}, \bibinfo {author} {\bibfnamefont {I.}~\bibnamefont {Hen}}, \bibinfo {author} {\bibfnamefont {A.~W.}\ \bibnamefont {Sandvik}}, \bibinfo {author} {\bibfnamefont {P.}~\bibnamefont {Shor}}, \bibinfo {author} {\bibfnamefont {A.~P.}\ \bibnamefont {Young}},\ and\ \bibinfo {author} {\bibfnamefont {F.}~\bibnamefont {Zamponi}},\ }\bibfield  {title} {\bibinfo {title} {Performance of the quantum adiabatic algorithm on random instances of two optimization problems on regular hypergraphs},\ }\href {https://doi.org/10.1103/PhysRevA.86.052334} {\bibfield  {journal} {\bibinfo  {journal} {Phys. Rev. A}\ }\textbf {\bibinfo {volume} {86}},\ \bibinfo {pages} {052334} (\bibinfo {year} {2012})}\BibitemShut {NoStop}%
\bibitem [{\citenamefont {Biroli}\ \emph {et~al.}(2010)\citenamefont {Biroli}, \citenamefont {Cugliandolo},\ and\ \citenamefont {Sicilia}}]{PhysRevE.81.050101}%
  \BibitemOpen
  \bibfield  {author} {\bibinfo {author} {\bibfnamefont {G.}~\bibnamefont {Biroli}}, \bibinfo {author} {\bibfnamefont {L.~F.}\ \bibnamefont {Cugliandolo}},\ and\ \bibinfo {author} {\bibfnamefont {A.}~\bibnamefont {Sicilia}},\ }\bibfield  {title} {\bibinfo {title} {Kibble-zurek mechanism and infinitely slow annealing through critical points},\ }\href {https://doi.org/10.1103/PhysRevE.81.050101} {\bibfield  {journal} {\bibinfo  {journal} {Phys. Rev. E}\ }\textbf {\bibinfo {volume} {81}},\ \bibinfo {pages} {050101} (\bibinfo {year} {2010})}\BibitemShut {NoStop}%
\bibitem [{\citenamefont {Farhi}\ \emph {et~al.}(2020)\citenamefont {Farhi}, \citenamefont {Gamarnik},\ and\ \citenamefont {Gutmann}}]{Farhi2020}%
  \BibitemOpen
  \bibfield  {author} {\bibinfo {author} {\bibfnamefont {E.}~\bibnamefont {Farhi}}, \bibinfo {author} {\bibfnamefont {D.}~\bibnamefont {Gamarnik}},\ and\ \bibinfo {author} {\bibfnamefont {S.}~\bibnamefont {Gutmann}},\ }\bibfield  {title} {\bibinfo {title} {The {Q}uantum {A}pproximate {O}ptimization {A}lgorithm {N}eeds to {S}ee the {W}hole {G}raph: {A} {T}ypical {C}ase},\ }\href {https://arxiv.org/abs/2004.09002} {\bibfield  {journal} {\bibinfo  {journal} {arXiv:2004.09002}\ } (\bibinfo {year} {2020})}\BibitemShut {NoStop}%
\bibitem [{\citenamefont {Zhou}\ \emph {et~al.}(2020)\citenamefont {Zhou}, \citenamefont {Wang}, \citenamefont {Choi}, \citenamefont {Pichler},\ and\ \citenamefont {Lukin}}]{PhysRevX.10.021067}%
  \BibitemOpen
  \bibfield  {author} {\bibinfo {author} {\bibfnamefont {L.}~\bibnamefont {Zhou}}, \bibinfo {author} {\bibfnamefont {S.-T.}\ \bibnamefont {Wang}}, \bibinfo {author} {\bibfnamefont {S.}~\bibnamefont {Choi}}, \bibinfo {author} {\bibfnamefont {H.}~\bibnamefont {Pichler}},\ and\ \bibinfo {author} {\bibfnamefont {M.~D.}\ \bibnamefont {Lukin}},\ }\bibfield  {title} {\bibinfo {title} {Quantum approximate optimization algorithm: Performance, mechanism, and implementation on near-term devices},\ }\href {https://doi.org/10.1103/PhysRevX.10.021067} {\bibfield  {journal} {\bibinfo  {journal} {Phys. Rev. X}\ }\textbf {\bibinfo {volume} {10}},\ \bibinfo {pages} {021067} (\bibinfo {year} {2020})}\BibitemShut {NoStop}%
\bibitem [{\citenamefont {Devitt}\ \emph {et~al.}(2013)\citenamefont {Devitt}, \citenamefont {Munro},\ and\ \citenamefont {Nemoto}}]{Devitt_2013}%
  \BibitemOpen
  \bibfield  {author} {\bibinfo {author} {\bibfnamefont {S.~J.}\ \bibnamefont {Devitt}}, \bibinfo {author} {\bibfnamefont {W.~J.}\ \bibnamefont {Munro}},\ and\ \bibinfo {author} {\bibfnamefont {K.}~\bibnamefont {Nemoto}},\ }\bibfield  {title} {\bibinfo {title} {Quantum error correction for beginners},\ }\href {https://doi.org/10.1088/0034-4885/76/7/076001} {\bibfield  {journal} {\bibinfo  {journal} {Rep. Prog. Phys.}\ }\textbf {\bibinfo {volume} {76}},\ \bibinfo {pages} {076001} (\bibinfo {year} {2013})}\BibitemShut {NoStop}%
\bibitem [{\citenamefont {Knill}\ \emph {et~al.}(2008)\citenamefont {Knill}, \citenamefont {Leibfried}, \citenamefont {Reichle}, \citenamefont {Britton}, \citenamefont {Blakestad}, \citenamefont {Jost}, \citenamefont {Langer}, \citenamefont {Ozeri}, \citenamefont {Seidelin},\ and\ \citenamefont {Wineland}}]{PhysRevA.77.012307}%
  \BibitemOpen
  \bibfield  {author} {\bibinfo {author} {\bibfnamefont {E.}~\bibnamefont {Knill}}, \bibinfo {author} {\bibfnamefont {D.}~\bibnamefont {Leibfried}}, \bibinfo {author} {\bibfnamefont {R.}~\bibnamefont {Reichle}}, \bibinfo {author} {\bibfnamefont {J.}~\bibnamefont {Britton}}, \bibinfo {author} {\bibfnamefont {R.~B.}\ \bibnamefont {Blakestad}}, \bibinfo {author} {\bibfnamefont {J.~D.}\ \bibnamefont {Jost}}, \bibinfo {author} {\bibfnamefont {C.}~\bibnamefont {Langer}}, \bibinfo {author} {\bibfnamefont {R.}~\bibnamefont {Ozeri}}, \bibinfo {author} {\bibfnamefont {S.}~\bibnamefont {Seidelin}},\ and\ \bibinfo {author} {\bibfnamefont {D.~J.}\ \bibnamefont {Wineland}},\ }\bibfield  {title} {\bibinfo {title} {Randomized benchmarking of quantum gates},\ }\href {https://doi.org/10.1103/PhysRevA.77.012307} {\bibfield  {journal} {\bibinfo  {journal} {Phys. Rev. A}\ }\textbf {\bibinfo {volume} {77}},\ \bibinfo {pages} {012307} (\bibinfo {year} {2008})}\BibitemShut {NoStop}%
\bibitem [{\citenamefont {Metropolis}\ \emph {et~al.}(2004)\citenamefont {Metropolis}, \citenamefont {Rosenbluth}, \citenamefont {Rosenbluth}, \citenamefont {Teller},\ and\ \citenamefont {Teller}}]{10.1063/1.1699114}%
  \BibitemOpen
  \bibfield  {author} {\bibinfo {author} {\bibfnamefont {N.}~\bibnamefont {Metropolis}}, \bibinfo {author} {\bibfnamefont {A.~W.}\ \bibnamefont {Rosenbluth}}, \bibinfo {author} {\bibfnamefont {M.~N.}\ \bibnamefont {Rosenbluth}}, \bibinfo {author} {\bibfnamefont {A.~H.}\ \bibnamefont {Teller}},\ and\ \bibinfo {author} {\bibfnamefont {E.}~\bibnamefont {Teller}},\ }\bibfield  {title} {\bibinfo {title} {{E}quation of {S}tate {C}alculations by {F}ast {C}omputing {M}achines},\ }\href {https://doi.org/10.1063/1.1699114} {\bibfield  {journal} {\bibinfo  {journal} {J. Chem. Phys.}\ }\textbf {\bibinfo {volume} {21}},\ \bibinfo {pages} {1087} (\bibinfo {year} {2004})}\BibitemShut {NoStop}%
\bibitem [{\citenamefont {Hastings}(1970)}]{10.1093/biomet/57.1.97}%
  \BibitemOpen
  \bibfield  {author} {\bibinfo {author} {\bibfnamefont {W.~K.}\ \bibnamefont {Hastings}},\ }\bibfield  {title} {\bibinfo {title} {Monte carlo sampling methods using {M}arkov chains and their applications},\ }\href {https://doi.org/10.1093/biomet/57.1.97} {\bibfield  {journal} {\bibinfo  {journal} {Biometrika}\ }\textbf {\bibinfo {volume} {57}},\ \bibinfo {pages} {97} (\bibinfo {year} {1970})}\BibitemShut {NoStop}%
\bibitem [{\citenamefont {Granville}\ \emph {et~al.}(1994)\citenamefont {Granville}, \citenamefont {Krivanek},\ and\ \citenamefont {Rasson}}]{295910}%
  \BibitemOpen
  \bibfield  {author} {\bibinfo {author} {\bibfnamefont {V.}~\bibnamefont {Granville}}, \bibinfo {author} {\bibfnamefont {M.}~\bibnamefont {Krivanek}},\ and\ \bibinfo {author} {\bibfnamefont {J.-P.}\ \bibnamefont {Rasson}},\ }\bibfield  {title} {\bibinfo {title} {Simulated annealing: a proof of convergence},\ }\href {https://doi.org/10.1109/34.295910} {\bibfield  {journal} {\bibinfo  {journal} {IEEE Trans. Pattern Anal. Mach. Intell.}\ }\textbf {\bibinfo {volume} {16}},\ \bibinfo {pages} {652} (\bibinfo {year} {1994})}\BibitemShut {NoStop}%
\bibitem [{\citenamefont {Nolte}\ and\ \citenamefont {Schrader}(1997)}]{10.1007/978-3-642-60744-8_32}%
  \BibitemOpen
  \bibfield  {author} {\bibinfo {author} {\bibfnamefont {A.}~\bibnamefont {Nolte}}\ and\ \bibinfo {author} {\bibfnamefont {R.}~\bibnamefont {Schrader}},\ }\bibfield  {title} {\bibinfo {title} {A {N}ote on the {F}inite {T}ime {B}ehaviour of {S}imulated {A}nnealing},\ }in\ \href@noop {} {\emph {\bibinfo {booktitle} {Operations {R}esearch {P}roceedings 1996}}}\ (\bibinfo  {publisher} {Springer Berlin Heidelberg},\ \bibinfo {address} {Berlin, Heidelberg},\ \bibinfo {year} {1997})\ pp.\ \bibinfo {pages} {175--180}\BibitemShut {NoStop}%
\bibitem [{\citenamefont {Hukushima}\ and\ \citenamefont {Nemoto}(1996)}]{doi:10.1143/JPSJ.65.1604}%
  \BibitemOpen
  \bibfield  {author} {\bibinfo {author} {\bibfnamefont {K.}~\bibnamefont {Hukushima}}\ and\ \bibinfo {author} {\bibfnamefont {K.}~\bibnamefont {Nemoto}},\ }\bibfield  {title} {\bibinfo {title} {Exchange {M}onte {C}arlo {M}ethod and {A}pplication to {S}pin {G}lass {S}imulations},\ }\href {https://doi.org/10.1143/JPSJ.65.1604} {\bibfield  {journal} {\bibinfo  {journal} {J. Phys. Soc. Jpn.}\ }\textbf {\bibinfo {volume} {65}},\ \bibinfo {pages} {1604} (\bibinfo {year} {1996})}\BibitemShut {NoStop}%
\bibitem [{\citenamefont {Dunning}\ \emph {et~al.}(2018)\citenamefont {Dunning}, \citenamefont {Gupta},\ and\ \citenamefont {Silberholz}}]{DunningEtAl2018}%
  \BibitemOpen
  \bibfield  {author} {\bibinfo {author} {\bibfnamefont {I.}~\bibnamefont {Dunning}}, \bibinfo {author} {\bibfnamefont {S.}~\bibnamefont {Gupta}},\ and\ \bibinfo {author} {\bibfnamefont {J.}~\bibnamefont {Silberholz}},\ }\bibfield  {title} {\bibinfo {title} {What works best when? a systematic evaluation of heuristics for max-cut and {QUBO}},\ }\href@noop {} {\bibfield  {journal} {\bibinfo  {journal} {{INFORMS} J. Comput.}\ }\textbf {\bibinfo {volume} {30}} (\bibinfo {year} {2018})}\BibitemShut {NoStop}%
\bibitem [{\citenamefont {Achterberg}\ \emph {et~al.}(2020)\citenamefont {Achterberg}, \citenamefont {Bixby}, \citenamefont {Gu}, \citenamefont {Rothberg},\ and\ \citenamefont {Weninger}}]{achterberg2020presolve}%
  \BibitemOpen
  \bibfield  {author} {\bibinfo {author} {\bibfnamefont {T.}~\bibnamefont {Achterberg}}, \bibinfo {author} {\bibfnamefont {R.~E.}\ \bibnamefont {Bixby}}, \bibinfo {author} {\bibfnamefont {Z.}~\bibnamefont {Gu}}, \bibinfo {author} {\bibfnamefont {E.}~\bibnamefont {Rothberg}},\ and\ \bibinfo {author} {\bibfnamefont {D.}~\bibnamefont {Weninger}},\ }\bibfield  {title} {\bibinfo {title} {Presolve reductions in mixed integer programming},\ }\href@noop {} {\bibfield  {journal} {\bibinfo  {journal} {INFORMS Journal on Computing}\ }\textbf {\bibinfo {volume} {32}},\ \bibinfo {pages} {473} (\bibinfo {year} {2020})}\BibitemShut {NoStop}%
\bibitem [{\citenamefont {Conforti}\ \emph {et~al.}(2014)\citenamefont {Conforti}, \citenamefont {Cornu{\'e}jols},\ and\ \citenamefont {Zambelli}}]{conforti2014integer}%
  \BibitemOpen
  \bibfield  {author} {\bibinfo {author} {\bibfnamefont {M.}~\bibnamefont {Conforti}}, \bibinfo {author} {\bibfnamefont {G.}~\bibnamefont {Cornu{\'e}jols}},\ and\ \bibinfo {author} {\bibfnamefont {G.}~\bibnamefont {Zambelli}},\ }\href@noop {} {\emph {\bibinfo {title} {Integer programming}}}\ (\bibinfo  {publisher} {Springer},\ \bibinfo {year} {2014})\BibitemShut {NoStop}%
\bibitem [{\citenamefont {Bixby}\ \emph {et~al.}(1999)\citenamefont {Bixby}, \citenamefont {Fenelon}, \citenamefont {Gu}, \citenamefont {Rothberg},\ and\ \citenamefont {Wunderling}}]{bixby1999mip}%
  \BibitemOpen
  \bibfield  {author} {\bibinfo {author} {\bibfnamefont {E.~R.}\ \bibnamefont {Bixby}}, \bibinfo {author} {\bibfnamefont {M.}~\bibnamefont {Fenelon}}, \bibinfo {author} {\bibfnamefont {Z.}~\bibnamefont {Gu}}, \bibinfo {author} {\bibfnamefont {E.}~\bibnamefont {Rothberg}},\ and\ \bibinfo {author} {\bibfnamefont {R.}~\bibnamefont {Wunderling}},\ }\bibfield  {title} {\bibinfo {title} {Mip: Theory and practice—closing the gap},\ }in\ \href@noop {} {\emph {\bibinfo {booktitle} {IFIP Conference on System Modeling and Optimization}}}\ (\bibinfo {organization} {Springer},\ \bibinfo {year} {1999})\ pp.\ \bibinfo {pages} {19--49}\BibitemShut {NoStop}%
\bibitem [{\citenamefont {Dupont}\ \emph {et~al.}(2023)\citenamefont {Dupont}, \citenamefont {Evert}, \citenamefont {Hodson}, \citenamefont {Sundar}, \citenamefont {Jeffrey}, \citenamefont {Yamaguchi}, \citenamefont {Feng}, \citenamefont {Maciejewski}, \citenamefont {Hadfield}, \citenamefont {Alam}, \citenamefont {Wang}, \citenamefont {Grabbe}, \citenamefont {Lott}, \citenamefont {Rieffel}, \citenamefont {Venturelli},\ and\ \citenamefont {Reagor}}]{Dupont2023}%
  \BibitemOpen
  \bibfield  {author} {\bibinfo {author} {\bibfnamefont {M.}~\bibnamefont {Dupont}}, \bibinfo {author} {\bibfnamefont {B.}~\bibnamefont {Evert}}, \bibinfo {author} {\bibfnamefont {M.~J.}\ \bibnamefont {Hodson}}, \bibinfo {author} {\bibfnamefont {B.}~\bibnamefont {Sundar}}, \bibinfo {author} {\bibfnamefont {S.}~\bibnamefont {Jeffrey}}, \bibinfo {author} {\bibfnamefont {Y.}~\bibnamefont {Yamaguchi}}, \bibinfo {author} {\bibfnamefont {D.}~\bibnamefont {Feng}}, \bibinfo {author} {\bibfnamefont {F.~B.}\ \bibnamefont {Maciejewski}}, \bibinfo {author} {\bibfnamefont {S.}~\bibnamefont {Hadfield}}, \bibinfo {author} {\bibfnamefont {M.~S.}\ \bibnamefont {Alam}}, \bibinfo {author} {\bibfnamefont {Z.}~\bibnamefont {Wang}}, \bibinfo {author} {\bibfnamefont {S.}~\bibnamefont {Grabbe}}, \bibinfo {author} {\bibfnamefont {P.~A.}\ \bibnamefont {Lott}}, \bibinfo {author} {\bibfnamefont {E.~G.}\ \bibnamefont {Rieffel}}, \bibinfo {author} {\bibfnamefont {D.}~\bibnamefont {Venturelli}},\ and\ \bibinfo {author} {\bibfnamefont
  {M.~J.}\ \bibnamefont {Reagor}},\ }\bibfield  {title} {\bibinfo {title} {Quantum-enhanced greedy combinatorial optimization solver},\ }\href {https://doi.org/10.1126/sciadv.adi0487} {\bibfield  {journal} {\bibinfo  {journal} {Sci. Adv.}\ }\textbf {\bibinfo {volume} {9}},\ \bibinfo {pages} {eadi0487} (\bibinfo {year} {2023})}\BibitemShut {NoStop}%
\bibitem [{\citenamefont {Williamson}\ and\ \citenamefont {Shmoys}(2011)}]{williamson2011}%
  \BibitemOpen
  \bibfield  {author} {\bibinfo {author} {\bibfnamefont {D.~P.}\ \bibnamefont {Williamson}}\ and\ \bibinfo {author} {\bibfnamefont {D.~B.}\ \bibnamefont {Shmoys}},\ }\href@noop {} {\emph {\bibinfo {title} {The {D}esign of {A}pproximation {A}lgorithms}}}\ (\bibinfo  {publisher} {Cambridge university press},\ \bibinfo {year} {2011})\BibitemShut {NoStop}%
\bibitem [{\citenamefont {Mohar}\ and\ \citenamefont {Poljak}(1990)}]{Mohar1990}%
  \BibitemOpen
  \bibfield  {author} {\bibinfo {author} {\bibfnamefont {B.}~\bibnamefont {Mohar}}\ and\ \bibinfo {author} {\bibfnamefont {S.}~\bibnamefont {Poljak}},\ }\bibfield  {title} {\bibinfo {title} {Eigenvalues and the max-cut problem},\ }\href {http://eudml.org/doc/13856} {\bibfield  {journal} {\bibinfo  {journal} {Czechoslov. Math. J.}\ }\textbf {\bibinfo {volume} {40}},\ \bibinfo {pages} {343} (\bibinfo {year} {1990})}\BibitemShut {NoStop}%
\bibitem [{\citenamefont {Delorme}\ and\ \citenamefont {Poljak}(1993{\natexlab{a}})}]{Delorme1993}%
  \BibitemOpen
  \bibfield  {author} {\bibinfo {author} {\bibfnamefont {C.}~\bibnamefont {Delorme}}\ and\ \bibinfo {author} {\bibfnamefont {S.}~\bibnamefont {Poljak}},\ }\bibfield  {title} {\bibinfo {title} {Laplacian eigenvalues and the maximum cut problem},\ }\href {https://doi.org/10.1007/BF01585184} {\bibfield  {journal} {\bibinfo  {journal} {Math. Program.}\ }\textbf {\bibinfo {volume} {62}},\ \bibinfo {pages} {557} (\bibinfo {year} {1993}{\natexlab{a}})}\BibitemShut {NoStop}%
\bibitem [{\citenamefont {Delorme}\ and\ \citenamefont {Poljak}(1993{\natexlab{b}})}]{DELORME1993145}%
  \BibitemOpen
  \bibfield  {author} {\bibinfo {author} {\bibfnamefont {C.}~\bibnamefont {Delorme}}\ and\ \bibinfo {author} {\bibfnamefont {S.}~\bibnamefont {Poljak}},\ }\bibfield  {title} {\bibinfo {title} {The performance of an eigenvalue bound on the max-cut problem in some classes of graphs},\ }\href {https://doi.org/https://doi.org/10.1016/0012-365X(93)90151-I} {\bibfield  {journal} {\bibinfo  {journal} {Discrete Math.}\ }\textbf {\bibinfo {volume} {111}},\ \bibinfo {pages} {145} (\bibinfo {year} {1993}{\natexlab{b}})}\BibitemShut {NoStop}%
\bibitem [{\citenamefont {Poljak}\ and\ \citenamefont {Rendl}(1995)}]{POLJAK1995249}%
  \BibitemOpen
  \bibfield  {author} {\bibinfo {author} {\bibfnamefont {S.}~\bibnamefont {Poljak}}\ and\ \bibinfo {author} {\bibfnamefont {F.}~\bibnamefont {Rendl}},\ }\bibfield  {title} {\bibinfo {title} {Solving the max-cut problem using eigenvalues},\ }\href {https://doi.org/https://doi.org/10.1016/0166-218X(94)00155-7} {\bibfield  {journal} {\bibinfo  {journal} {Discret. Appl. Math.}\ }\textbf {\bibinfo {volume} {62}},\ \bibinfo {pages} {249} (\bibinfo {year} {1995})}\BibitemShut {NoStop}%
\bibitem [{\citenamefont {O'Donoghue}\ \emph {et~al.}(2016)\citenamefont {O'Donoghue}, \citenamefont {Chu}, \citenamefont {Parikh},\ and\ \citenamefont {Boyd}}]{ocpb:16}%
  \BibitemOpen
  \bibfield  {author} {\bibinfo {author} {\bibfnamefont {B.}~\bibnamefont {O'Donoghue}}, \bibinfo {author} {\bibfnamefont {E.}~\bibnamefont {Chu}}, \bibinfo {author} {\bibfnamefont {N.}~\bibnamefont {Parikh}},\ and\ \bibinfo {author} {\bibfnamefont {S.}~\bibnamefont {Boyd}},\ }\bibfield  {title} {\bibinfo {title} {Conic optimization via operator splitting and homogeneous self-dual embedding},\ }\href {http://stanford.edu/~boyd/papers/scs.html} {\bibfield  {journal} {\bibinfo  {journal} {J. Optim. Theory Appl.}\ }\textbf {\bibinfo {volume} {169}},\ \bibinfo {pages} {1042} (\bibinfo {year} {2016})}\BibitemShut {NoStop}%
\bibitem [{\citenamefont {O'Donoghue}(2021)}]{odonoghue:21}%
  \BibitemOpen
  \bibfield  {author} {\bibinfo {author} {\bibfnamefont {B.}~\bibnamefont {O'Donoghue}},\ }\bibfield  {title} {\bibinfo {title} {Operator splitting for a homogeneous embedding of the linear complementarity problem},\ }\href@noop {} {\bibfield  {journal} {\bibinfo  {journal} {{SIAM} J. Optim.}\ }\textbf {\bibinfo {volume} {31}},\ \bibinfo {pages} {1999} (\bibinfo {year} {2021})}\BibitemShut {NoStop}%
\bibitem [{\citenamefont {Farhi}\ \emph {et~al.}(2014{\natexlab{b}})\citenamefont {Farhi}, \citenamefont {Goldstone},\ and\ \citenamefont {Gutmann}}]{Farhi2014b}%
  \BibitemOpen
  \bibfield  {author} {\bibinfo {author} {\bibfnamefont {E.}~\bibnamefont {Farhi}}, \bibinfo {author} {\bibfnamefont {J.}~\bibnamefont {Goldstone}},\ and\ \bibinfo {author} {\bibfnamefont {S.}~\bibnamefont {Gutmann}},\ }\bibfield  {title} {\bibinfo {title} {{A} {Q}uantum {A}pproximate {O}ptimization {A}lgorithm {A}pplied to a {B}ounded {O}ccurrence {C}onstraint {P}roblem},\ }\href {https://arxiv.org/abs/1412.6062} {\bibfield  {journal} {\bibinfo  {journal} {arXiv:1412.6062}\ } (\bibinfo {year} {2014}{\natexlab{b}})}\BibitemShut {NoStop}%
\bibitem [{\citenamefont {Farhi}\ and\ \citenamefont {Harrow}(2016)}]{Farhi2016}%
  \BibitemOpen
  \bibfield  {author} {\bibinfo {author} {\bibfnamefont {E.}~\bibnamefont {Farhi}}\ and\ \bibinfo {author} {\bibfnamefont {A.~W.}\ \bibnamefont {Harrow}},\ }\bibfield  {title} {\bibinfo {title} {{Q}uantum {S}upremacy through the {Q}uantum {A}pproximate {O}ptimization {A}lgorithm},\ }\href {https://arxiv.org/abs/1602.07674} {\bibfield  {journal} {\bibinfo  {journal} {arXiv:1602.07674}\ } (\bibinfo {year} {2016})}\BibitemShut {NoStop}%
\bibitem [{\citenamefont {Wurtz}\ and\ \citenamefont {Love}(2022)}]{Wurtz2022}%
  \BibitemOpen
  \bibfield  {author} {\bibinfo {author} {\bibfnamefont {J.}~\bibnamefont {Wurtz}}\ and\ \bibinfo {author} {\bibfnamefont {P.~J.}\ \bibnamefont {Love}},\ }\bibfield  {title} {\bibinfo {title} {Counterdiabaticity and the quantum approximate optimization algorithm},\ }\href {https://doi.org/10.22331/q-2022-01-27-635} {\bibfield  {journal} {\bibinfo  {journal} {{Quantum}}\ }\textbf {\bibinfo {volume} {6}},\ \bibinfo {pages} {635} (\bibinfo {year} {2022})}\BibitemShut {NoStop}%
\bibitem [{\citenamefont {Wang}\ \emph {et~al.}(2018)\citenamefont {Wang}, \citenamefont {Hadfield}, \citenamefont {Jiang},\ and\ \citenamefont {Rieffel}}]{PhysRevA.97.022304}%
  \BibitemOpen
  \bibfield  {author} {\bibinfo {author} {\bibfnamefont {Z.}~\bibnamefont {Wang}}, \bibinfo {author} {\bibfnamefont {S.}~\bibnamefont {Hadfield}}, \bibinfo {author} {\bibfnamefont {Z.}~\bibnamefont {Jiang}},\ and\ \bibinfo {author} {\bibfnamefont {E.~G.}\ \bibnamefont {Rieffel}},\ }\bibfield  {title} {\bibinfo {title} {Quantum approximate optimization algorithm for maxcut: A fermionic view},\ }\href {https://doi.org/10.1103/PhysRevA.97.022304} {\bibfield  {journal} {\bibinfo  {journal} {Phys. Rev. A}\ }\textbf {\bibinfo {volume} {97}},\ \bibinfo {pages} {022304} (\bibinfo {year} {2018})}\BibitemShut {NoStop}%
\bibitem [{\citenamefont {Sherrington}\ and\ \citenamefont {Kirkpatrick}(1975)}]{PhysRevLett.35.1792}%
  \BibitemOpen
  \bibfield  {author} {\bibinfo {author} {\bibfnamefont {D.}~\bibnamefont {Sherrington}}\ and\ \bibinfo {author} {\bibfnamefont {S.}~\bibnamefont {Kirkpatrick}},\ }\bibfield  {title} {\bibinfo {title} {Solvable {M}odel of a {S}pin-{G}lass},\ }\href {https://doi.org/10.1103/PhysRevLett.35.1792} {\bibfield  {journal} {\bibinfo  {journal} {Phys. Rev. Lett.}\ }\textbf {\bibinfo {volume} {35}},\ \bibinfo {pages} {1792} (\bibinfo {year} {1975})}\BibitemShut {NoStop}%
\bibitem [{\citenamefont {Farhi}\ \emph {et~al.}(2022)\citenamefont {Farhi}, \citenamefont {Goldstone}, \citenamefont {Gutmann},\ and\ \citenamefont {Zhou}}]{Farhi2022}%
  \BibitemOpen
  \bibfield  {author} {\bibinfo {author} {\bibfnamefont {E.}~\bibnamefont {Farhi}}, \bibinfo {author} {\bibfnamefont {J.}~\bibnamefont {Goldstone}}, \bibinfo {author} {\bibfnamefont {S.}~\bibnamefont {Gutmann}},\ and\ \bibinfo {author} {\bibfnamefont {L.}~\bibnamefont {Zhou}},\ }\bibfield  {title} {\bibinfo {title} {The {Q}uantum {A}pproximate {O}ptimization {A}lgorithm and the {S}herrington-{K}irkpatrick {M}odel at {I}nfinite {S}ize},\ }\href {https://doi.org/10.22331/q-2022-07-07-759} {\bibfield  {journal} {\bibinfo  {journal} {{Quantum}}\ }\textbf {\bibinfo {volume} {6}},\ \bibinfo {pages} {759} (\bibinfo {year} {2022})}\BibitemShut {NoStop}%
\bibitem [{\citenamefont {McClean}\ \emph {et~al.}(2018)\citenamefont {McClean}, \citenamefont {Boixo}, \citenamefont {Smelyanskiy}, \citenamefont {Babbush},\ and\ \citenamefont {Neven}}]{McClean2018}%
  \BibitemOpen
  \bibfield  {author} {\bibinfo {author} {\bibfnamefont {J.~R.}\ \bibnamefont {McClean}}, \bibinfo {author} {\bibfnamefont {S.}~\bibnamefont {Boixo}}, \bibinfo {author} {\bibfnamefont {V.~N.}\ \bibnamefont {Smelyanskiy}}, \bibinfo {author} {\bibfnamefont {R.}~\bibnamefont {Babbush}},\ and\ \bibinfo {author} {\bibfnamefont {H.}~\bibnamefont {Neven}},\ }\bibfield  {title} {\bibinfo {title} {Barren plateaus in quantum neural network training landscapes},\ }\href {https://doi.org/10.1038/s41467-018-07090-4} {\bibfield  {journal} {\bibinfo  {journal} {Nat. Comm.}\ }\textbf {\bibinfo {volume} {9}},\ \bibinfo {pages} {4812} (\bibinfo {year} {2018})}\BibitemShut {NoStop}%
\bibitem [{\citenamefont {Lotshaw}\ \emph {et~al.}(2021)\citenamefont {Lotshaw}, \citenamefont {Humble}, \citenamefont {Herrman}, \citenamefont {Ostrowski},\ and\ \citenamefont {Siopsis}}]{Lotshaw2021}%
  \BibitemOpen
  \bibfield  {author} {\bibinfo {author} {\bibfnamefont {P.~C.}\ \bibnamefont {Lotshaw}}, \bibinfo {author} {\bibfnamefont {T.~S.}\ \bibnamefont {Humble}}, \bibinfo {author} {\bibfnamefont {R.}~\bibnamefont {Herrman}}, \bibinfo {author} {\bibfnamefont {J.}~\bibnamefont {Ostrowski}},\ and\ \bibinfo {author} {\bibfnamefont {G.}~\bibnamefont {Siopsis}},\ }\bibfield  {title} {\bibinfo {title} {Empirical performance bounds for quantum approximate optimization},\ }\href {https://doi.org/10.1007/s11128-021-03342-3} {\bibfield  {journal} {\bibinfo  {journal} {Quantum Inf. Process.}\ }\textbf {\bibinfo {volume} {20}},\ \bibinfo {pages} {403} (\bibinfo {year} {2021})}\BibitemShut {NoStop}%
\bibitem [{\citenamefont {Lee}\ \emph {et~al.}(2023)\citenamefont {Lee}, \citenamefont {Xie}, \citenamefont {Cai}, \citenamefont {Saito},\ and\ \citenamefont {Asai}}]{Lee2022}%
  \BibitemOpen
  \bibfield  {author} {\bibinfo {author} {\bibfnamefont {X.}~\bibnamefont {Lee}}, \bibinfo {author} {\bibfnamefont {N.}~\bibnamefont {Xie}}, \bibinfo {author} {\bibfnamefont {D.}~\bibnamefont {Cai}}, \bibinfo {author} {\bibfnamefont {Y.}~\bibnamefont {Saito}},\ and\ \bibinfo {author} {\bibfnamefont {N.}~\bibnamefont {Asai}},\ }\bibfield  {title} {\bibinfo {title} {A {D}epth-{P}rogressive {I}nitialization {S}trategy for {Q}uantum {A}pproximate {O}ptimization {A}lgorithm},\ }\bibfield  {journal} {\bibinfo  {journal} {Mathematics}\ }\textbf {\bibinfo {volume} {11}},\ \href {https://doi.org/10.3390/math11092176} {10.3390/math11092176} (\bibinfo {year} {2023})\BibitemShut {NoStop}%
\bibitem [{\citenamefont {Sud}\ \emph {et~al.}(2024)\citenamefont {Sud}, \citenamefont {Hadfield}, \citenamefont {Rieffel}, \citenamefont {Tubman},\ and\ \citenamefont {Hogg}}]{Sud2022}%
  \BibitemOpen
  \bibfield  {author} {\bibinfo {author} {\bibfnamefont {J.}~\bibnamefont {Sud}}, \bibinfo {author} {\bibfnamefont {S.}~\bibnamefont {Hadfield}}, \bibinfo {author} {\bibfnamefont {E.}~\bibnamefont {Rieffel}}, \bibinfo {author} {\bibfnamefont {N.}~\bibnamefont {Tubman}},\ and\ \bibinfo {author} {\bibfnamefont {T.}~\bibnamefont {Hogg}},\ }\bibfield  {title} {\bibinfo {title} {Parameter-setting heuristic for the quantum alternating operator ansatz},\ }\href {https://doi.org/10.1103/PhysRevResearch.6.023171} {\bibfield  {journal} {\bibinfo  {journal} {Phys. Rev. Res.}\ }\textbf {\bibinfo {volume} {6}},\ \bibinfo {pages} {023171} (\bibinfo {year} {2024})}\BibitemShut {NoStop}%
\bibitem [{\citenamefont {Lee}\ \emph {et~al.}(2024)\citenamefont {Lee}, \citenamefont {Yan}, \citenamefont {Xie}, \citenamefont {Cai}, \citenamefont {Saito},\ and\ \citenamefont {Asai}}]{Lee2023}%
  \BibitemOpen
  \bibfield  {author} {\bibinfo {author} {\bibfnamefont {X.}~\bibnamefont {Lee}}, \bibinfo {author} {\bibfnamefont {X.}~\bibnamefont {Yan}}, \bibinfo {author} {\bibfnamefont {N.}~\bibnamefont {Xie}}, \bibinfo {author} {\bibfnamefont {D.}~\bibnamefont {Cai}}, \bibinfo {author} {\bibfnamefont {Y.}~\bibnamefont {Saito}},\ and\ \bibinfo {author} {\bibfnamefont {N.}~\bibnamefont {Asai}},\ }\bibfield  {title} {\bibinfo {title} {Iterative layerwise training for the quantum approximate optimization algorithm},\ }\href {https://doi.org/10.1103/PhysRevA.109.052406} {\bibfield  {journal} {\bibinfo  {journal} {Phys. Rev. A}\ }\textbf {\bibinfo {volume} {109}},\ \bibinfo {pages} {052406} (\bibinfo {year} {2024})}\BibitemShut {NoStop}%
\bibitem [{\citenamefont {Li}\ \emph {et~al.}(2024)\citenamefont {Li}, \citenamefont {Li}, \citenamefont {Song}, \citenamefont {Qin}, \citenamefont {Wen},\ and\ \citenamefont {Gao}}]{Li2023}%
  \BibitemOpen
  \bibfield  {author} {\bibinfo {author} {\bibfnamefont {L.}~\bibnamefont {Li}}, \bibinfo {author} {\bibfnamefont {J.}~\bibnamefont {Li}}, \bibinfo {author} {\bibfnamefont {Y.}~\bibnamefont {Song}}, \bibinfo {author} {\bibfnamefont {S.}~\bibnamefont {Qin}}, \bibinfo {author} {\bibfnamefont {Q.}~\bibnamefont {Wen}},\ and\ \bibinfo {author} {\bibfnamefont {F.}~\bibnamefont {Gao}},\ }\bibfield  {title} {\bibinfo {title} {An efficient quantum proactive incremental learning algorithm},\ }\href {https://doi.org/10.1007/s11433-024-2501-4} {\bibfield  {journal} {\bibinfo  {journal} {Sci. China: Phys. Mech. Astron.}\ }\textbf {\bibinfo {volume} {68}},\ \bibinfo {pages} {210313} (\bibinfo {year} {2024})}\BibitemShut {NoStop}%
\bibitem [{\citenamefont {STEGER}\ and\ \citenamefont {WORMALD}(1999)}]{STEGER_WORMALD_1999}%
  \BibitemOpen
  \bibfield  {author} {\bibinfo {author} {\bibfnamefont {A.}~\bibnamefont {STEGER}}\ and\ \bibinfo {author} {\bibfnamefont {N.~C.}\ \bibnamefont {WORMALD}},\ }\bibfield  {title} {\bibinfo {title} {Generating {R}andom {R}egular {G}raphs {Q}uickly},\ }\href {https://doi.org/10.1017/S0963548399003867} {\bibfield  {journal} {\bibinfo  {journal} {Comb. Probab. Comput.}\ }\textbf {\bibinfo {volume} {8}},\ \bibinfo {pages} {377–396} (\bibinfo {year} {1999})}\BibitemShut {NoStop}%
\bibitem [{\citenamefont {Kim}\ and\ \citenamefont {Vu}(2003)}]{10.1145/780542.780576}%
  \BibitemOpen
  \bibfield  {author} {\bibinfo {author} {\bibfnamefont {J.~H.}\ \bibnamefont {Kim}}\ and\ \bibinfo {author} {\bibfnamefont {V.~H.}\ \bibnamefont {Vu}},\ }\bibfield  {title} {\bibinfo {title} {Generating {R}andom {R}egular {G}raphs},\ }in\ \href {https://doi.org/10.1145/780542.780576} {\emph {\bibinfo {booktitle} {Proceedings of the Thirty-Fifth Annual ACM Symposium on Theory of Computing}}},\ \bibinfo {series and number} {STOC '03}\ (\bibinfo  {publisher} {Association for Computing Machinery},\ \bibinfo {address} {New York, NY, USA},\ \bibinfo {year} {2003})\ p.\ \bibinfo {pages} {213–222}\BibitemShut {NoStop}%
\bibitem [{\citenamefont {{OEIS Foundation Inc.}}(2024)}]{oeis}%
  \BibitemOpen
  \bibfield  {author} {\bibinfo {author} {\bibnamefont {{OEIS Foundation Inc.}}},\ }\href@noop {} {\bibinfo {title} {The {O}n-{L}ine {E}ncyclopedia of {I}nteger {S}equences}} (\bibinfo {year} {2024}),\ \bibinfo {note} {published electronically at \url{http://oeis.org}}\BibitemShut {NoStop}%
\bibitem [{\citenamefont {Hagberg}\ \emph {et~al.}(2008)\citenamefont {Hagberg}, \citenamefont {Schult},\ and\ \citenamefont {Swart}}]{SciPyProceedings_11}%
  \BibitemOpen
  \bibfield  {author} {\bibinfo {author} {\bibfnamefont {A.~A.}\ \bibnamefont {Hagberg}}, \bibinfo {author} {\bibfnamefont {D.~A.}\ \bibnamefont {Schult}},\ and\ \bibinfo {author} {\bibfnamefont {P.~J.}\ \bibnamefont {Swart}},\ }\bibfield  {title} {\bibinfo {title} {Exploring {N}etwork {S}tructure, {D}ynamics, and {F}unction using {N}etworkx},\ }in\ \href@noop {} {\emph {\bibinfo {booktitle} {Proceedings of the 7th Python in Science Conference}}},\ \bibinfo {editor} {edited by\ \bibinfo {editor} {\bibfnamefont {G.}~\bibnamefont {Varoquaux}}, \bibinfo {editor} {\bibfnamefont {T.}~\bibnamefont {Vaught}},\ and\ \bibinfo {editor} {\bibfnamefont {J.}~\bibnamefont {Millman}}}\ (\bibinfo {address} {Pasadena, CA USA},\ \bibinfo {year} {2008})\ pp.\ \bibinfo {pages} {11 -- 15}\BibitemShut {NoStop}%
\bibitem [{\citenamefont {Wormald}(1981)}]{WORMALD1981168}%
  \BibitemOpen
  \bibfield  {author} {\bibinfo {author} {\bibfnamefont {N.~C.}\ \bibnamefont {Wormald}},\ }\bibfield  {title} {\bibinfo {title} {The asymptotic distribution of short cycles in random regular graphs},\ }\href {https://doi.org/https://doi.org/10.1016/S0095-8956(81)80022-6} {\bibfield  {journal} {\bibinfo  {journal} {J. Comb. Theory, Ser. B}\ }\textbf {\bibinfo {volume} {31}},\ \bibinfo {pages} {168} (\bibinfo {year} {1981})}\BibitemShut {NoStop}%
\bibitem [{\citenamefont {McKay}\ \emph {et~al.}(2004)\citenamefont {McKay}, \citenamefont {Wormald},\ and\ \citenamefont {Wysocka}}]{McKay2004}%
  \BibitemOpen
  \bibfield  {author} {\bibinfo {author} {\bibfnamefont {B.~D.}\ \bibnamefont {McKay}}, \bibinfo {author} {\bibfnamefont {N.~C.}\ \bibnamefont {Wormald}},\ and\ \bibinfo {author} {\bibfnamefont {B.}~\bibnamefont {Wysocka}},\ }\bibfield  {title} {\bibinfo {title} {Short {C}ycles in {R}andom {R}egular {G}raphs},\ }\bibfield  {journal} {\bibinfo  {journal} {Electron. J. Comb.}\ }\textbf {\bibinfo {volume} {11}},\ \href {https://doi.org/https://doi.org/10.37236/1819} {https://doi.org/10.37236/1819} (\bibinfo {year} {2004})\BibitemShut {NoStop}%
\bibitem [{\citenamefont {Bodlaender}\ and\ \citenamefont {Koster}(2010)}]{BODLAENDER2010259}%
  \BibitemOpen
  \bibfield  {author} {\bibinfo {author} {\bibfnamefont {H.~L.}\ \bibnamefont {Bodlaender}}\ and\ \bibinfo {author} {\bibfnamefont {A.~M.}\ \bibnamefont {Koster}},\ }\bibfield  {title} {\bibinfo {title} {Treewidth computations {I}. {U}pper bounds},\ }\href {https://doi.org/https://doi.org/10.1016/j.ic.2009.03.008} {\bibfield  {journal} {\bibinfo  {journal} {Inf. Comput.}\ }\textbf {\bibinfo {volume} {208}},\ \bibinfo {pages} {259} (\bibinfo {year} {2010})}\BibitemShut {NoStop}%
\bibitem [{\citenamefont {Shervashidze}\ \emph {et~al.}(2011)\citenamefont {Shervashidze}, \citenamefont {Schweitzer}, \citenamefont {van Leeuwen}, \citenamefont {Mehlhorn},\ and\ \citenamefont {Borgwardt}}]{JMLR:v12:shervashidze11a}%
  \BibitemOpen
  \bibfield  {author} {\bibinfo {author} {\bibfnamefont {N.}~\bibnamefont {Shervashidze}}, \bibinfo {author} {\bibfnamefont {P.}~\bibnamefont {Schweitzer}}, \bibinfo {author} {\bibfnamefont {E.~J.}\ \bibnamefont {van Leeuwen}}, \bibinfo {author} {\bibfnamefont {K.}~\bibnamefont {Mehlhorn}},\ and\ \bibinfo {author} {\bibfnamefont {K.~M.}\ \bibnamefont {Borgwardt}},\ }\bibfield  {title} {\bibinfo {title} {Weisfeiler-{L}ehman {G}raph {K}ernels},\ }\href {http://jmlr.org/papers/v12/shervashidze11a.html} {\bibfield  {journal} {\bibinfo  {journal} {J. Mach. Learn. Res.}\ }\textbf {\bibinfo {volume} {12}},\ \bibinfo {pages} {2539} (\bibinfo {year} {2011})}\BibitemShut {NoStop}%
\bibitem [{\citenamefont {Babai}(2015)}]{Laszlo2015}%
  \BibitemOpen
  \bibfield  {author} {\bibinfo {author} {\bibfnamefont {L.}~\bibnamefont {Babai}},\ }\bibfield  {title} {\bibinfo {title} {{G}raph {I}somorphism in {Q}uasipolynomial {T}ime},\ }\href {https://arxiv.org/abs/1512.03547} {\bibfield  {journal} {\bibinfo  {journal} {arXiv:1512.03547}\ } (\bibinfo {year} {2015})}\BibitemShut {NoStop}%
\bibitem [{\citenamefont {Helfgott}\ \emph {et~al.}(2017)\citenamefont {Helfgott}, \citenamefont {Bajpai},\ and\ \citenamefont {Dona}}]{Andres2017}%
  \BibitemOpen
  \bibfield  {author} {\bibinfo {author} {\bibfnamefont {H.~A.}\ \bibnamefont {Helfgott}}, \bibinfo {author} {\bibfnamefont {J.}~\bibnamefont {Bajpai}},\ and\ \bibinfo {author} {\bibfnamefont {D.}~\bibnamefont {Dona}},\ }\bibfield  {title} {\bibinfo {title} {Graph isomorphisms in quasi-polynomial time},\ }\href {https://arxiv.org/abs/1710.04574} {\bibfield  {journal} {\bibinfo  {journal} {arXiv:1710.04574}\ } (\bibinfo {year} {2017})}\BibitemShut {NoStop}%
\bibitem [{\citenamefont {Matsuo}\ \emph {et~al.}(2023)\citenamefont {Matsuo}, \citenamefont {Yamashita},\ and\ \citenamefont {Egger}}]{MATSUO20232022EAP1159}%
  \BibitemOpen
  \bibfield  {author} {\bibinfo {author} {\bibfnamefont {A.}~\bibnamefont {Matsuo}}, \bibinfo {author} {\bibfnamefont {S.}~\bibnamefont {Yamashita}},\ and\ \bibinfo {author} {\bibfnamefont {D.~J.}\ \bibnamefont {Egger}},\ }\bibfield  {title} {\bibinfo {title} {A {SAT} {A}pproach to the {I}nitial {M}apping {P}roblem in {S}wap {G}ate {I}nsertion for {C}ommuting {G}ates},\ }\href {https://doi.org/10.1587/transfun.2022EAP1159} {\bibfield  {journal} {\bibinfo  {journal} {IEICE Trans. Fundam. Electron. Commun. Comput. Sci.}\ }\textbf {\bibinfo {volume} {E106.A}},\ \bibinfo {pages} {1424} (\bibinfo {year} {2023})}\BibitemShut {NoStop}%
\bibitem [{\citenamefont {Audemard}\ and\ \citenamefont {Simon}(2018)}]{doi:10.1142/S0218213018400018}%
  \BibitemOpen
  \bibfield  {author} {\bibinfo {author} {\bibfnamefont {G.}~\bibnamefont {Audemard}}\ and\ \bibinfo {author} {\bibfnamefont {L.}~\bibnamefont {Simon}},\ }\bibfield  {title} {\bibinfo {title} {On the {G}lucose {SAT} solver},\ }\href {https://doi.org/10.1142/S0218213018400018} {\bibfield  {journal} {\bibinfo  {journal} {Int. J. Artif. Intell. Tools}\ }\textbf {\bibinfo {volume} {27}},\ \bibinfo {pages} {1840001} (\bibinfo {year} {2018})}\BibitemShut {NoStop}%
\bibitem [{\citenamefont {E{\'e}n}\ and\ \citenamefont {S{\"o}rensson}(2004)}]{10.1007/978-3-540-24605-3_37}%
  \BibitemOpen
  \bibfield  {author} {\bibinfo {author} {\bibfnamefont {N.}~\bibnamefont {E{\'e}n}}\ and\ \bibinfo {author} {\bibfnamefont {N.}~\bibnamefont {S{\"o}rensson}},\ }\bibfield  {title} {\bibinfo {title} {An {E}xtensible {SAT}-solver},\ }in\ \href@noop {} {\emph {\bibinfo {booktitle} {Theory and {A}pplications of {S}atisfiability {T}esting}}},\ \bibinfo {editor} {edited by\ \bibinfo {editor} {\bibfnamefont {E.}~\bibnamefont {Giunchiglia}}\ and\ \bibinfo {editor} {\bibfnamefont {A.}~\bibnamefont {Tacchella}}}\ (\bibinfo  {publisher} {Springer Berlin Heidelberg},\ \bibinfo {address} {Berlin, Heidelberg},\ \bibinfo {year} {2004})\ pp.\ \bibinfo {pages} {502--518}\BibitemShut {NoStop}%
\bibitem [{\citenamefont {Ignatiev}\ \emph {et~al.}(2018)\citenamefont {Ignatiev}, \citenamefont {Morgado},\ and\ \citenamefont {Marques{-}Silva}}]{imms-sat18}%
  \BibitemOpen
  \bibfield  {author} {\bibinfo {author} {\bibfnamefont {A.}~\bibnamefont {Ignatiev}}, \bibinfo {author} {\bibfnamefont {A.}~\bibnamefont {Morgado}},\ and\ \bibinfo {author} {\bibfnamefont {J.}~\bibnamefont {Marques{-}Silva}},\ }\bibfield  {title} {\bibinfo {title} {{PySAT:} {A} {Python} {T}oolkit for {P}rototyping with {SAT} {O}racles},\ }in\ \href {https://doi.org/10.1007/978-3-319-94144-8_26} {\emph {\bibinfo {booktitle} {SAT}}}\ (\bibinfo {year} {2018})\ pp.\ \bibinfo {pages} {428--437}\BibitemShut {NoStop}%
\bibitem [{\citenamefont {Weidenfeller}\ \emph {et~al.}(2022)\citenamefont {Weidenfeller}, \citenamefont {Valor}, \citenamefont {Gacon}, \citenamefont {Tornow}, \citenamefont {Bello}, \citenamefont {Woerner},\ and\ \citenamefont {Egger}}]{Weidenfeller2022scalingofquantum}%
  \BibitemOpen
  \bibfield  {author} {\bibinfo {author} {\bibfnamefont {J.}~\bibnamefont {Weidenfeller}}, \bibinfo {author} {\bibfnamefont {L.~C.}\ \bibnamefont {Valor}}, \bibinfo {author} {\bibfnamefont {J.}~\bibnamefont {Gacon}}, \bibinfo {author} {\bibfnamefont {C.}~\bibnamefont {Tornow}}, \bibinfo {author} {\bibfnamefont {L.}~\bibnamefont {Bello}}, \bibinfo {author} {\bibfnamefont {S.}~\bibnamefont {Woerner}},\ and\ \bibinfo {author} {\bibfnamefont {D.~J.}\ \bibnamefont {Egger}},\ }\bibfield  {title} {\bibinfo {title} {Scaling of the quantum approximate optimization algorithm on superconducting qubit based hardware},\ }\href {https://doi.org/10.22331/q-2022-12-07-870} {\bibfield  {journal} {\bibinfo  {journal} {{Quantum}}\ }\textbf {\bibinfo {volume} {6}},\ \bibinfo {pages} {870} (\bibinfo {year} {2022})}\BibitemShut {NoStop}%
\bibitem [{\citenamefont {Sack}\ and\ \citenamefont {Egger}(2024{\natexlab{b}})}]{Sack2023}%
  \BibitemOpen
  \bibfield  {author} {\bibinfo {author} {\bibfnamefont {S.~H.}\ \bibnamefont {Sack}}\ and\ \bibinfo {author} {\bibfnamefont {D.~J.}\ \bibnamefont {Egger}},\ }\bibfield  {title} {\bibinfo {title} {Large-scale quantum approximate optimization on nonplanar graphs with machine learning noise mitigation},\ }\href {https://doi.org/10.1103/PhysRevResearch.6.013223} {\bibfield  {journal} {\bibinfo  {journal} {Phys. Rev. Res.}\ }\textbf {\bibinfo {volume} {6}},\ \bibinfo {pages} {013223} (\bibinfo {year} {2024}{\natexlab{b}})}\BibitemShut {NoStop}%
\bibitem [{\citenamefont {Wallman}\ and\ \citenamefont {Emerson}(2016)}]{PhysRevA.94.052325}%
  \BibitemOpen
  \bibfield  {author} {\bibinfo {author} {\bibfnamefont {J.~J.}\ \bibnamefont {Wallman}}\ and\ \bibinfo {author} {\bibfnamefont {J.}~\bibnamefont {Emerson}},\ }\bibfield  {title} {\bibinfo {title} {Noise tailoring for scalable quantum computation via randomized compiling},\ }\href {https://doi.org/10.1103/PhysRevA.94.052325} {\bibfield  {journal} {\bibinfo  {journal} {Phys. Rev. A}\ }\textbf {\bibinfo {volume} {94}},\ \bibinfo {pages} {052325} (\bibinfo {year} {2016})}\BibitemShut {NoStop}%
\bibitem [{\citenamefont {Sete}\ \emph {et~al.}(2024)\citenamefont {Sete}, \citenamefont {Tripathi}, \citenamefont {Valery}, \citenamefont {Lidar},\ and\ \citenamefont {Mutus}}]{Sete2024}%
  \BibitemOpen
  \bibfield  {author} {\bibinfo {author} {\bibfnamefont {E.~A.}\ \bibnamefont {Sete}}, \bibinfo {author} {\bibfnamefont {V.}~\bibnamefont {Tripathi}}, \bibinfo {author} {\bibfnamefont {J.~A.}\ \bibnamefont {Valery}}, \bibinfo {author} {\bibfnamefont {D.}~\bibnamefont {Lidar}},\ and\ \bibinfo {author} {\bibfnamefont {J.~Y.}\ \bibnamefont {Mutus}},\ }\bibfield  {title} {\bibinfo {title} {Error budget of a parametric resonance entangling gate with a tunable coupler},\ }\href {https://doi.org/10.1103/PhysRevApplied.22.014059} {\bibfield  {journal} {\bibinfo  {journal} {Phys. Rev. Appl.}\ }\textbf {\bibinfo {volume} {22}},\ \bibinfo {pages} {014059} (\bibinfo {year} {2024})}\BibitemShut {NoStop}%
\bibitem [{\citenamefont {Cai}\ and\ \citenamefont {Benjamin}(2019)}]{Cai2019}%
  \BibitemOpen
  \bibfield  {author} {\bibinfo {author} {\bibfnamefont {Z.}~\bibnamefont {Cai}}\ and\ \bibinfo {author} {\bibfnamefont {S.~C.}\ \bibnamefont {Benjamin}},\ }\bibfield  {title} {\bibinfo {title} {Constructing smaller pauli twirling sets for arbitrary error channels},\ }\href {https://doi.org/10.1038/s41598-019-46722-7} {\bibfield  {journal} {\bibinfo  {journal} {Sci. Rep.}\ }\textbf {\bibinfo {volume} {9}},\ \bibinfo {pages} {11281} (\bibinfo {year} {2019})}\BibitemShut {NoStop}%
\bibitem [{\citenamefont {Hashim}\ \emph {et~al.}(2021)\citenamefont {Hashim}, \citenamefont {Naik}, \citenamefont {Morvan}, \citenamefont {Ville}, \citenamefont {Mitchell}, \citenamefont {Kreikebaum}, \citenamefont {Davis}, \citenamefont {Smith}, \citenamefont {Iancu}, \citenamefont {O'Brien}, \citenamefont {Hincks}, \citenamefont {Wallman}, \citenamefont {Emerson},\ and\ \citenamefont {Siddiqi}}]{PhysRevX.11.041039}%
  \BibitemOpen
  \bibfield  {author} {\bibinfo {author} {\bibfnamefont {A.}~\bibnamefont {Hashim}}, \bibinfo {author} {\bibfnamefont {R.~K.}\ \bibnamefont {Naik}}, \bibinfo {author} {\bibfnamefont {A.}~\bibnamefont {Morvan}}, \bibinfo {author} {\bibfnamefont {J.-L.}\ \bibnamefont {Ville}}, \bibinfo {author} {\bibfnamefont {B.}~\bibnamefont {Mitchell}}, \bibinfo {author} {\bibfnamefont {J.~M.}\ \bibnamefont {Kreikebaum}}, \bibinfo {author} {\bibfnamefont {M.}~\bibnamefont {Davis}}, \bibinfo {author} {\bibfnamefont {E.}~\bibnamefont {Smith}}, \bibinfo {author} {\bibfnamefont {C.}~\bibnamefont {Iancu}}, \bibinfo {author} {\bibfnamefont {K.~P.}\ \bibnamefont {O'Brien}}, \bibinfo {author} {\bibfnamefont {I.}~\bibnamefont {Hincks}}, \bibinfo {author} {\bibfnamefont {J.~J.}\ \bibnamefont {Wallman}}, \bibinfo {author} {\bibfnamefont {J.}~\bibnamefont {Emerson}},\ and\ \bibinfo {author} {\bibfnamefont {I.}~\bibnamefont {Siddiqi}},\ }\bibfield  {title} {\bibinfo {title} {Randomized compiling for scalable quantum computing on a noisy
  superconducting quantum processor},\ }\href {https://doi.org/10.1103/PhysRevX.11.041039} {\bibfield  {journal} {\bibinfo  {journal} {Phys. Rev. X}\ }\textbf {\bibinfo {volume} {11}},\ \bibinfo {pages} {041039} (\bibinfo {year} {2021})}\BibitemShut {NoStop}%
\bibitem [{\citenamefont {Beale}\ \emph {et~al.}(2020)\citenamefont {Beale}, \citenamefont {Boone}, \citenamefont {Carignan-Dugas}, \citenamefont {Chytros}, \citenamefont {Dahlen}, \citenamefont {Dawkins}, \citenamefont {Emerson}, \citenamefont {Ferracin}, \citenamefont {Frey}, \citenamefont {Hincks}, \citenamefont {Hufnagel}, \citenamefont {Iyer}, \citenamefont {Jain}, \citenamefont {Kolbush}, \citenamefont {Ospadov}, \citenamefont {Pino}, \citenamefont {Qassim}, \citenamefont {Saunders}, \citenamefont {Skanes-Norman}, \citenamefont {Stasiuk}, \citenamefont {Wallman}, \citenamefont {Winick},\ and\ \citenamefont {Wright}}]{beale_2020_3945250}%
  \BibitemOpen
  \bibfield  {author} {\bibinfo {author} {\bibfnamefont {S.~J.}\ \bibnamefont {Beale}}, \bibinfo {author} {\bibfnamefont {K.}~\bibnamefont {Boone}}, \bibinfo {author} {\bibfnamefont {A.}~\bibnamefont {Carignan-Dugas}}, \bibinfo {author} {\bibfnamefont {A.}~\bibnamefont {Chytros}}, \bibinfo {author} {\bibfnamefont {D.}~\bibnamefont {Dahlen}}, \bibinfo {author} {\bibfnamefont {H.}~\bibnamefont {Dawkins}}, \bibinfo {author} {\bibfnamefont {J.}~\bibnamefont {Emerson}}, \bibinfo {author} {\bibfnamefont {S.}~\bibnamefont {Ferracin}}, \bibinfo {author} {\bibfnamefont {V.}~\bibnamefont {Frey}}, \bibinfo {author} {\bibfnamefont {I.}~\bibnamefont {Hincks}}, \bibinfo {author} {\bibfnamefont {D.}~\bibnamefont {Hufnagel}}, \bibinfo {author} {\bibfnamefont {P.}~\bibnamefont {Iyer}}, \bibinfo {author} {\bibfnamefont {A.}~\bibnamefont {Jain}}, \bibinfo {author} {\bibfnamefont {J.}~\bibnamefont {Kolbush}}, \bibinfo {author} {\bibfnamefont {E.}~\bibnamefont {Ospadov}}, \bibinfo {author} {\bibfnamefont {J.~L.}\ \bibnamefont
  {Pino}}, \bibinfo {author} {\bibfnamefont {H.}~\bibnamefont {Qassim}}, \bibinfo {author} {\bibfnamefont {J.}~\bibnamefont {Saunders}}, \bibinfo {author} {\bibfnamefont {J.}~\bibnamefont {Skanes-Norman}}, \bibinfo {author} {\bibfnamefont {A.}~\bibnamefont {Stasiuk}}, \bibinfo {author} {\bibfnamefont {J.~J.}\ \bibnamefont {Wallman}}, \bibinfo {author} {\bibfnamefont {A.}~\bibnamefont {Winick}},\ and\ \bibinfo {author} {\bibfnamefont {E.}~\bibnamefont {Wright}},\ }\href {https://doi.org/10.5281/zenodo.3945250} {\bibinfo {title} {True-q}} (\bibinfo {year} {2020})\BibitemShut {NoStop}%
\bibitem [{\citenamefont {Arute}\ \emph {et~al.}(2019)\citenamefont {Arute}, \citenamefont {Arya}, \citenamefont {Babbush}, \citenamefont {Bacon}, \citenamefont {Bardin}, \citenamefont {Barends}, \citenamefont {Biswas}, \citenamefont {Boixo}, \citenamefont {Brandao}, \citenamefont {Buell}, \citenamefont {Burkett}, \citenamefont {Chen}, \citenamefont {Chen}, \citenamefont {Chiaro}, \citenamefont {Collins}, \citenamefont {Courtney}, \citenamefont {Dunsworth}, \citenamefont {Farhi}, \citenamefont {Foxen}, \citenamefont {Fowler}, \citenamefont {Gidney}, \citenamefont {Giustina}, \citenamefont {Graff}, \citenamefont {Guerin}, \citenamefont {Habegger}, \citenamefont {Harrigan}, \citenamefont {Hartmann}, \citenamefont {Ho}, \citenamefont {Hoffmann}, \citenamefont {Huang}, \citenamefont {Humble}, \citenamefont {Isakov}, \citenamefont {Jeffrey}, \citenamefont {Jiang}, \citenamefont {Kafri}, \citenamefont {Kechedzhi}, \citenamefont {Kelly}, \citenamefont {Klimov}, \citenamefont {Knysh}, \citenamefont {Korotkov},
  \citenamefont {Kostritsa}, \citenamefont {Landhuis}, \citenamefont {Lindmark}, \citenamefont {Lucero}, \citenamefont {Lyakh}, \citenamefont {Mandr{\`a}}, \citenamefont {McClean}, \citenamefont {McEwen}, \citenamefont {Megrant}, \citenamefont {Mi}, \citenamefont {Michielsen}, \citenamefont {Mohseni}, \citenamefont {Mutus}, \citenamefont {Naaman}, \citenamefont {Neeley}, \citenamefont {Neill}, \citenamefont {Niu}, \citenamefont {Ostby}, \citenamefont {Petukhov}, \citenamefont {Platt}, \citenamefont {Quintana}, \citenamefont {Rieffel}, \citenamefont {Roushan}, \citenamefont {Rubin}, \citenamefont {Sank}, \citenamefont {Satzinger}, \citenamefont {Smelyanskiy}, \citenamefont {Sung}, \citenamefont {Trevithick}, \citenamefont {Vainsencher}, \citenamefont {Villalonga}, \citenamefont {White}, \citenamefont {Yao}, \citenamefont {Yeh}, \citenamefont {Zalcman}, \citenamefont {Neven},\ and\ \citenamefont {Martinis}}]{Arute2019}%
  \BibitemOpen
  \bibfield  {author} {\bibinfo {author} {\bibfnamefont {F.}~\bibnamefont {Arute}}, \bibinfo {author} {\bibfnamefont {K.}~\bibnamefont {Arya}}, \bibinfo {author} {\bibfnamefont {R.}~\bibnamefont {Babbush}}, \bibinfo {author} {\bibfnamefont {D.}~\bibnamefont {Bacon}}, \bibinfo {author} {\bibfnamefont {J.~C.}\ \bibnamefont {Bardin}}, \bibinfo {author} {\bibfnamefont {R.}~\bibnamefont {Barends}}, \bibinfo {author} {\bibfnamefont {R.}~\bibnamefont {Biswas}}, \bibinfo {author} {\bibfnamefont {S.}~\bibnamefont {Boixo}}, \bibinfo {author} {\bibfnamefont {F.~G. S.~L.}\ \bibnamefont {Brandao}}, \bibinfo {author} {\bibfnamefont {D.~A.}\ \bibnamefont {Buell}}, \bibinfo {author} {\bibfnamefont {B.}~\bibnamefont {Burkett}}, \bibinfo {author} {\bibfnamefont {Y.}~\bibnamefont {Chen}}, \bibinfo {author} {\bibfnamefont {Z.}~\bibnamefont {Chen}}, \bibinfo {author} {\bibfnamefont {B.}~\bibnamefont {Chiaro}}, \bibinfo {author} {\bibfnamefont {R.}~\bibnamefont {Collins}}, \bibinfo {author} {\bibfnamefont {W.}~\bibnamefont
  {Courtney}}, \bibinfo {author} {\bibfnamefont {A.}~\bibnamefont {Dunsworth}}, \bibinfo {author} {\bibfnamefont {E.}~\bibnamefont {Farhi}}, \bibinfo {author} {\bibfnamefont {B.}~\bibnamefont {Foxen}}, \bibinfo {author} {\bibfnamefont {A.}~\bibnamefont {Fowler}}, \bibinfo {author} {\bibfnamefont {C.}~\bibnamefont {Gidney}}, \bibinfo {author} {\bibfnamefont {M.}~\bibnamefont {Giustina}}, \bibinfo {author} {\bibfnamefont {R.}~\bibnamefont {Graff}}, \bibinfo {author} {\bibfnamefont {K.}~\bibnamefont {Guerin}}, \bibinfo {author} {\bibfnamefont {S.}~\bibnamefont {Habegger}}, \bibinfo {author} {\bibfnamefont {M.~P.}\ \bibnamefont {Harrigan}}, \bibinfo {author} {\bibfnamefont {M.~J.}\ \bibnamefont {Hartmann}}, \bibinfo {author} {\bibfnamefont {A.}~\bibnamefont {Ho}}, \bibinfo {author} {\bibfnamefont {M.}~\bibnamefont {Hoffmann}}, \bibinfo {author} {\bibfnamefont {T.}~\bibnamefont {Huang}}, \bibinfo {author} {\bibfnamefont {T.~S.}\ \bibnamefont {Humble}}, \bibinfo {author} {\bibfnamefont {S.~V.}\ \bibnamefont
  {Isakov}}, \bibinfo {author} {\bibfnamefont {E.}~\bibnamefont {Jeffrey}}, \bibinfo {author} {\bibfnamefont {Z.}~\bibnamefont {Jiang}}, \bibinfo {author} {\bibfnamefont {D.}~\bibnamefont {Kafri}}, \bibinfo {author} {\bibfnamefont {K.}~\bibnamefont {Kechedzhi}}, \bibinfo {author} {\bibfnamefont {J.}~\bibnamefont {Kelly}}, \bibinfo {author} {\bibfnamefont {P.~V.}\ \bibnamefont {Klimov}}, \bibinfo {author} {\bibfnamefont {S.}~\bibnamefont {Knysh}}, \bibinfo {author} {\bibfnamefont {A.}~\bibnamefont {Korotkov}}, \bibinfo {author} {\bibfnamefont {F.}~\bibnamefont {Kostritsa}}, \bibinfo {author} {\bibfnamefont {D.}~\bibnamefont {Landhuis}}, \bibinfo {author} {\bibfnamefont {M.}~\bibnamefont {Lindmark}}, \bibinfo {author} {\bibfnamefont {E.}~\bibnamefont {Lucero}}, \bibinfo {author} {\bibfnamefont {D.}~\bibnamefont {Lyakh}}, \bibinfo {author} {\bibfnamefont {S.}~\bibnamefont {Mandr{\`a}}}, \bibinfo {author} {\bibfnamefont {J.~R.}\ \bibnamefont {McClean}}, \bibinfo {author} {\bibfnamefont {M.}~\bibnamefont
  {McEwen}}, \bibinfo {author} {\bibfnamefont {A.}~\bibnamefont {Megrant}}, \bibinfo {author} {\bibfnamefont {X.}~\bibnamefont {Mi}}, \bibinfo {author} {\bibfnamefont {K.}~\bibnamefont {Michielsen}}, \bibinfo {author} {\bibfnamefont {M.}~\bibnamefont {Mohseni}}, \bibinfo {author} {\bibfnamefont {J.}~\bibnamefont {Mutus}}, \bibinfo {author} {\bibfnamefont {O.}~\bibnamefont {Naaman}}, \bibinfo {author} {\bibfnamefont {M.}~\bibnamefont {Neeley}}, \bibinfo {author} {\bibfnamefont {C.}~\bibnamefont {Neill}}, \bibinfo {author} {\bibfnamefont {M.~Y.}\ \bibnamefont {Niu}}, \bibinfo {author} {\bibfnamefont {E.}~\bibnamefont {Ostby}}, \bibinfo {author} {\bibfnamefont {A.}~\bibnamefont {Petukhov}}, \bibinfo {author} {\bibfnamefont {J.~C.}\ \bibnamefont {Platt}}, \bibinfo {author} {\bibfnamefont {C.}~\bibnamefont {Quintana}}, \bibinfo {author} {\bibfnamefont {E.~G.}\ \bibnamefont {Rieffel}}, \bibinfo {author} {\bibfnamefont {P.}~\bibnamefont {Roushan}}, \bibinfo {author} {\bibfnamefont {N.~C.}\ \bibnamefont {Rubin}},
  \bibinfo {author} {\bibfnamefont {D.}~\bibnamefont {Sank}}, \bibinfo {author} {\bibfnamefont {K.~J.}\ \bibnamefont {Satzinger}}, \bibinfo {author} {\bibfnamefont {V.}~\bibnamefont {Smelyanskiy}}, \bibinfo {author} {\bibfnamefont {K.~J.}\ \bibnamefont {Sung}}, \bibinfo {author} {\bibfnamefont {M.~D.}\ \bibnamefont {Trevithick}}, \bibinfo {author} {\bibfnamefont {A.}~\bibnamefont {Vainsencher}}, \bibinfo {author} {\bibfnamefont {B.}~\bibnamefont {Villalonga}}, \bibinfo {author} {\bibfnamefont {T.}~\bibnamefont {White}}, \bibinfo {author} {\bibfnamefont {Z.~J.}\ \bibnamefont {Yao}}, \bibinfo {author} {\bibfnamefont {P.}~\bibnamefont {Yeh}}, \bibinfo {author} {\bibfnamefont {A.}~\bibnamefont {Zalcman}}, \bibinfo {author} {\bibfnamefont {H.}~\bibnamefont {Neven}},\ and\ \bibinfo {author} {\bibfnamefont {J.~M.}\ \bibnamefont {Martinis}},\ }\bibfield  {title} {\bibinfo {title} {Quantum supremacy using a programmable superconducting processor},\ }\href {https://doi.org/10.1038/s41586-019-1666-5} {\bibfield
  {journal} {\bibinfo  {journal} {Nature}\ }\textbf {\bibinfo {volume} {574}},\ \bibinfo {pages} {505} (\bibinfo {year} {2019})}\BibitemShut {NoStop}%
\bibitem [{\citenamefont {Wu}\ \emph {et~al.}(2021)\citenamefont {Wu}, \citenamefont {Bao}, \citenamefont {Cao}, \citenamefont {Chen}, \citenamefont {Chen}, \citenamefont {Chen}, \citenamefont {Chung}, \citenamefont {Deng}, \citenamefont {Du}, \citenamefont {Fan}, \citenamefont {Gong}, \citenamefont {Guo}, \citenamefont {Guo}, \citenamefont {Guo}, \citenamefont {Han}, \citenamefont {Hong}, \citenamefont {Huang}, \citenamefont {Huo}, \citenamefont {Li}, \citenamefont {Li}, \citenamefont {Li}, \citenamefont {Li}, \citenamefont {Liang}, \citenamefont {Lin}, \citenamefont {Lin}, \citenamefont {Qian}, \citenamefont {Qiao}, \citenamefont {Rong}, \citenamefont {Su}, \citenamefont {Sun}, \citenamefont {Wang}, \citenamefont {Wang}, \citenamefont {Wu}, \citenamefont {Xu}, \citenamefont {Yan}, \citenamefont {Yang}, \citenamefont {Yang}, \citenamefont {Ye}, \citenamefont {Yin}, \citenamefont {Ying}, \citenamefont {Yu}, \citenamefont {Zha}, \citenamefont {Zhang}, \citenamefont {Zhang}, \citenamefont {Zhang}, \citenamefont
  {Zhang}, \citenamefont {Zhao}, \citenamefont {Zhao}, \citenamefont {Zhou}, \citenamefont {Zhu}, \citenamefont {Lu}, \citenamefont {Peng}, \citenamefont {Zhu},\ and\ \citenamefont {Pan}}]{PhysRevLett.127.180501}%
  \BibitemOpen
  \bibfield  {author} {\bibinfo {author} {\bibfnamefont {Y.}~\bibnamefont {Wu}}, \bibinfo {author} {\bibfnamefont {W.-S.}\ \bibnamefont {Bao}}, \bibinfo {author} {\bibfnamefont {S.}~\bibnamefont {Cao}}, \bibinfo {author} {\bibfnamefont {F.}~\bibnamefont {Chen}}, \bibinfo {author} {\bibfnamefont {M.-C.}\ \bibnamefont {Chen}}, \bibinfo {author} {\bibfnamefont {X.}~\bibnamefont {Chen}}, \bibinfo {author} {\bibfnamefont {T.-H.}\ \bibnamefont {Chung}}, \bibinfo {author} {\bibfnamefont {H.}~\bibnamefont {Deng}}, \bibinfo {author} {\bibfnamefont {Y.}~\bibnamefont {Du}}, \bibinfo {author} {\bibfnamefont {D.}~\bibnamefont {Fan}}, \bibinfo {author} {\bibfnamefont {M.}~\bibnamefont {Gong}}, \bibinfo {author} {\bibfnamefont {C.}~\bibnamefont {Guo}}, \bibinfo {author} {\bibfnamefont {C.}~\bibnamefont {Guo}}, \bibinfo {author} {\bibfnamefont {S.}~\bibnamefont {Guo}}, \bibinfo {author} {\bibfnamefont {L.}~\bibnamefont {Han}}, \bibinfo {author} {\bibfnamefont {L.}~\bibnamefont {Hong}}, \bibinfo {author} {\bibfnamefont
  {H.-L.}\ \bibnamefont {Huang}}, \bibinfo {author} {\bibfnamefont {Y.-H.}\ \bibnamefont {Huo}}, \bibinfo {author} {\bibfnamefont {L.}~\bibnamefont {Li}}, \bibinfo {author} {\bibfnamefont {N.}~\bibnamefont {Li}}, \bibinfo {author} {\bibfnamefont {S.}~\bibnamefont {Li}}, \bibinfo {author} {\bibfnamefont {Y.}~\bibnamefont {Li}}, \bibinfo {author} {\bibfnamefont {F.}~\bibnamefont {Liang}}, \bibinfo {author} {\bibfnamefont {C.}~\bibnamefont {Lin}}, \bibinfo {author} {\bibfnamefont {J.}~\bibnamefont {Lin}}, \bibinfo {author} {\bibfnamefont {H.}~\bibnamefont {Qian}}, \bibinfo {author} {\bibfnamefont {D.}~\bibnamefont {Qiao}}, \bibinfo {author} {\bibfnamefont {H.}~\bibnamefont {Rong}}, \bibinfo {author} {\bibfnamefont {H.}~\bibnamefont {Su}}, \bibinfo {author} {\bibfnamefont {L.}~\bibnamefont {Sun}}, \bibinfo {author} {\bibfnamefont {L.}~\bibnamefont {Wang}}, \bibinfo {author} {\bibfnamefont {S.}~\bibnamefont {Wang}}, \bibinfo {author} {\bibfnamefont {D.}~\bibnamefont {Wu}}, \bibinfo {author} {\bibfnamefont
  {Y.}~\bibnamefont {Xu}}, \bibinfo {author} {\bibfnamefont {K.}~\bibnamefont {Yan}}, \bibinfo {author} {\bibfnamefont {W.}~\bibnamefont {Yang}}, \bibinfo {author} {\bibfnamefont {Y.}~\bibnamefont {Yang}}, \bibinfo {author} {\bibfnamefont {Y.}~\bibnamefont {Ye}}, \bibinfo {author} {\bibfnamefont {J.}~\bibnamefont {Yin}}, \bibinfo {author} {\bibfnamefont {C.}~\bibnamefont {Ying}}, \bibinfo {author} {\bibfnamefont {J.}~\bibnamefont {Yu}}, \bibinfo {author} {\bibfnamefont {C.}~\bibnamefont {Zha}}, \bibinfo {author} {\bibfnamefont {C.}~\bibnamefont {Zhang}}, \bibinfo {author} {\bibfnamefont {H.}~\bibnamefont {Zhang}}, \bibinfo {author} {\bibfnamefont {K.}~\bibnamefont {Zhang}}, \bibinfo {author} {\bibfnamefont {Y.}~\bibnamefont {Zhang}}, \bibinfo {author} {\bibfnamefont {H.}~\bibnamefont {Zhao}}, \bibinfo {author} {\bibfnamefont {Y.}~\bibnamefont {Zhao}}, \bibinfo {author} {\bibfnamefont {L.}~\bibnamefont {Zhou}}, \bibinfo {author} {\bibfnamefont {Q.}~\bibnamefont {Zhu}}, \bibinfo {author} {\bibfnamefont {C.-Y.}\
  \bibnamefont {Lu}}, \bibinfo {author} {\bibfnamefont {C.-Z.}\ \bibnamefont {Peng}}, \bibinfo {author} {\bibfnamefont {X.}~\bibnamefont {Zhu}},\ and\ \bibinfo {author} {\bibfnamefont {J.-W.}\ \bibnamefont {Pan}},\ }\bibfield  {title} {\bibinfo {title} {Strong quantum computational advantage using a superconducting quantum processor},\ }\href {https://doi.org/10.1103/PhysRevLett.127.180501} {\bibfield  {journal} {\bibinfo  {journal} {Phys. Rev. Lett.}\ }\textbf {\bibinfo {volume} {127}},\ \bibinfo {pages} {180501} (\bibinfo {year} {2021})}\BibitemShut {NoStop}%
\bibitem [{\citenamefont {Kim}\ \emph {et~al.}(2023)\citenamefont {Kim}, \citenamefont {Eddins}, \citenamefont {Anand}, \citenamefont {Wei}, \citenamefont {van~den Berg}, \citenamefont {Rosenblatt}, \citenamefont {Nayfeh}, \citenamefont {Wu}, \citenamefont {Zaletel}, \citenamefont {Temme},\ and\ \citenamefont {Kandala}}]{Kim2023}%
  \BibitemOpen
  \bibfield  {author} {\bibinfo {author} {\bibfnamefont {Y.}~\bibnamefont {Kim}}, \bibinfo {author} {\bibfnamefont {A.}~\bibnamefont {Eddins}}, \bibinfo {author} {\bibfnamefont {S.}~\bibnamefont {Anand}}, \bibinfo {author} {\bibfnamefont {K.~X.}\ \bibnamefont {Wei}}, \bibinfo {author} {\bibfnamefont {E.}~\bibnamefont {van~den Berg}}, \bibinfo {author} {\bibfnamefont {S.}~\bibnamefont {Rosenblatt}}, \bibinfo {author} {\bibfnamefont {H.}~\bibnamefont {Nayfeh}}, \bibinfo {author} {\bibfnamefont {Y.}~\bibnamefont {Wu}}, \bibinfo {author} {\bibfnamefont {M.}~\bibnamefont {Zaletel}}, \bibinfo {author} {\bibfnamefont {K.}~\bibnamefont {Temme}},\ and\ \bibinfo {author} {\bibfnamefont {A.}~\bibnamefont {Kandala}},\ }\bibfield  {title} {\bibinfo {title} {Evidence for the utility of quantum computing before fault tolerance},\ }\href {https://doi.org/10.1038/s41586-023-06096-3} {\bibfield  {journal} {\bibinfo  {journal} {Nature}\ }\textbf {\bibinfo {volume} {618}},\ \bibinfo {pages} {500} (\bibinfo {year}
  {2023})}\BibitemShut {NoStop}%
\bibitem [{\citenamefont {Nielsen}\ and\ \citenamefont {Chuang}(2010)}]{Nielsen2011}%
  \BibitemOpen
  \bibfield  {author} {\bibinfo {author} {\bibfnamefont {M.~A.}\ \bibnamefont {Nielsen}}\ and\ \bibinfo {author} {\bibfnamefont {I.~L.}\ \bibnamefont {Chuang}},\ }\href@noop {} {\emph {\bibinfo {title} {Quantum Computation and Quantum Information: 10th Anniversary Edition}}},\ \bibinfo {edition} {10th}\ ed.\ (\bibinfo  {publisher} {Cambridge University Press},\ \bibinfo {year} {2010})\BibitemShut {NoStop}%
\bibitem [{\citenamefont {Harris}\ \emph {et~al.}(2020)\citenamefont {Harris}, \citenamefont {Millman}, \citenamefont {van~der Walt}, \citenamefont {Gommers}, \citenamefont {Virtanen}, \citenamefont {Cournapeau}, \citenamefont {Wieser}, \citenamefont {Taylor}, \citenamefont {Berg}, \citenamefont {Smith}, \citenamefont {Kern}, \citenamefont {Picus}, \citenamefont {Hoyer}, \citenamefont {van Kerkwijk}, \citenamefont {Brett}, \citenamefont {Haldane}, \citenamefont {del R{\'{i}}o}, \citenamefont {Wiebe}, \citenamefont {Peterson}, \citenamefont {G{\'{e}}rard-Marchant}, \citenamefont {Sheppard}, \citenamefont {Reddy}, \citenamefont {Weckesser}, \citenamefont {Abbasi}, \citenamefont {Gohlke},\ and\ \citenamefont {Oliphant}}]{harris2020array}%
  \BibitemOpen
  \bibfield  {author} {\bibinfo {author} {\bibfnamefont {C.~R.}\ \bibnamefont {Harris}}, \bibinfo {author} {\bibfnamefont {K.~J.}\ \bibnamefont {Millman}}, \bibinfo {author} {\bibfnamefont {S.~J.}\ \bibnamefont {van~der Walt}}, \bibinfo {author} {\bibfnamefont {R.}~\bibnamefont {Gommers}}, \bibinfo {author} {\bibfnamefont {P.}~\bibnamefont {Virtanen}}, \bibinfo {author} {\bibfnamefont {D.}~\bibnamefont {Cournapeau}}, \bibinfo {author} {\bibfnamefont {E.}~\bibnamefont {Wieser}}, \bibinfo {author} {\bibfnamefont {J.}~\bibnamefont {Taylor}}, \bibinfo {author} {\bibfnamefont {S.}~\bibnamefont {Berg}}, \bibinfo {author} {\bibfnamefont {N.~J.}\ \bibnamefont {Smith}}, \bibinfo {author} {\bibfnamefont {R.}~\bibnamefont {Kern}}, \bibinfo {author} {\bibfnamefont {M.}~\bibnamefont {Picus}}, \bibinfo {author} {\bibfnamefont {S.}~\bibnamefont {Hoyer}}, \bibinfo {author} {\bibfnamefont {M.~H.}\ \bibnamefont {van Kerkwijk}}, \bibinfo {author} {\bibfnamefont {M.}~\bibnamefont {Brett}}, \bibinfo {author} {\bibfnamefont
  {A.}~\bibnamefont {Haldane}}, \bibinfo {author} {\bibfnamefont {J.~F.}\ \bibnamefont {del R{\'{i}}o}}, \bibinfo {author} {\bibfnamefont {M.}~\bibnamefont {Wiebe}}, \bibinfo {author} {\bibfnamefont {P.}~\bibnamefont {Peterson}}, \bibinfo {author} {\bibfnamefont {P.}~\bibnamefont {G{\'{e}}rard-Marchant}}, \bibinfo {author} {\bibfnamefont {K.}~\bibnamefont {Sheppard}}, \bibinfo {author} {\bibfnamefont {T.}~\bibnamefont {Reddy}}, \bibinfo {author} {\bibfnamefont {W.}~\bibnamefont {Weckesser}}, \bibinfo {author} {\bibfnamefont {H.}~\bibnamefont {Abbasi}}, \bibinfo {author} {\bibfnamefont {C.}~\bibnamefont {Gohlke}},\ and\ \bibinfo {author} {\bibfnamefont {T.~E.}\ \bibnamefont {Oliphant}},\ }\bibfield  {title} {\bibinfo {title} {Array programming with {NumPy}},\ }\href {https://doi.org/10.1038/s41586-020-2649-2} {\bibfield  {journal} {\bibinfo  {journal} {Nature}\ }\textbf {\bibinfo {volume} {585}},\ \bibinfo {pages} {357} (\bibinfo {year} {2020})}\BibitemShut {NoStop}%
\bibitem [{\citenamefont {Lam}\ \emph {et~al.}(2015)\citenamefont {Lam}, \citenamefont {Pitrou},\ and\ \citenamefont {Seibert}}]{lam2015numba}%
  \BibitemOpen
  \bibfield  {author} {\bibinfo {author} {\bibfnamefont {S.~K.}\ \bibnamefont {Lam}}, \bibinfo {author} {\bibfnamefont {A.}~\bibnamefont {Pitrou}},\ and\ \bibinfo {author} {\bibfnamefont {S.}~\bibnamefont {Seibert}},\ }\bibfield  {title} {\bibinfo {title} {Numba: {A} llvm-based python jit compiler},\ }in\ \href@noop {} {\emph {\bibinfo {booktitle} {Proceedings of the Second Workshop on the LLVM Compiler Infrastructure in HPC}}}\ (\bibinfo {year} {2015})\ pp.\ \bibinfo {pages} {1--6}\BibitemShut {NoStop}%
\bibitem [{\citenamefont {Chen}\ \emph {et~al.}(2021)\citenamefont {Chen}, \citenamefont {Chi}, \citenamefont {Fan},\ and\ \citenamefont {Ma}}]{Chen2021}%
  \BibitemOpen
  \bibfield  {author} {\bibinfo {author} {\bibfnamefont {Y.}~\bibnamefont {Chen}}, \bibinfo {author} {\bibfnamefont {Y.}~\bibnamefont {Chi}}, \bibinfo {author} {\bibfnamefont {J.}~\bibnamefont {Fan}},\ and\ \bibinfo {author} {\bibfnamefont {C.}~\bibnamefont {Ma}},\ }\bibfield  {title} {\bibinfo {title} {Spectral {M}ethods for {D}ata {S}cience: {A} {S}tatistical {P}erspective},\ }\href {https://doi.org/10.1561/2200000079} {\bibfield  {journal} {\bibinfo  {journal} {Found. Trends Mach. Learn.}\ }\textbf {\bibinfo {volume} {14}},\ \bibinfo {pages} {566} (\bibinfo {year} {2021})}\BibitemShut {NoStop}%
\bibitem [{\citenamefont {Bravyi}\ \emph {et~al.}(2020)\citenamefont {Bravyi}, \citenamefont {Kliesch}, \citenamefont {Koenig},\ and\ \citenamefont {Tang}}]{PhysRevLett.125.260505}%
  \BibitemOpen
  \bibfield  {author} {\bibinfo {author} {\bibfnamefont {S.}~\bibnamefont {Bravyi}}, \bibinfo {author} {\bibfnamefont {A.}~\bibnamefont {Kliesch}}, \bibinfo {author} {\bibfnamefont {R.}~\bibnamefont {Koenig}},\ and\ \bibinfo {author} {\bibfnamefont {E.}~\bibnamefont {Tang}},\ }\bibfield  {title} {\bibinfo {title} {Obstacles to variational quantum optimization from symmetry protection},\ }\href {https://doi.org/10.1103/PhysRevLett.125.260505} {\bibfield  {journal} {\bibinfo  {journal} {Phys. Rev. Lett.}\ }\textbf {\bibinfo {volume} {125}},\ \bibinfo {pages} {260505} (\bibinfo {year} {2020})}\BibitemShut {NoStop}%
\bibitem [{\citenamefont {Brady}\ and\ \citenamefont {Hadfield}(2024)}]{Brady2023}%
  \BibitemOpen
  \bibfield  {author} {\bibinfo {author} {\bibfnamefont {L.~T.}\ \bibnamefont {Brady}}\ and\ \bibinfo {author} {\bibfnamefont {S.}~\bibnamefont {Hadfield}},\ }\bibfield  {title} {\bibinfo {title} {Iterative quantum algorithms for maximum independent set},\ }\href {https://doi.org/10.1103/PhysRevA.110.052435} {\bibfield  {journal} {\bibinfo  {journal} {Phys. Rev. A}\ }\textbf {\bibinfo {volume} {110}},\ \bibinfo {pages} {052435} (\bibinfo {year} {2024})}\BibitemShut {NoStop}%
\bibitem [{\citenamefont {Karalekas}\ \emph {et~al.}(2020)\citenamefont {Karalekas}, \citenamefont {Tezak}, \citenamefont {Peterson}, \citenamefont {Ryan}, \citenamefont {da~Silva},\ and\ \citenamefont {Smith}}]{Karalekas_2020}%
  \BibitemOpen
  \bibfield  {author} {\bibinfo {author} {\bibfnamefont {P.~J.}\ \bibnamefont {Karalekas}}, \bibinfo {author} {\bibfnamefont {N.~A.}\ \bibnamefont {Tezak}}, \bibinfo {author} {\bibfnamefont {E.~C.}\ \bibnamefont {Peterson}}, \bibinfo {author} {\bibfnamefont {C.~A.}\ \bibnamefont {Ryan}}, \bibinfo {author} {\bibfnamefont {M.~P.}\ \bibnamefont {da~Silva}},\ and\ \bibinfo {author} {\bibfnamefont {R.~S.}\ \bibnamefont {Smith}},\ }\bibfield  {title} {\bibinfo {title} {A quantum-classical cloud platform optimized for variational hybrid algorithms},\ }\href {https://doi.org/10.1088/2058-9565/ab7559} {\bibfield  {journal} {\bibinfo  {journal} {Quantum Sci. Technol.}\ }\textbf {\bibinfo {volume} {5}},\ \bibinfo {pages} {024003} (\bibinfo {year} {2020})}\BibitemShut {NoStop}%
\bibitem [{\citenamefont {Berman}\ and\ \citenamefont {Karpinski}(1999)}]{10.1007/3-540-48523-6_17}%
  \BibitemOpen
  \bibfield  {author} {\bibinfo {author} {\bibfnamefont {P.}~\bibnamefont {Berman}}\ and\ \bibinfo {author} {\bibfnamefont {M.}~\bibnamefont {Karpinski}},\ }\bibfield  {title} {\bibinfo {title} {On {S}ome {T}ighter {I}napproximability {R}esults ({E}xtended {A}bstract)},\ }in\ \href@noop {} {\emph {\bibinfo {booktitle} {Automata, Languages and Programming}}},\ \bibinfo {editor} {edited by\ \bibinfo {editor} {\bibfnamefont {J.}~\bibnamefont {Wiedermann}}, \bibinfo {editor} {\bibfnamefont {P.}~\bibnamefont {van Emde~Boas}},\ and\ \bibinfo {editor} {\bibfnamefont {M.}~\bibnamefont {Nielsen}}}\ (\bibinfo  {publisher} {Springer Berlin Heidelberg},\ \bibinfo {address} {Berlin, Heidelberg},\ \bibinfo {year} {1999})\ pp.\ \bibinfo {pages} {200--209}\BibitemShut {NoStop}%
\bibitem [{\citenamefont {Halperin}\ \emph {et~al.}(2004)\citenamefont {Halperin}, \citenamefont {Livnat},\ and\ \citenamefont {Zwick}}]{HALPERIN2004169}%
  \BibitemOpen
  \bibfield  {author} {\bibinfo {author} {\bibfnamefont {E.}~\bibnamefont {Halperin}}, \bibinfo {author} {\bibfnamefont {D.}~\bibnamefont {Livnat}},\ and\ \bibinfo {author} {\bibfnamefont {U.}~\bibnamefont {Zwick}},\ }\bibfield  {title} {\bibinfo {title} {{MAX CUT} in cubic graphs},\ }\href {https://doi.org/https://doi.org/10.1016/j.jalgor.2004.06.001} {\bibfield  {journal} {\bibinfo  {journal} {J. Algorithms}\ }\textbf {\bibinfo {volume} {53}},\ \bibinfo {pages} {169} (\bibinfo {year} {2004})}\BibitemShut {NoStop}%
\end{thebibliography}%
\let\addcontentsline\oldaddcontentsline

\end{document}